\documentclass[]{aa}  

\usepackage{graphicx}

\usepackage{txfonts}
\usepackage{hyperref}

\usepackage{amsmath}
\usepackage{comment}
\usepackage{cleveref}
\usepackage{subfigure}
\usepackage{xcolor}
\usepackage{placeins}
\usepackage{capt-of}
\usepackage{graphicx}
\usepackage[percent]{overpic}
\usepackage{tikz}

\usepackage{savesym}
\savesymbol{tablenum}
\usepackage{siunitx}
\restoresymbol{SIX}{tablenum}

\newcommand{\revision}[1]{{#1}}

\newcommand{\moltwelveco}{$^{12}$CO}
\newcommand{\molthirteenco}{$^{13}$CO}
\newcommand{\molceighteeno}{C$^{18}$O}

\def\app#1#2{%
  \mathrel{%
    \setbox0=\hbox{$#1\sim$}%
    \setbox2=\hbox{%
      \rlap{\hbox{$#1\propto$}}%
      \lower1.1\ht0\box0%
    }%
    \raise0.25\ht2\box2%
  }%
}
\def\approxprop{\mathpalette\app\relax}

\begin{document}

    \title{Spiral excitation in protoplanetary disks through gap-edge illumination}

   \subtitle{Distinctive kinematic signatures in CO isotopologues}

   \author{Dhruv Muley \inst{1, 2}
        \and León-Alexander Hühn \inst{3}
        \and Haochang Jiang \inst{1}
        \and David Melon Fuksman \inst{1}}

   \institute{Max-Planck-Institut f\"ur Astronomie, Königstuhl 17, Heidelberg, DE 69117
    \and
    Max-Planck-Institut f\"ur Astrophysik, Karl-Schwarzschild-Straße 1, Garching bei M\"unchen, DE 85748\\
    \email{\href{mailto:dmuley@mpa-garching.mpg.de}{dmuley@mpa-garching.mpg.de}}
    \and
    Institut für Theoretische Astrophysik, Ruprecht-Karls Universität Heidelberg, Albert-Ueberle-Straße 2, Heidelberg, DE 69120\\
    \email{\href{mailto:huehn@uni-heidelberg.de}{huehn@uni-heidelberg.de}}
              }
   \date{Received \revision{2 November 2025}; accepted 27 November 2025}


  \abstract{High-resolution, near-infrared observations have revealed prominent, two-armed spirals in a multitude of systems, such as MWC~758, SAO~206462, and V1247~Ori. Alongside the classical theory of disk-companion interaction, shadow-based driving has come into vogue as a potential explanation for such large-scale substructures. How might these two mechanisms be distinguished from one another in observations? To investigate this question, we ran a pair of hydrodynamical simulations with \texttt{PLUTO}. One, with full radiation hydrodynamics and gas-grain collision, was designed to develop shadow-driven spirals at the outer gap edge of a sub-thermal, Saturn-mass planet. The other, with parametrized $\beta$-cooling, was set up to capture the more standard view of spiral wave excitation by a super-thermal, multi-Jupiter-mass, exterior planetary companion. Post-processing of these simulations with the Monte Carlo radiative transfer (MCRT) code \texttt{RADMC3D} revealed that strong vertical velocities in the shadow-driven case create a prominent two-armed feature in the moment-1 CO maps, particularly when the disk is viewed face-on in optically thicker isotopologues; such a feature is not seen in the standard planet-driven case. Conversely, the presence or absence of such signatures in two-armed spiral systems would distinguish those potentially driven by exterior, multi-Jupiter-mass companions, and thus help identify promising targets for future direct-imaging campaigns.}


   \maketitle
%

\section{Introduction}

Detailed, multiwavelength observations of circumstellar disks around forming stars---enabled by instruments such as ALMA and VLT/SPHERE---have demonstrated that they contain a rich diversity of substructure, including gaps, rings \citep[e.g.,][]{Andrews2018} and spirals \citep{Teague2019,Teague2022} in both their gas and dust components. The traditional explanation, which has been the subject of numerous analytical \citep[e.g.,][]{Goldreich1978,Goldreich1979,Goldreich1980} and numerical \citep{FSC14,Fung2015,Zhu2015,Zhang2024} studies over the decades, is that such features result from gravitational interaction between the disk and nascent planetary companions: the spirals from the supersonic motion of the planetary potential through the disk gas, and the gaps and rings the cumulative result of angular momentum exchange through these spirals over thousands of planetary orbits. Among other mechanisms, such as photoevaporation \citep{Ercolano2022}, magnetohydrodynamical zonal flows \citep{Hu2022}, infalling material \citep{Kuznetsova2022,Calcino2025}, and the gravitational instability \citep[e.g.,][]{Dong2015,Bethune2021}, recent simulations \citep[among others, the work of][]{Montesinos2016,Montesinos2018,Su2024}, including some with full radiation hydrodynamics \citep[e.g.,][]{Zhang2024_1,Ziampras2025}, have demonstrated that the temperature variations created by shadows on the disk are also capable of driving substructure. Indeed, disk-shadow interaction follows an analytical theory which bears many similarities to that of disk-planet interaction \citep{Zhu2025}.

Of particular interest in this context are the various observed double-armed spirals in near-infrared (NIR) scattered light \citep[e.g.,][]{Shuai2022}, exemplified by systems such as MWC 758 \citep{Grady2013}, SAO 206462/HD 135344B \citep{Muto2012}, and V1247 Ori \citep{Ren2024}. According to the standard theory of disk-planet interaction, these two-armed spirals would indicate the presence of an exterior, multi-Jupiter-mass driver \citep{Dong2017,Bae2018a,Bae2018b}. In a companion paper to this work \citep{Muley2024}, however, we have demonstrated with 3D radiation-hydrodynamical simulations that shadowing effects at the outer edges of planet-carved gaps are also capable of driving large-scale, two-armed spirals. For this alternative hypothesis, the \revision{minimum} masses (Saturn-to-Jupiter mass) and semimajor axes (interior, rather than exterior, to the observed spirals) are far smaller than required classically.

Advances in near-infrared angular differential imaging (ADI) \citep{AsensioTorres2021, Desidera2021}, mid-infrared \textit{James Webb Space Telescope} (JWST) MIRI \citep[e.g.,][]{Wagner2019,Wagner2024,Cugno2024}, and spectroscopy of accretion tracers such as $H\alpha$ and $Pa\beta$ \citep{Follette2023,Biddle2024} mean that the massive, external companions predicted by the classical theory ought to be relatively accessible to observations. \revision{Despite putative detections \citep[e.g.,][]{Wagner2023,Xie2024}} \revision{and circumstantial support from spiral pattern speeds \citep[e.g.,][]{Ren2018,Ren2020,Ren2024}}, conclusive evidence for such companions alongside two-armed spirals remains elusive. \revision{T}his adds appeal to the shadow-driven spiral hypothesis, for which the required companions are small enough to hide inside the contrast curves of current direct-imaging techniques. Nevertheless, factors such as extinction \citep{Wagner2023,Cugno2025} and disk contamination in ADI mean that even high-mass companions may evade detection.

How, given this uncertainty, might we differentiate between planet- and shadow-driven two-armed spirals? Kinematics may be the key. The velocity perturbations generated by the shadow-driven spirals in \cite{Muley2024} were strongest in the disk atmosphere, whereas those from the classical planet-driven spirals were strongest in the midplane. Moreover, the azimuthal temperature gradient caused by a shadow causes significant vertical motion \citep{Zhang2024_1}, as disk columns passing through fall and rise in search of vertical hydrostatic equilibrium. These flow patterns can be studied using the spectral lines of various chemical tracers entrained in the gas, a technique which has seen previous observational application in the search for localized \revision{``kinks''} \citep{Pinte2018,Pinte2019} \revision{or} ``Doppler flips'' \revision{\citep{Cassasus2019}} associated with planets, one-armed spiral substructure \citep{Teague2019,Teague2022}, and the deviations from Keplerian rotation associated with gaps and rings in the disk \citep{Teague2021}. From the theoretical side, a number of authors have post-processed simulation snapshots with Monte Carlo radiative transfer (MCRT) codes, with the aim of modeling how disk-planet interaction \citep[e.g.,][]{Bae21}, hydrodynamical \citep[e.g.,][]{BarrazaAlfaro2021,BarrazaAlfaro2023}, and magnetohydrodynamical \citep[e.g.,][]{Flock2015} instabilities. \citep{Zhang2024_1} have taken a first step to applying this technique to simulations shadow-driven spirals, using MCRT to compute the optical surface of the \moltwelveco $J = 3-2$ \ transition and taking velocity components in their simulations at that location.

Drawing inspiration from these works, we revisit the \texttt{PLUTO} hydrodynamical simulations in \cite{Muley2024}, and post-process the final snapshots using the \texttt{RADMC3D} MCRT code. We construct mock observations in the NIR $H$-band ($\lambda_H = 1.62 \ \mu$m)---which traces starlight scattered off of the small dust grains well-coupled with the gas---as well as in the $J = 2-1$ and $J = 3-2$ transitions of three CO isotopologues (\moltwelveco, \ \molthirteenco, \ \molceighteeno), which trace gas properties in the upper, intermediate, and midplane regions of the disk, respectively. In Section \ref{sec:methods}, we describe our methods, while in Section \ref{sec:results}, we present and analyze our mock NIR observations and kinematic maps, finding significant morphological differences which can be used to distinguish between planet- and shadow-driven spirals. In Section \ref{sec:conclusion}, we present our conclusions and chart paths for future work.

\begin{figure*}
    \centering
    \includegraphics[width=1.0\textwidth]{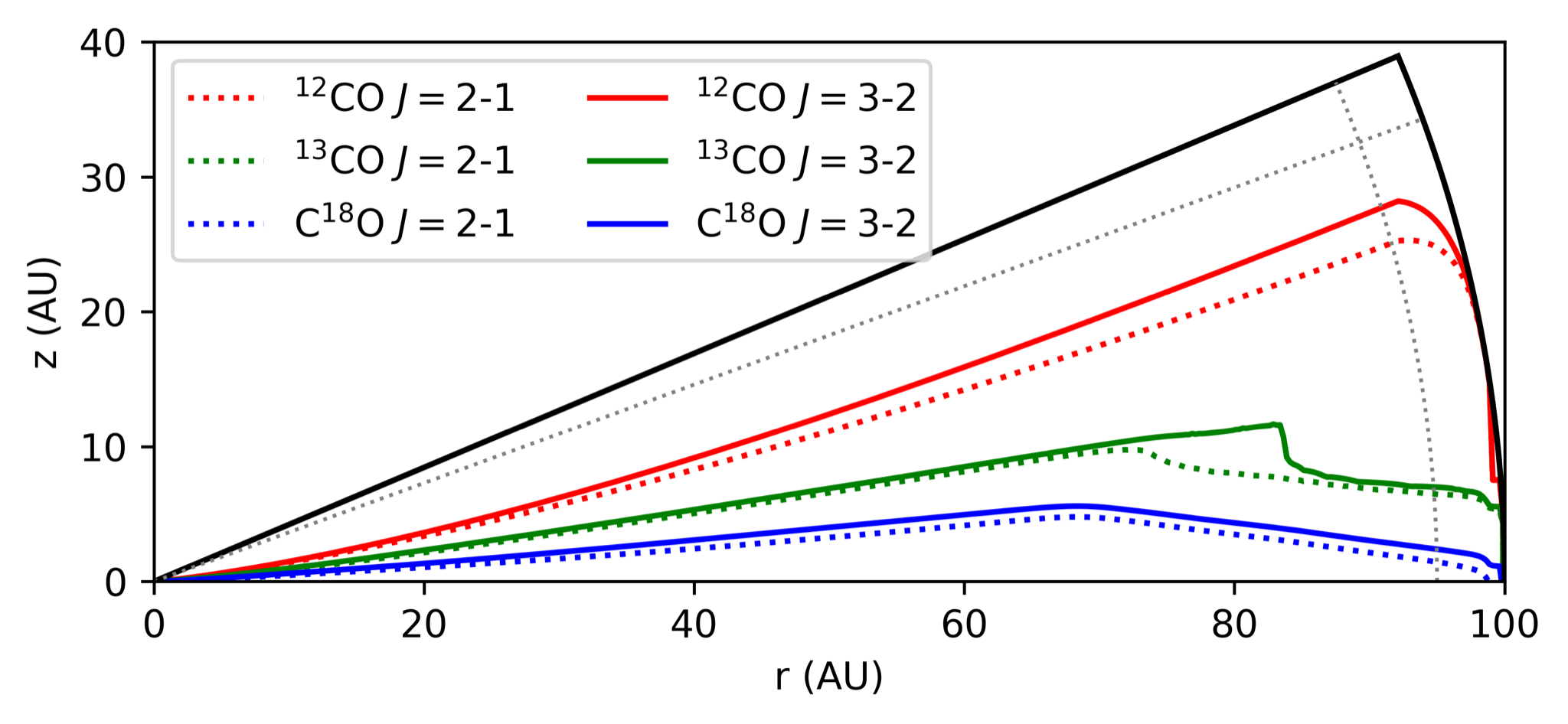}
    \caption{Plots of the vertical $\tau_z = 1$ emission surfaces for the two transitions ($J = 2-1, J = 3-2$) of each of the three isotopologues (\moltwelveco,\molthirteenco,\molceighteeno) in the initial condition for our fiducial protoplanetary disk. \revision{Black solid} lines indicate simulation boundaries (in this case, $r_{\rm out} = 100 {\rm \ au}$), while dashed \revision{grey} lines indicate wave-damping regions in the hydrodynamical simulations. The sharp transition in the \molthirteenco $J = 3-2$ \ $\tau_z = 1$ surface is caused by freezeout of the immediately overlying material at large radii.}
    \label{fig:tau1_surface}
\end{figure*}
\section{Methods}\label{sec:methods}
\subsection{PLUTO radiation hydrodynamics}
We run simulations of disk-planet interaction using the \texttt{PLUTO} hydrodynamics code \citep{Mignone2007}, equipped with an M1 radiation transport module \citep{MelonFuksman2019,MelonFuksman2021} and a ``three-temperature'' (3T) scheme that captures the non-equilibrium energy exchange between gas, dust, and radiation \citep{Muley2023}. One simulation, designed to give rise to shadow-driven spirals, makes use of the full 3T scheme, whereas another, aimed at reproducing a more classical picture of disk-planet interaction, uses parametrized Newtonian cooling to a fixed background condition (so-called ``$\beta$-cooling''). Our temperature-stratified initial condition was obtained iteratively, by alternating computations of radiative transfer and hydrostatic equilibrium following the procedure in \cite{MelonFuksman2022}. All opacity and gas-grain collisions were held to come from small grains well-mixed with the gas at a mass fraction $f_{\rm dg} = 10^{-3}$; for the dust  grains, we used opacities from \cite{Krieger2020}, whereas the gas itself had no intrinsic opacity. In general, our setups were identical to those in the companion paper \citep{Muley2024b}, but with several important distinctions:

\begin{enumerate}
    \item Within the wave-damping zones near the boundaries, we opt for a shorter damping time of $t_{\rm damp} = 0.01 \times 2\pi\Omega_K^{-1}$ to better contain mass within the disk. We opt not to damp the density $\rho$, in order to prevent artificial addition or removal of material in the disk atmosphere by the significant vertical motions associated with the shadow-driven spirals. This does not qualitatively impact the development of shadow-driven spirals seen in the companion paper.
    \item In the $\beta$-cooling simulations, we extend the outer edge of the domain to $r_{\rm out} = 140$ \ au, and shift the planet to an $a_p = 100$ \ au. To maintain the original cell size, we adjust the resolution of these simulations is $158 (r) \times 58 (\theta) \times 460(\phi)$, as opposed than $134 (r) \times 58 (\theta) \times 460(\phi)$ as in the 3T simulations. As in the companion paper, we run \revision{these simulations for 250000 y (${\sim}$1000 orbits at 40 au), and then} an additional \revision{2500 y (${\sim}$10 } orbits \revision{at 40 au)} at doubled resolution in each case.
    \item We change the planet-to-star mass ratio in the $\beta$-cooling case to $q_p = 4 \times 10^{-3}$ (corresponding to roughly $4 \ M_J$, or a thermal mass $q_{\rm thermal} \equiv (M_p/M_*)h_p^{-3} \simeq 10$, given the typical disk aspect ratio $h_p=0.075$ at $a_p = 100 au$), while the planet in the 3T case remains at an $a_p = 40$ \ au and $q_p = 3 \times 10^{-4}$ (roughly Saturn-mass, $q_{\rm thermal} \simeq 1.1$ at $a_p = 40$ \ au).
\end{enumerate}
The latter two changes in particular enable a direct comparison between the traditional picture of a multi-Jupiter-mass, exterior driver \citep[such as investigated by, e.g.,][in the context of MWC 758]{Dong2015b}, and our shadow-driven hypothesis, as explanations for observed two-armed spirals.

\subsection{RADMC3D post-processing}
To obtain mock gas and dust images from our disk simulations, we post-process them using the \texttt{RADMC3D} Monte Carlo Radiative transfer code. We avoid unphysical illumination of the simulation inner boundary by padding each simulated disk to $r_{\rm in} = 0.4$ au, using fluid field values from the full 2.5D ($r$ and $\theta$-dependent, axisymmetric in $\phi$) initial condition. We do not compute temperatures directly from \texttt{RADMC3D} using the \texttt{mctherm} routine, but rather use the temperatures from the radiation-hydro simulations, as the latter include contributions from $P dV$ work on the gas which a pure radiative-equilibrium calculation would obviate.

We compute datacubes for two rovibrational transitions ($J~=~2-1$, $J~=~3-2$) in each of three carbon monoxide isotopologues (\moltwelveco, \ \molthirteenco, \ and \molceighteeno). As shown in Figure \ref{fig:tau1_surface}, the emitting surfaces of these isotopologues are located in the upper, middle, and lower layers of the disk respectively. We assume a fiducial mass ratio $f_{\rm 12CO} = 10^{-4}$ between \moltwelveco \ and the H/He mixture simulated in the hydrodynamics, and set $f_{\rm 13CO} = (1/77)f_{\rm 12CO}$ and $f_{\rm C18O} = (1/560)f_{\rm 12CO}$, following \cite{BarrazaAlfaro2023}. We set the freeze-out temperature $T_{\rm CO} = 20 $K, weighting the nominal mass ratio of CO in each grid cell by a factor
\begin{equation}
    w_{\rm freeze} = (1 + \exp(-(T - T_{\rm CO})/\Delta T_{\rm CO}))^{-1}
\end{equation}
where $\Delta T_{\rm CO} = 0.2 \ $K. This ensures that there are no rapid jumps in CO-isotopologue concentration from one cell to the next, keeping the emitting surfaces smooth.

As shown in Figure \ref{fig:tau1_surface}, these isotopologues provide coverage of the upper, middle, and lower layers of the disk respectively. In addition, we follow the procedure described in \cite{Muley2024b} to create ray-traced images in the near-infrared $H$-band ($\lambda_H = 1.62 \mu$m), observable with VLT/SPHERE. These observations trace the small dust grains, which are well-mixed with the gas; as in our hydrodynamic simulations, we assume perfect mixing of these small grains with the gas, i.e. a constant small-dust-to-gas ratio $f_{\rm sg} = 10^{-3}$. 

In making our mock images, we assume a notional distance of $d = 100 {\rm \ pc}$ to each object; for comparison, HD 163296 is 101 au from Earth, and the DSHARP disks are between 100-200 pc away \citep{Andrews2018}. For the mock molecular-line observations, we assume a Gaussian beam with FWHM $0.19'' \times 0.17''$, taking care to apply this convolution to the raw datacube before computing moments. For the mock \textit{H}-band observations, we apply a typical VLT/SPHERE beam of $0.06'' \times 0.06''$, as in \cite{Muley2024b}.

\section{Results}\label{sec:results}

\begin{figure*}
    \centering
    \begin{overpic}[width=0.9174364896\textwidth,origin=c]{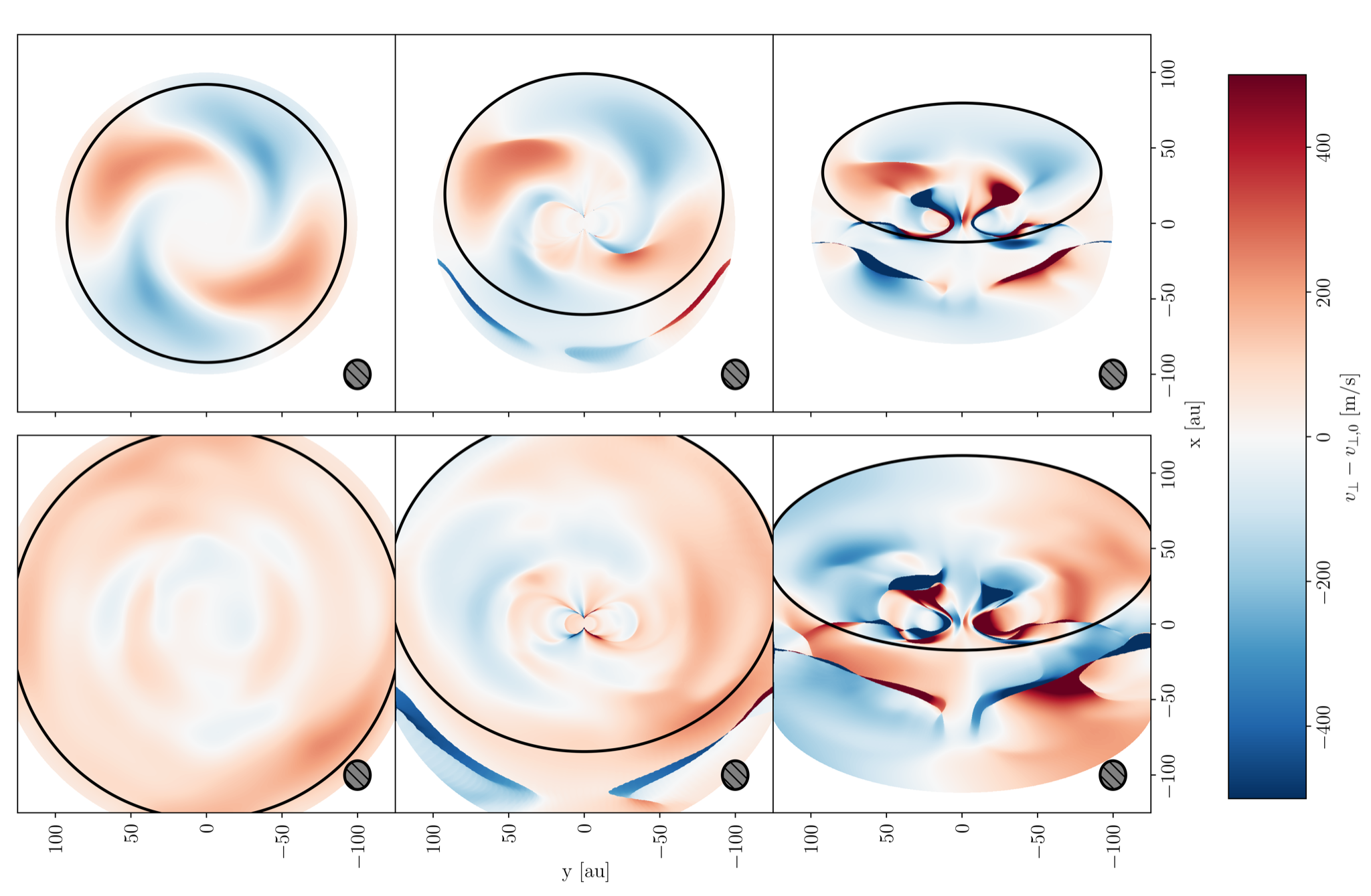}
    \put (3,30) {\Large Planet-driven}
    \put (3,59) {\Large Shadow-driven}
    \put (3,37) {\large $0^{\circ}$}
    \put (30.25,37) {\large $30^{\circ}$}
    \put (57.5,37) {\large $60^{\circ}$}
    \end{overpic}
    \caption{Kinematic moment-1 (line-of-sight velocity) maps for the shadow-driven spirals, above, and the planet-driven spirals, below, in the $J=2-1$ transition of the \moltwelveco \ isotopologue; \revision{in each case, the moment-1 map of the initial condition is subtracted from that at simulation end}. From left to right, we plot disk inclinations of 0$^\circ$, 30$^\circ$, and 60$^\circ$. \revision{The FWHM of the fiducial ALMA beam ($0.19'' \times 0.17''$) is indicated by a grey hatched ellipse}. The spiral morphology is much more prominent in the shadow-driven \revision{case than in the spiral-driven case}. }
    \label{fig:moment_1_0}
\end{figure*}
\begin{figure*}
    \centering
    \begin{overpic}[width=0.9174364896\textwidth,origin=c]{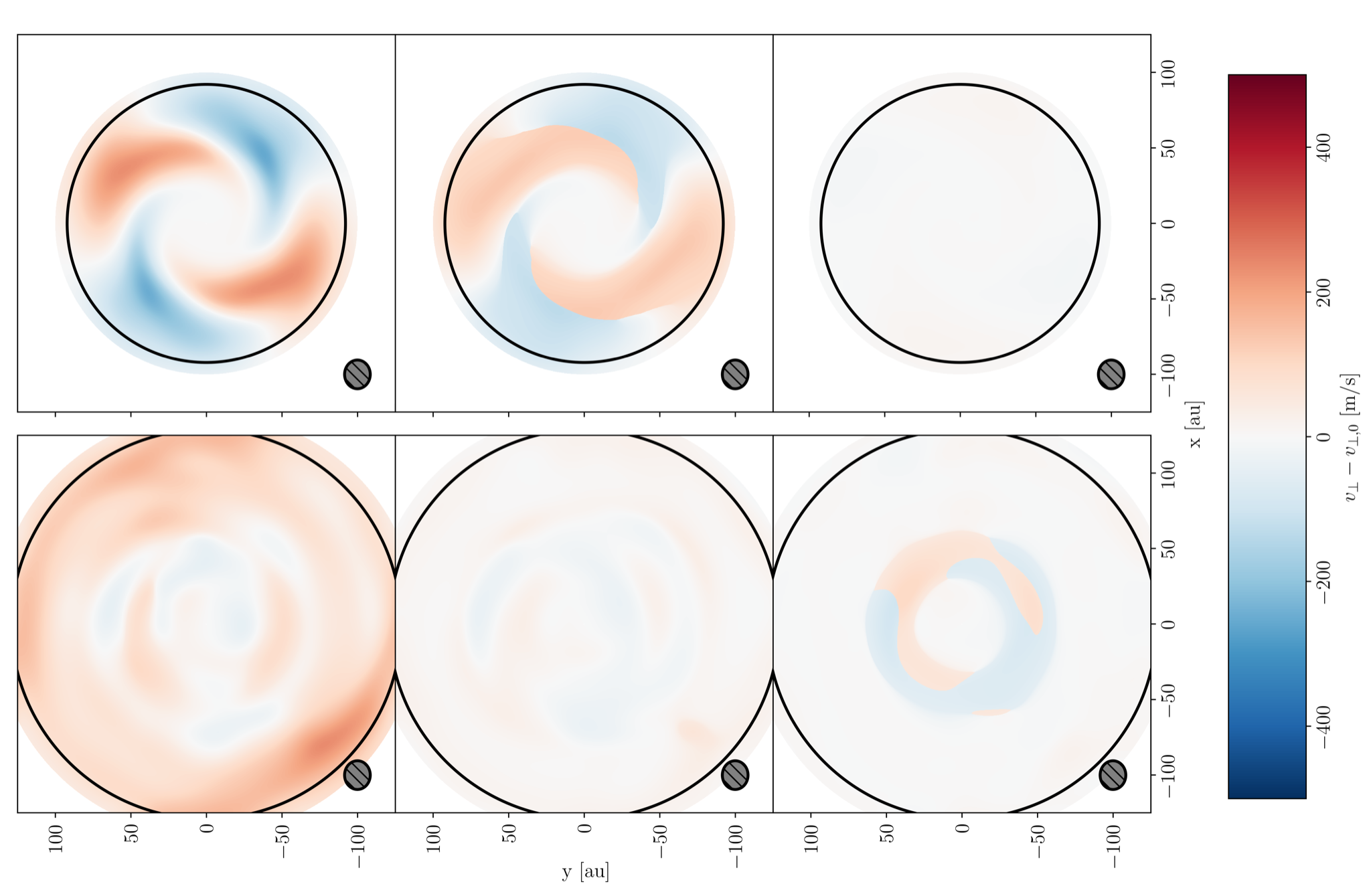}
    \put (3,30) {\Large Planet-driven}
    \put (3,59) {\Large Shadow-driven}
    \put (3,37) {\large \moltwelveco}
    \put (30.25,37) {\large \molthirteenco}
    \put (57.5,37) {\large \molceighteeno}
    \end{overpic}
    \caption{Kinematic moment-1 (line-of-sight velocity) maps \revision{at simulation end, with moment-1 from the initial condition subtracted off, as in Figure \ref{fig:moment_1_0}.}  In contrast to the plots in Figure \ref{fig:moment_1_0}, we fix the disk inclination to 0$^\circ$ (face-on), and plot, from left to right, the maps corresponding to the $J=2-1$ transitions of the \moltwelveco \ , \molthirteenco \ , and \molceighteeno \ isotopologues, which trace the upper, intermediate, and midplane layers of the disk respectively. In  \molceighteeno, the two-armed signature of the shadow-driven spiral is virtually absent, but for the standard planet-driven case the inner Lindlbad spirals remain visible.}
    \label{fig:moment_1_1}
\end{figure*}
\subsection{Spiral structure}
We compute moment maps both for our final simulations, and for the initial condition, using the \texttt{bettermoments} method \citep{Teague2018}. Such a background model assumes perfect knowledge about the disk properties, eliminating errors associated with prescribed, semi-analytical $\tau_z = 1$ surfaces, density, and velocity fields which would manifest in fits of real observations. Nevertheless, it retains the drawback that large-scale changes to disk structure during the simulation would alter the $\tau_z = 1$ surface. Thus, the differential moment maps we plot include contributions from both the change in velocity at each point in the disk over the simulation lifetime, and from the shift in the emitting layer whose velocities are sampled. In Figure \ref{fig:moment_1_0}, we plot the modeled line-of-sight velocity in each of our simulations, at various inclinations, in the $J = 2-1$ transition of the \moltwelveco \ isotopologue. In Figure \ref{fig:moment_1_1}, we show line-of-sight velocities in the $J =2-1$ transitions of all three isotopologues (\moltwelveco, \ \molthirteenco, \ \molceighteeno) at a fixed 0$^{\circ}$ inclination. In all cases, the moment-1 maps from the final simulations are regularized by subtracting the moment-1 computed from the initial condition. We present our full library of moment-1 maps in the Appendix (see \revision{Appendix} \ref{sec:app_library_desc} for \revision{a} description, and \revision{the following sections for images}).

We find that the three-temperature simulations manifest a clear two-armed, shadow-driven kinematic spiral, strongest in the \moltwelveco \ lines tracing the upper disk atmosphere and becoming progressively weaker in the deeper layers probed by \molthirteenco \ and \molceighteeno \ respectively. This is consistent with the fact that in our simulations, the shadow is excited at the $\tau_r = 1$ surface for stellar irradiation, and would thus not be expected to generate significant motion in the midplane \citep{Muley2024b}.

These shadow-driven spirals are strongest ($v_{\perp} \approx 
\SI{200}{\meter\per\second}$) in the face-on case, but remain visible in \moltwelveco at a 30$^\circ$ inclination. At 60$^\circ$, they become all but invisible, particularly when beam convolution is considered. This geometrical effect evidences the strong vertical motions in the shadowed disk, driven by disk columns attempting to relax to hydrostatic equilibrium as the (sub-)Keplerian flow transports them into and out of shadowed regions \citep{Muley2024b}. Mathematically, this effect can be accounted for including in-plane advection of vertical momentum---principally in the azimuthal direction---in the evaluation of hydrostatic equilibrium:
\begin{equation}
    \frac{\partial P}{\partial z} - \rho_g \frac{\partial\Phi}{\partial z} + \Omega \frac{\partial(\rho_g v_z)}{\partial \phi} \approx 0 .
\end{equation}
This term was the subject of extensive discussion in \cite{Zhang2024_1}, who demonstrated its importance in the context of shadow-driven spirals and discussed implications for kinematic observations.

The physical picture is very different for our multi-Jupiter-mass disk-planet interaction setup, in which we used $\beta$-cooling to an azimuthally symmetric background structure to preclude the development of shadow-driven spirals. In this case, there is no clear, open, two-armed kinematic spiral in any isotopologue, but rather \revision{a comparatively weak ($v_{\perp} \lesssim 
\SI{100}{\meter\per\second}$) signal from the tightly-wound, inner planetary spirals}. These vertical-motion spirals are strongest in \moltwelveco, and as shown in \cite{Muley2024}, can also be attributed to the effect of in-plane advection on hydrostatic equilibrium. However, given that the typical wavenumber of planet-driven Lindblad spirals ($m \approx (h/r)^{-1}$) is much larger than that of shadow-driven spirals in this case ($m = 2$), the former are significantly thinner\revision{,} and \revision{lose their distinctive morphology} under beam convolution. Unlike in \cite{Bae21}, where a highly-evolved, well-settled grain size distribution leads to long cooling times---and consequently, robust buoyancy spirals---at the \moltwelveco \ emission surface, the constant dust-to-gas ratio in our simulations means that collisional cooling is faster, and buoyancy-related features consequently weaker.

Such a high-mass planet opens a wide and deep gap, which is visible in the optically thinner \molthirteenco \ and \molceighteeno \ isotopologues (see Appendices \ref{sec:app_b} and \ref{sec:app_c}) , whose emitting surfaces are closer to the midplane. In particular, the face-on \molthirteenco \ images show vertical motions (${\sim}$50-100 m/s) at the inner gap edge, potentially attributable to the meridional flow patterns in \cite{Fung2016} or \cite{Bi2023}. The polar accretion flow onto the planet \citep[e.g.,][]{Fung2015b,Schulik2019}, of comparable magnitude, is also visible. At a 30-degree inclination these features become less prominent, but a prominent ``Doppler flip'' \citep[e.g.,][]{Cassasus2019,Pinte2022} associated with the in-plane motions generated by the planet comes into view (further discussed in Appendix \ref{sec:app_library_desc}). All of these features are notably absent in our shadow-driven spiral model, given that the planet is of significantly lower mass in that case.

\begin{figure}
    \centering
    \includegraphics[width=0.5\textwidth]{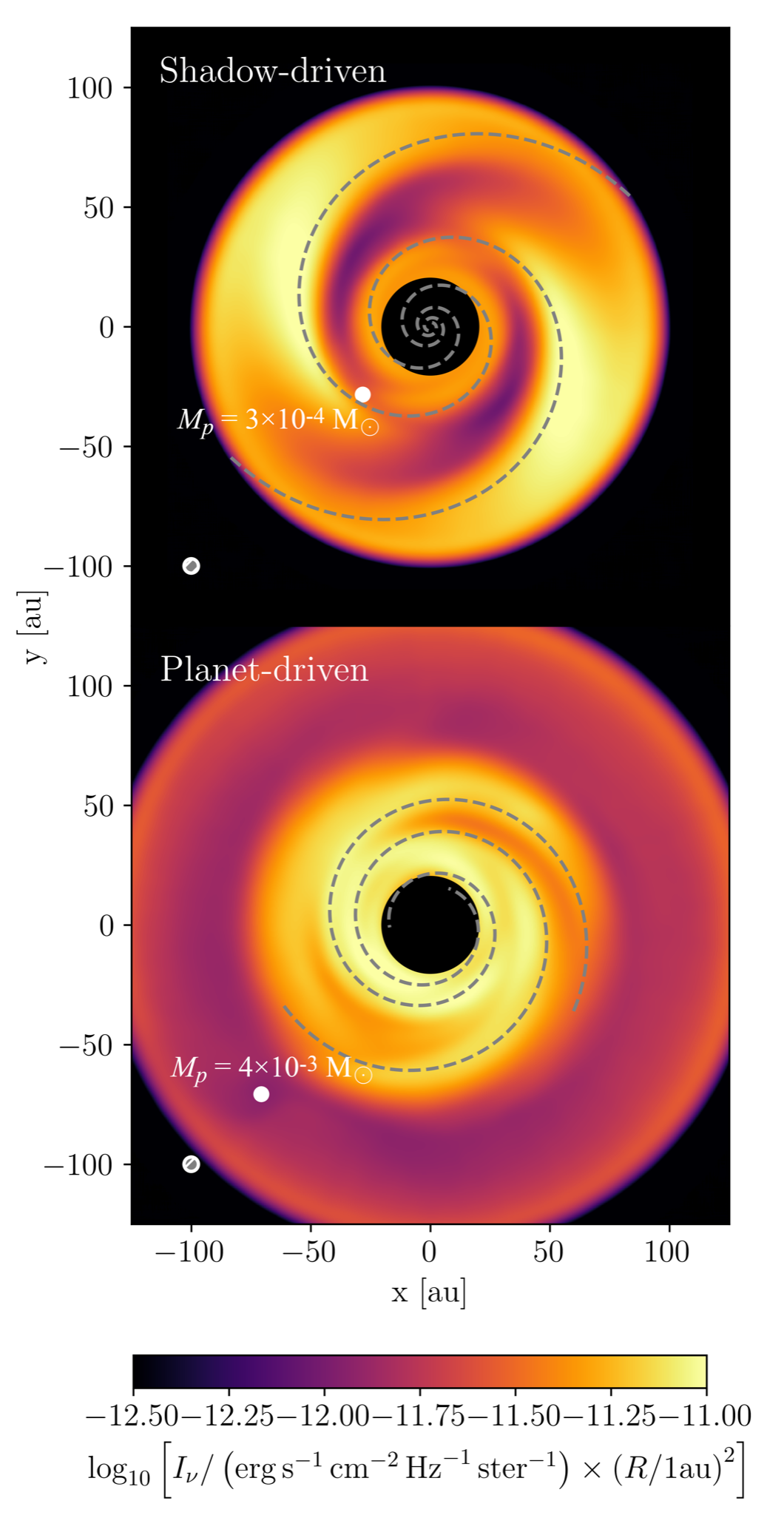}
    \caption{Modeled near-infrared (NIR) image in the \textit{H}-band ($\lambda_H = 1.62 \ \mu$m of both the shadow-driven (above) and planet-driven (below) spiral cases. In each panel, the location of the planet is identified with a white dot, and an Archimedian spiral fit in grey dashed lines. The beam is indicated with a grey hatched ellipse, while the effect of a coronagraph is mimicked by the black circle 20 au (0.2'') in radius. The standard planetary Lindblad spirals are weakened in the disk atmosphere---traced by NIR scattering off of the entrained dust grains---by wave refraction at the temperature transition corresponding to the disk's $\tau_r$ = 1 surface. }
    \label{fig:scattered_light}
\end{figure}

Our mock VLT/SPHERE \textit{H}-band images, shown in Figure \ref{fig:scattered_light}, with their high angular resolution making them a useful complement to the kinematic maps. The planet-driven case builds on the work of \cite{Dong2017}, which sought to reproduce the substructures in the MWC 758 system using a model with similar properties to ours. In the present work, our use of a realistic, vertically-stratified temperature structure \citep{Juhasz2018} causes wave refraction, and the non-isothermal equation of state reduces the density contrast across the spiral \citep[e.g.,][]{Zhu2015,Muley2024}, contributing to a weakening of the planet-driven spiral---particularly in the disk atmosphere, which is the region probed by NIR observations. A simple visual fit using Archimedian spirals, overplotted on the mock images, shows that the shadow-driven spirals are both more open (pitch angle $\alpha_s \approx 0.24$ vs. 0.14) and rotationally symmetric (angular separation $\Delta \phi_{s} \approx \pi/2$ vs. $2\pi/3$) than the planet-driven spirals. 

Taken together, our results imply that a near-infrared double-armed spiral with a strong kinematic counterpart in the upper layer (\moltwelveco \ or \molthirteenco), but weak when closer to the midplane (\molceighteeno), is likely shadow-driven, with the upper layers of the disk experiencing significant vertical expansion and contraction as their orbits take them through regions of greater or less illumination. By contrast, a near-infrared spiral with comparatively weaker kinematic spirals in the upper layer (such as \moltwelveco \ or \molthirteenco), is consistent with planetary driving, and thus a good candidate for direct-imaging searches. We grant, however, that depletion of CO would alter the height of each isotopologue's emitting surface \citep{Paneque2025} and thus the disk layer whose properties it traces.

\subsection{Line broadening}
\begin{figure}
    \centering
    \begin{overpic}[width=1\linewidth]{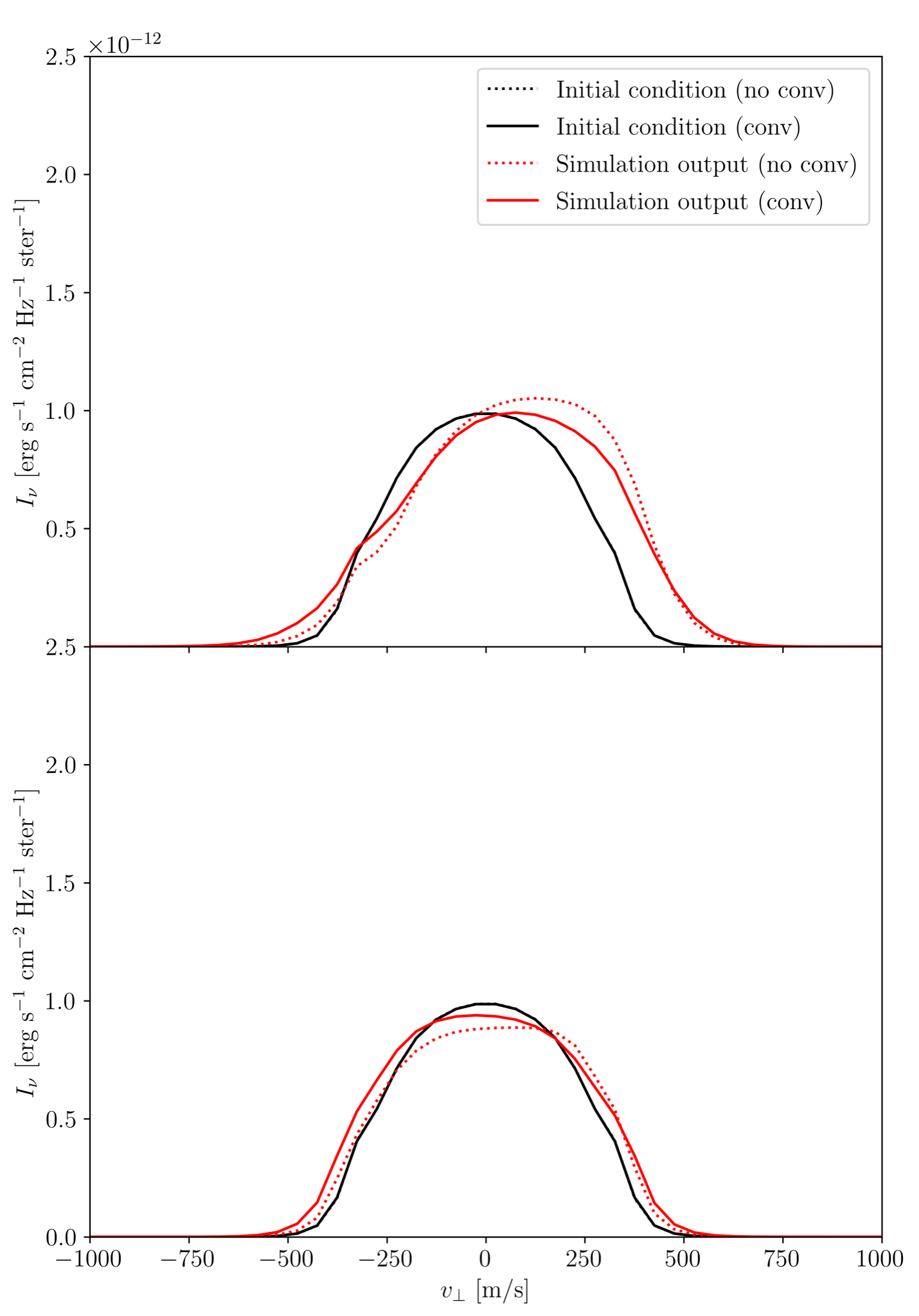}
    \put (9, 90) {\Large Shadow-driven}
    \put (9, 45) {\Large Planet-driven}
    \end{overpic}
    \caption{Plot of \moltwelveco $J = 2-1$ lines, for the planet-driven spiral models \textit{(above)} and the shadow-driven models \textit{(below)}, \revision{at a position $R = 60$ au and $\phi = 0$ with respect to the $x$-axes of Figure \ref{fig:scattered_light} (and, equivalently, of the qualitative maps in Figure \ref{fig:moment_2_maps}).} In the initial condition, line broadening results exclusively from thermal motion, and is largely unaffected by beam convolution. In evolved disks, however, the line profile is shifted and broadened by \revision{vertical motions within spiral arms, with beam convolution over regions of differing velocity dampening the overall directional shift, while enhancing broadening.}}
    \label{fig:line_broadening}
\end{figure}

In Figure \ref{fig:line_broadening}, we plot simulated line profiles in the optically thick $J = 2-1$ transition of the \moltwelveco \ isotopologue, taken at a representative radial position $R = 60$ au and angular position $\phi = 0$ with respect to the $x$-axis \revision{of Figure \ref{fig:scattered_light}}. In the upper panel, we plot the line from our shadow-driven spiral simulation, whereas in the lower panel we show the line from the planet-driven simulation, considering in both cases a face-on orientation. In the initial condition, these lines appear to be symmetric around $v_{\perp} = 0$ and only minimally affected by beam convolution; this is in line with expectations, because the average line-of sight velocity is zero for an unperturbed face-on disk, and the only contribution to line width is thermal. Because $T \approxprop R^{-1/2}$, and $v_{\rm th} \approxprop T^{1/2}$, $v_{\rm th} \approxprop R^{-1/4}$ and varies only weakly (${\sim}10\%$) over the beam's length scale at the fiducial $R_{\rm line}$. Even this difference is significantly suppressed at first order by the approximate symmetry of the convolution kernel around $R_{\rm line}$.

In an evolved disk, the presence of net upward and downward velocities at the emission surface moves the line center away from $v_{\perp} = 0$. This lowers the optical depths at the peak rest-frame wavelengths of the lines, allowing light from deeper layers---less perturbed by the spiral structure---to shine through, leading to a profile more complex than a simple, shifted initial condition. Both of these effects are more pronounced in the shadow-driven than in the planet-driven case. Beam convolution, by averaging over regions of both positive and negative $v_{\perp}$, contributes to additional line broadening (moment-2) while damping the measured $v_{\perp}$ (or, equivalently, moment-1) to zero.

\begin{figure*}
    \centering
    \begin{overpic}[width=0.45\linewidth]{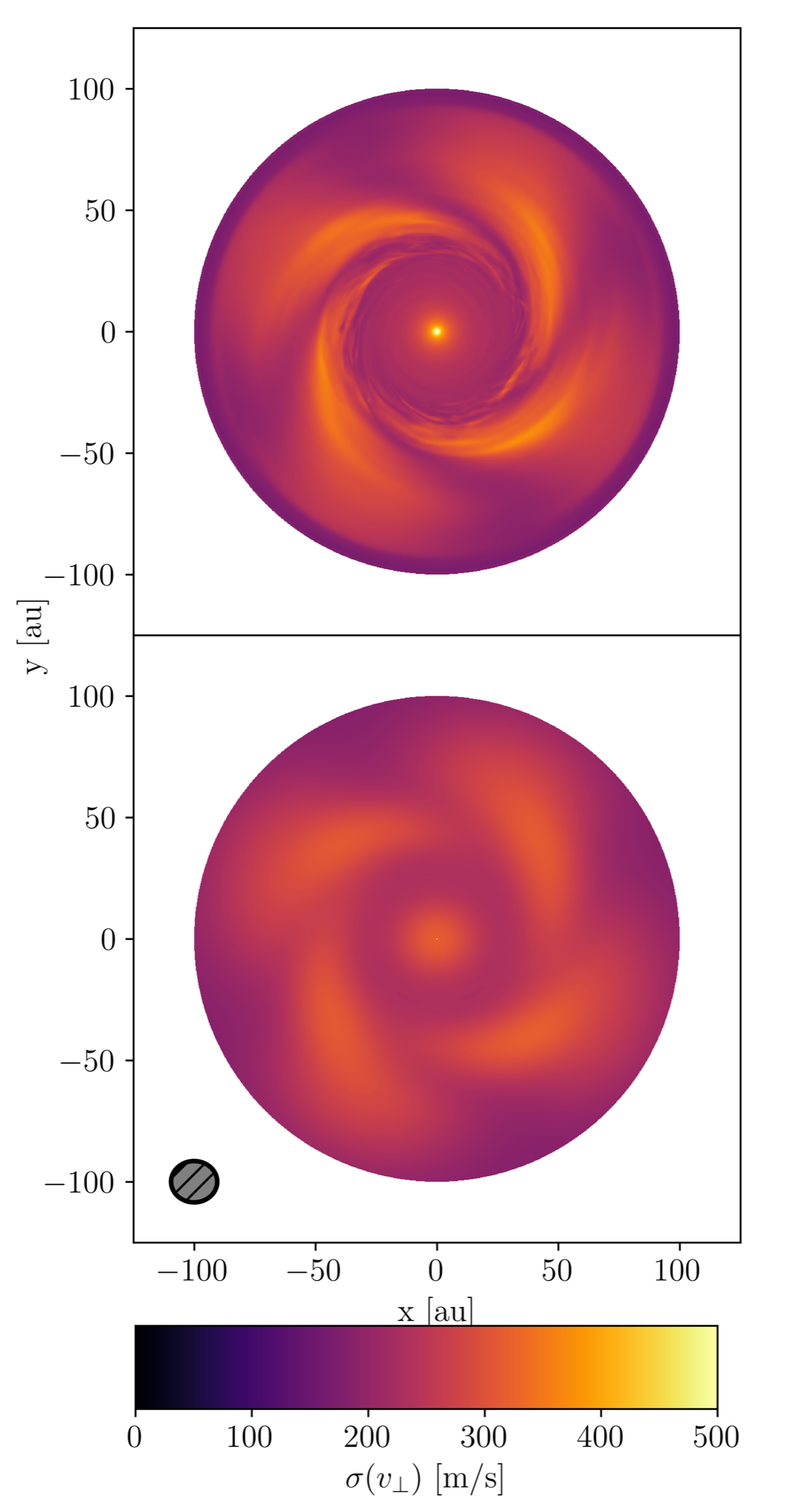}
    \put (12, 95) {\Large Shadow-driven}
    \end{overpic}
    \begin{overpic}[width=0.45\linewidth]{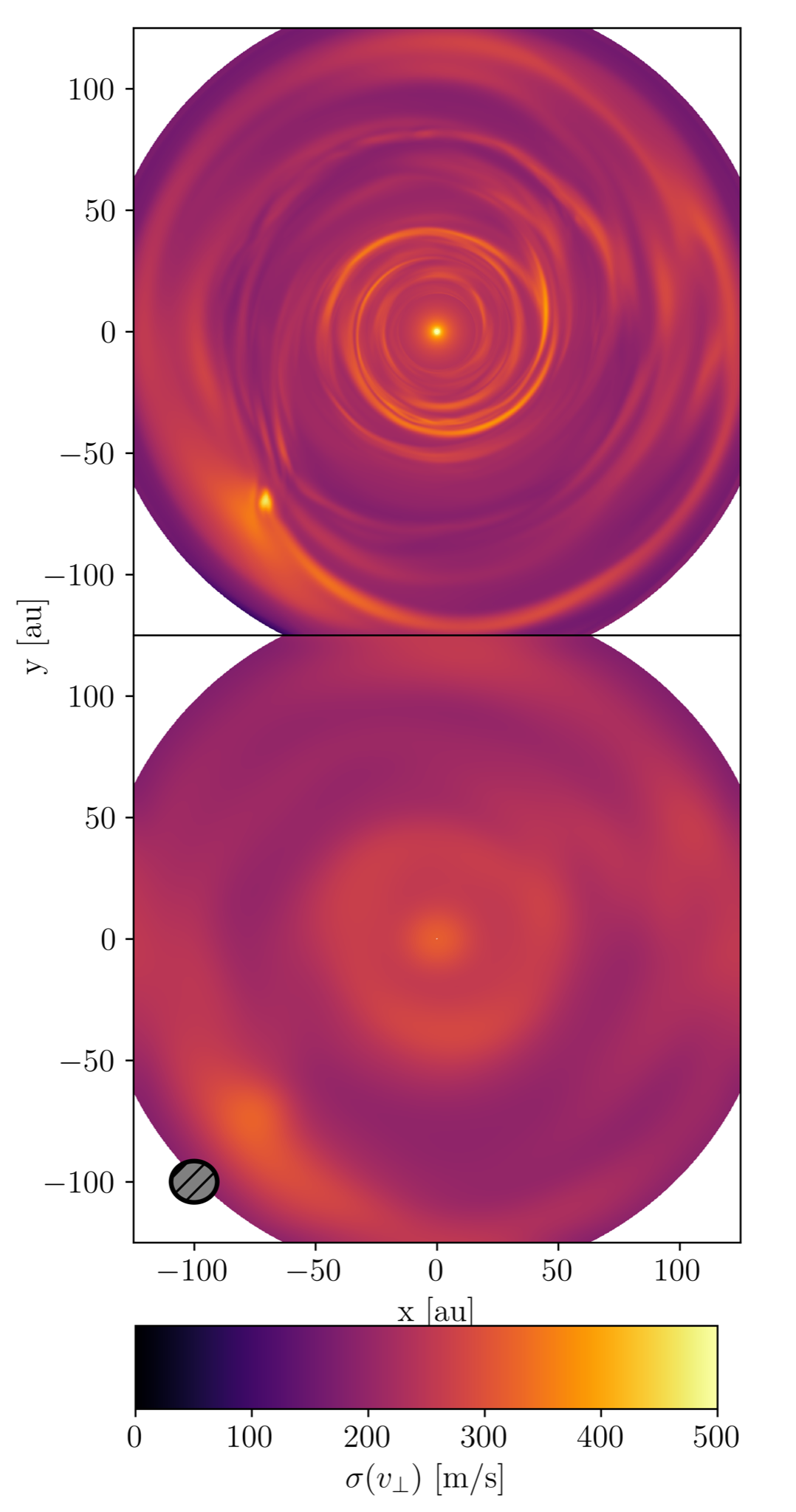}
    \put (12, 95) {\Large Planet-driven}
    \end{overpic}
    \caption{Maps of the typical line-of-sight velocity dispersion along each line of sight in the $J = 2-1$ transition of \moltwelveco, for the shadow-driven case at left and the planet-driven case at right. Upper panels are computed from the raw datacubes, while lower panels are computed from datacubes convolved with the fiducial beam. In the planet-driven case, we see a strong localized increase in velocity dispersion near the planet location, as well as in the interior of the disk; in the shadow-driven case we see four separate peaks along the outer edge of the gap, corresponding to the launching points of the spirals.}
    \label{fig:moment_2_maps}
\end{figure*}

The moment-2 maps of both the shadow- and planet-driven spirals in Figure \ref{fig:moment_2_maps} present a more qualitative view of the situation. In the planet-driven case, we observe a strong signal in the vicinity of the planet, corresponding to the hot polar accretion flows feeding the circumplanetary disk; the effects of this are visible to a lesser extent throughout the entire planetary corotation region. In the inner regions of the disk, where the planet-driven spirals become more tightly wound, we also find an elevated velocity dispersion. For the shadow-driven spiral, the morphology is altogether different, with the planetary gap surrounded by four separate regions of elevated velocity dispersion; these correspond to regions of strong shadowing where the kinematic spirals are launched. Beam convolution preserves this basic structure, although it smears out the specific details. A similar picture holds in the optically thinner \molthirteenco \ and \molceighteeno \ isotopologues, not shown here, although as in the moment-1 case, signals tend to be weaker. These differences in morphology provide yet another means of distinguishing planet-driven from shadow-driven sprial features.

Our analysis is similar to that of \cite{Dong2019}, who likewise post-processed simulations of disk-planet interaction to study the line broadening caused by planet-induced flows. We build on their work by using a temperature structure self-consistently obtained using radiation hydrodynamics, rather than setting a constant, unstratified background temperature. Moreover, we include the effects of beam convolution in order to facilitate comparison with actual observations. Although both works include a case of equivalent mass ($M_p = 4 \  M_J$), the disk aspect ratio is much smaller in their simulations ($h_p = 0.05$) than in ours ($h_p \approx 0.075$), meaning their high-mass planet has a greater thermal mass. Consequently, the planet-carved gap is both deeper and narrower in their simulations \citep[e.g.,][]{Dong2016}, with fast and unsteady flows having established them throughout the gap. In radiative-transfer post-processing, this leads to significant line broadening throughout the gap, unlike in our work where such rapid flows, and consequently line broadening, are largely confined to the circumplanetary region.

\begin{figure*}
    \centering
    \begin{overpic}[width=0.6\textwidth,angle=90,origin=c]{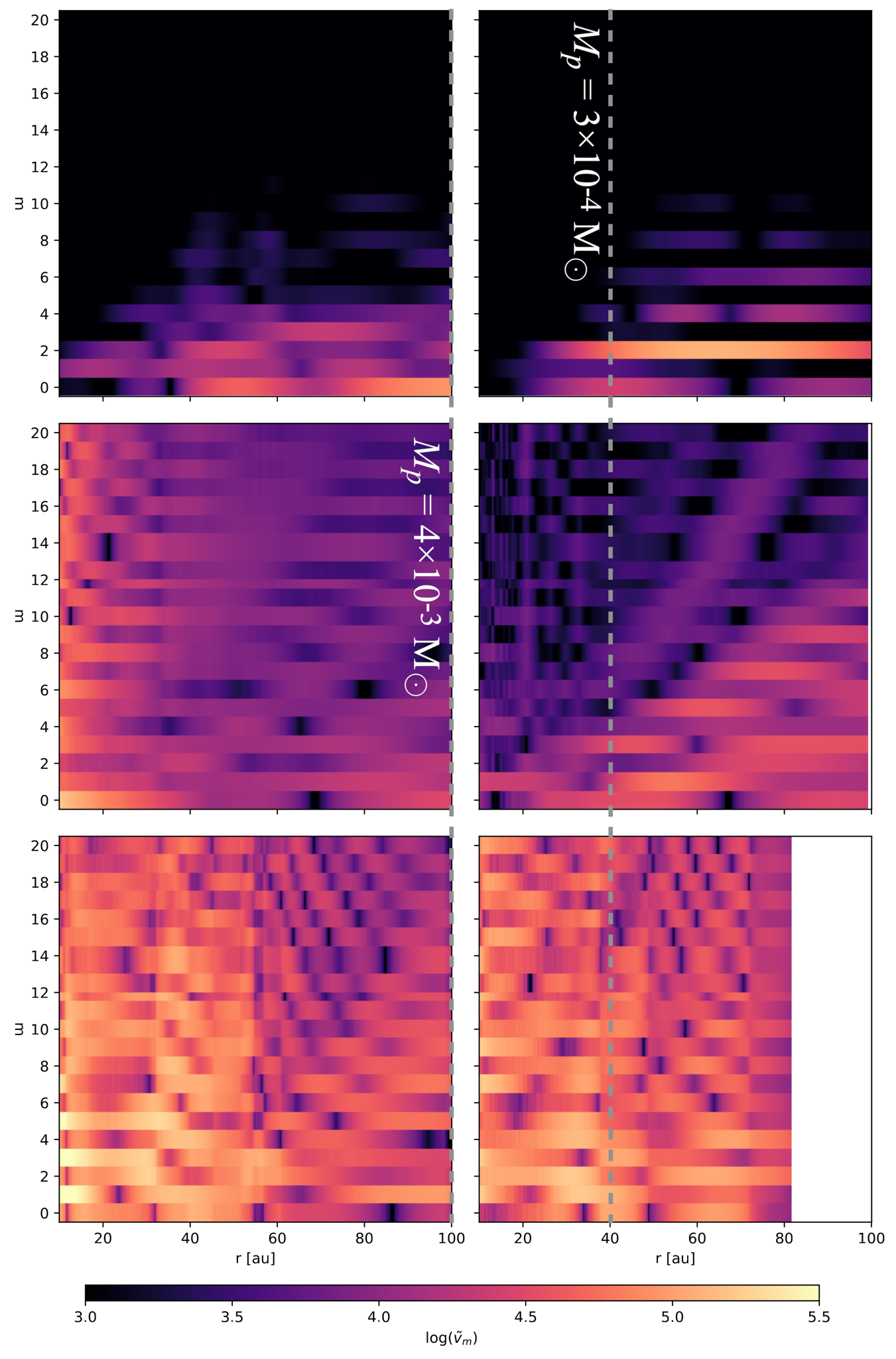}
    \put (3,46) {\color{white} \Large Planet-driven}
    \put (3,77) {\color{white} \Large Shadow-driven}
    \put (3,53) {\color{white} \large $0^{\circ}$}
    \put (33,53) {\color{white} \large $30^{\circ}$}
    \put (63,53) {\color{white} \large $60^{\circ}$}
    \end{overpic}
    \vspace*{-16ex}
    \caption{Modal plots for moment-1 maps constructed from beam-convolved \moltwelveco \ $J = 2-1$ datacubes. As in Figure \ref{fig:moment_1_0}, we plot the shadow-driven case in the upper row, and the more standard planet-driven case in the lower, with inclinations of 0$^\circ$, 30$^\circ$, and 60$^\circ$ from left to right. We indicate the planet's radial location in each case with a dotted line. For the shadow-driven spiral, power is most concentrated in even-numbered modes, a feature absent from the classical case; however, this distinction becomes less apparent in the inclined cases. To avoid contamination of the mode structure due to emission from the (unphysically situated) outer radial boundary of the simulation, we clip large radii from our plot for the $i_d = 30^\circ$ and $ 60^\circ$ cases.}
    \label{fig:modal_analysis_0}
\end{figure*}

\begin{figure*}
    \centering
    \begin{overpic}[width=0.6\textwidth,angle=90,origin=c]{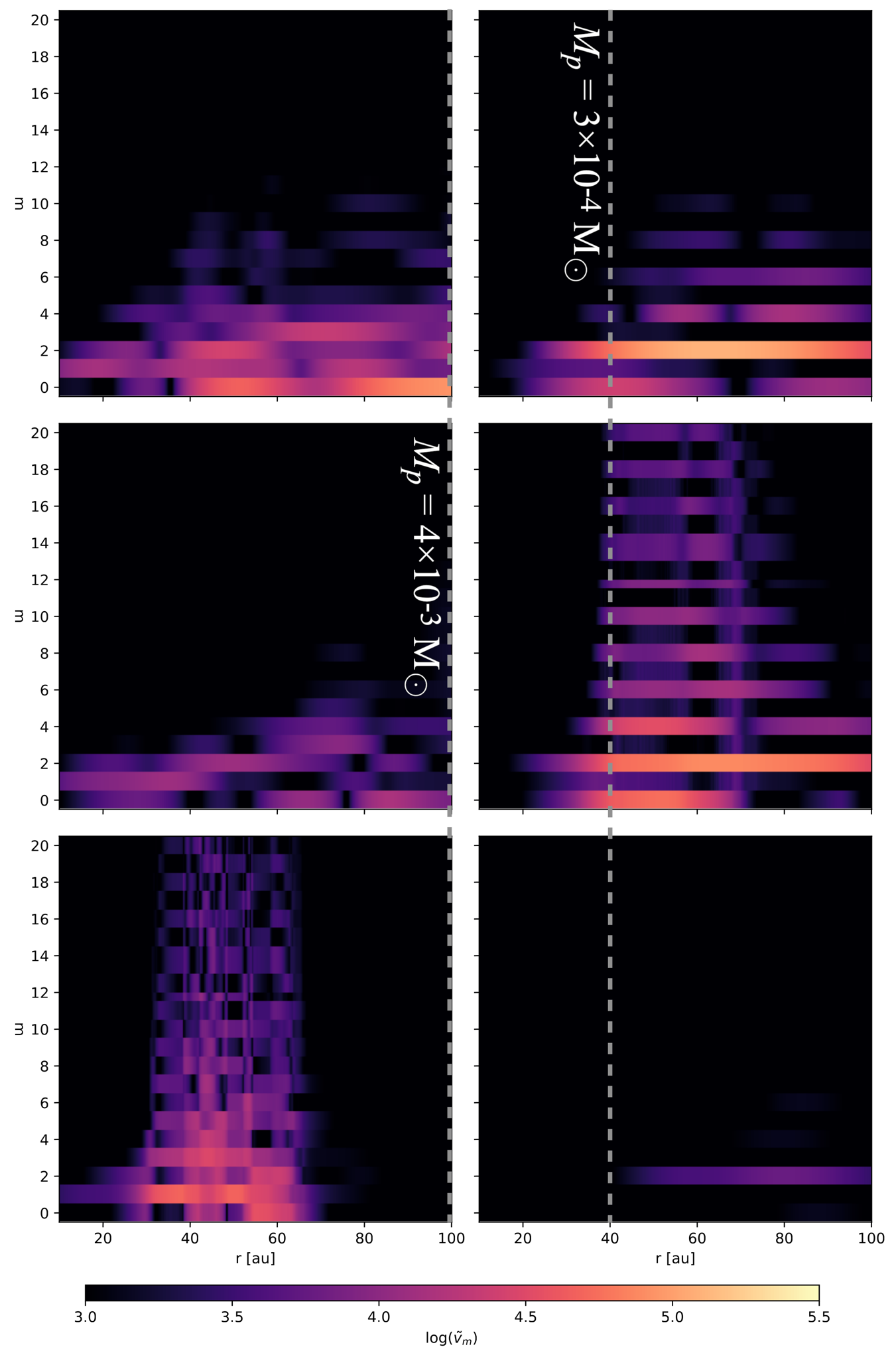}
    \put (3,46) {\color{white} \Large Planet-driven}
    \put (3,77) {\color{white} \Large Shadow-driven}
    \put (3,53) {\color{white} \large \moltwelveco}
    \put (33,53) {\color{white} \large \molthirteenco}
    \put (63,53) {\color{white} \large \molceighteeno}
    \end{overpic}
    \vspace*{-16ex}
    \caption{Modal plots for moment-1 maps constructed from beam-convolved datacubes. From left to right, we show modal plots associated with the $J=2-1$ transitions of \moltwelveco, \molthirteenco, and \molceighteeno, respectively, fixing the inclination to 0$^\circ$.  We indicate the planet's radial location in each case with a dotted line. As in Figure \ref{fig:modal_analysis_0}, the shadow-driven case is above, while the planet-driven case is below. Shadow-driven spirals show an $m = 2$ mode steadily decreasing in strength with deeper and deeper isotopologues.}
    \label{fig:modal_analysis_1}
\end{figure*}

\subsection{Modal analysis}
Fourier analysis can be used to provide an alternative view of shadow-driven spirals, emphasizing information that is difficult to discern from images themselves. This exercise was carried out by \cite{Calcino2025}, in order to characterize the spiral arms generated by infalling material in a protoplanetary disk. We adapt this technique for our purposes by Fourier-transforming our mock observables (specifically, the moment-1 maps) rather than the simulated fluid quantities directly, and by considering more general disk inclinations than face-on. We deproject our mock observations using the following relation, assuming zero position angle:
\begin{subequations}
\begin{equation}
\begin{split}
    x - x_c &= r \cos \alpha_{\rm iso} \cos i_{d}  \sin \phi \\
y &= r \cos \alpha_{\rm iso} \sin \phi
\end{split}
\end{equation}
where
\begin{equation}
    \begin{split}
            x_c &= r \sin \alpha_{\rm iso} \sin i_{d}
    \end{split}
\end{equation}
\end{subequations}
and $\phi$ represents the azimuthal angle with respect to the $x$-axis.

As stated previously, we use disk inclinations $i_{d} = \{0^\circ, 30^\circ, 60^\circ \}$ for our mock images. $\alpha_{\rm iso}$ is the angular elevation of the emitting layer of the transition in question, which we hold to be $\alpha_{\rm iso} = \{0.24, 0.13, 0.07\}$ for the $J = 2-1$ transitions of \moltwelveco, \ \molthirteenco, \ and \molceighteeno \ respectively, and $\alpha_{\rm iso} = \{0.275, 0.145, 0.085\}$ for the $J = 3-2$ transitions of the same isotopologues.

In Figure \ref{fig:modal_analysis_0}, we present a moment-1 power spectrum computed from beam-convolved, $J = 2-1$ \moltwelveco \ mock observations. As in the corresponding Figure \ref{fig:moment_1_0}, the top row represents shadow-driven spirals and the bottom row planet-driven spirals, with the inclination increasing from left to right. In the face-on case, power is concentrated in the $m = 2$ modes (and more generally, in even-numbered modes) in the shadow-driven case; in the standard planet-driven case, the power shows no such bias toward even wavenumbers. \revision{We caution, however, that this result holds only for small mass ratios between the primary and companion; for larger ratios, particularly when the system is an equal binary, the $m = 2$ mode can come to dominate the potential and thus, the shape of the companion-driven spirals \citep{Artymowicz1994}.}

As the disk inclination increases, this distinction fades, as the difference in sky-projected area between the near and far sides of the disk, as well as the contribution of the $v_r$ and $v_{\phi}$ to the line-of-sight velocity, introduces odd frequency content even in the shadow-driven case. The $v_{\phi}$ component in particular, which represents fast, quasi-Keplerian background flow, introduces artifacts in moment-1 maps whenever it varies on a length scale shorter than the beam-convolution scale. These artifacts are most prominent in the inner disk---growing with increasing inclination due to sky-projection effects---and in Fourier space, manifest as high-frequency content. Computing the true first moment of the spectrum---as opposed to \revision{using} \texttt{bettermoments}, which estimates the line centroid---would mitigate these issues, but at the cost of suppressing features in the outer disk.

In Figure \ref{fig:modal_analysis_1}, we present power spectra for the $J = 2-1$ transitions of all three of the CO isotopologues in our work, following the pattern of Figure \ref{fig:moment_1_1}; all presented disks are face-on. As in Fig. \ref{fig:modal_analysis_0}, we see that even-numbered modes in general, and the $m = 2$ mode in particular, are strong for the shadow-driven case. The power in $m = 2$ is weaker for \molthirteenco \ than for \moltwelveco, and for \molceighteeno \ weaker still. \revision{As previously mentioned, the} planet-driven case differs \revision{from the shadow-driven case} in that its power spectrum contains odd frequency content; \revision{moreover, we find there to be} no clear trend in mode strength \revision{with respect to the altitude of the optical surface of the relevant transition.} We attribute the higher-order harmonics in the power spectra of \molthirteenco \ in the shadow-driven case, and \molceighteeno \ in the spiral-driven case, to the sharp interfaces created by \texttt{bettermoments} between approaching and receding regions.

\section{Conclusions}\label{sec:conclusion}
With the \texttt{PLUTO} hydrodynamics code, we have re-run one of the radiative-hydrodynamics simulations in \cite{Muley2024b}, in which two-armed, shadow-driven spirals were seen to develop exterior to sufficiently deep planet-carved gaps in protoplanetary disks with. For comparison, we also ran a simulation with $\beta$-cooling and a widely-separated multi-Jupiter-mass planet, which drove two-armed interior spirals according to the classical theory of disk-planet interaction \citep{Goldreich1978,Goldreich1979,Bae2018a}. Using the \texttt{RADMC3D} MCRT code, we have post-processed these to obtain mock images and datacubes in wavelengths tracing both gas and small, entrained dust grains. These have been analyzed to produce moment maps, line profiles, and modal decompositions, which encode telltale signs of the underlying spiral driving mechanism.

Specifically, we have found that the double-armed, shadow-driven spirals in \cite{Muley2024b} are associated with strong kinematic signatures, particularly when viewed face-on, due to the substantial vertical motion associated with shadow-driven spirals. By contrast, the double-armed inner Lindblad spirals produced in the classical picture of disk-planet interaction are of comparatively limited prominence in the kinematics. Beyond intrinsic differences in sky-projected velocity profiles, the higher wavenumbers of the planet-driven spirals with respect to the shadow-driven ones ($m \approx 1/h$ versus $m = 2$) mean their angular widths are smaller, meaning that they can be more easily washed out by beam convolution. In Fourier space, the moment-1 maps of \revision{spirals driven by shadows and by planetary-mass companions} can be distinguished in face-on observations by the predominantly even frequency content of the \revision{former}. For inclined disks, the differing angles of the optical surfaces on either side of the disk with respect to the plane of the sky create a bilateral asymmetry in the projected sizes of disk features, introducing odd frequency content even \revision{in} the shadow-driven case.

\revision{The kinematic signatures we identify would be accessible to surveys such as exoALMA \citep[][and subsequent papers in series]{Teague2025}} provide a useful heuristic to distinguish between potentially shadow-driven \revision{ and classically} planet-driven cases in \revision{systems with two-armed spiral structure in scattered light}. \revision{The exterior, multi-Jupiter-mass planets required to trigger the latter mechanism would, in turn, be particularly promising targets for direct-imaging campaigns. \footnote{\revision{Interior planets associated with shadow-driven, gap-edge spirals would be less detectable due to lower contrast, and because their masses need not be super-Jupiter. In exceptionally high-mass cases, however \citep[for instance, the putative detection in SAO 206462 by][]{Maio2025}, they may nevertheless be detectable.}}}

We envision our study as a starting point for a wide range of potential future investigations. Besides shadows and classical disk-planet interaction, other mechanisms such as as infalling gas \citep[e.g.,][]{Kuznetsova2022,Huehn2025}, stellar flybys \citep[e.g.,]{Smallwood2023}, and the gravitational instability \citep[e.g.,][]{Dong2015} are also capable of creating $m = 2$ spirals, and it would be worthwhile to use simulations and MCRT post-processing to study their distinguishing features in molecular lines and near-infrared scattered light. Within the existing scope, using such as short characteristics \citep{Davis2012}, discrete ordinates \citep{Jiang2021}, or half-moment methods \citep{MelonFuksman2025} are capable of accommodating crossing light beams, making the disk's temperature structure more accurate, improving the handling of shadows, and enabling the incorporation of accreting planets alongside shadow-driven spirals. Additionally, one could vary the dust-to-gas ratio or grain size distribution, and with it, the gas-grain thermal coupling time. Not only might this affect the shadow-driving mechanism---dependent as it is on the response of the disk to radiation---but also disk-planet interaction itself. For long thermal relaxation times, planets are known to excite spirals at buoyancy resonances \citep{Zhu2012,Lubow2014}, typified by strong vertical motions near the planet's orbital radius. Such spirals, distinct in morphology from shadow-driven vertical motions, would produce their own kinematic signatures \citep{Bae21} and serve as signposts of the planet's existence.

In post-processing, testing different isotopologue ratios and concentrations would change the emitting layers, providing better coverage of the full parameter space spanned by observed systems. More ambitiously, one could use a radiative-transfer code capable of computing non-LTE level populations from the radiation field, such as \texttt{SPARX}\footnote{\href{https://charms.asiaa.sinica.edu.tw/sparx/}{https://charms.asiaa.sinica.edu.tw/sparx/}} or \texttt{SKIRT} \citep{Matsumoto2023}. One could also move beyond convolving mock images with a Gaussian beam, and make use of the ALMA-specific \texttt{CASA} pipeline to construct more realistic mock observations. This would be of particular value in the context of our modal analysis; our current approach necessarily restricts us to Fourier transforming the image-plane moment-1 maps, but it may be worthwhile to see if the telltale spectral patterns of shadow- and planet-driven spirals would be visible in the $uv$-plane. In addition, use of such tools would offer insight into the telescope configurations and on-source time required to identify features of interest \citep[e.g.,][]{Speedie2022}.

The present study underscores the importance of using disk observations at different wavelengths, tracing different layers and components, when trying to constrain the underlying physical scenario giving rise to its morphology. A natural extension would be to also model the millimeter/submillimeter emission, which traces the large-dust population, and has been of significant interest to observational campaigns \citep[e.g., DSHARP][]{Andrews2018} aimed at understanding disk substructure. The simplest approach would be to take the hydrodynamical snapshot, and use equilibrium prescriptions for the vertical settling and radial accumulation of large grains to compute their density field for radiative-transfer post-processing \citep[e.g.,][]{Hashimoto2015,Dullemond2018,Muley2019,Brown2021}. However, the fact that disk-planet interaction is a dynamical problem makes it highly desirable to self-consistently incorporate the kinematic and thermal coupling between gas, dust, and radiation during the hydrodynamical simulations themselves. While schemes for multispecies dust dynamics exist \citep[e.g.,][]{BenitezLlambay2019,Huang2022,Krapp2024}, multispecies thermodynamics would require generalizing the existing three-temperature scheme of \cite{Muley2023}. This is a substantial undertaking requiring the development and testing of new numerical methods, which we defer to future studies.

\begin{acknowledgements}
We thank Jiaqing Bi, Til Birnstiel, Mario Flock, Margot Leemker, Anna Penzlin, Matthias Samland, Volker Springel, and Alexandros Ziampras for useful discussions which improved the quality of this work. \revision{Furthermore, we thank the referee for a helpful and informative report.} Numerical simulations were carried out on the Raven cluster of the Max-Planck-Gesellschaft and the Vera Cluster of the Max-Planck-Institut f\"ur Astronomie, both hosted by the Max Planck Computing and Data Facility (MPCDF) in Garching bei München.

\end{acknowledgements}

\bibliography{threetemp_spiral_aanda}

@ARTICLE{Speedie2022,
       author = {{Speedie}, Jessica and {Booth}, Richard A. and {Dong}, Ruobing},
        title = "{Observing Planet-driven Dust Spirals with ALMA}",
      journal = {\apj},
         year = 2022,
        month = may,
       volume = {930},
       number = {1},
          eid = {40},
        pages = {40},
          doi = {10.3847/1538-4357/ac5cc0},
archivePrefix = {arXiv},
       eprint = {2203.00692},
 primaryClass = {astro-ph.EP},
       adsurl = {https://ui.adsabs.harvard.edu/abs/2022ApJ...930...40S}
}

@ARTICLE{MelonFuksman2021,
       author = {{Melon Fuksman}, Julio David and {Klahr}, Hubert and {Flock}, Mario and {Mignone}, Andrea},
        title = "{A Two-moment Radiation Hydrodynamics Scheme Applicable to Simulations of Planet Formation in Circumstellar Disks}",
      journal = {\apj},
         year = 2021,
        month = jan,
       volume = {906},
       number = {2},
          eid = {78},
        pages = {78},
          doi = {10.3847/1538-4357/abc879},
archivePrefix = {arXiv},
       eprint = {2005.01785},
 primaryClass = {astro-ph.EP},
       adsurl = {https://ui.adsabs.harvard.edu/abs/2021ApJ...906...78M}
}

@ARTICLE{MelonFuksman2022,
       author = {{Melon Fuksman}, Julio David and {Klahr}, Hubert},
        title = "{No Self-shadowing Instability in 2D Radiation Hydrodynamical Models of Irradiated Protoplanetary Disks}",
      journal = {\apj},
         year = 2022,
        month = sep,
       volume = {936},
       number = {1},
          eid = {16},
        pages = {16},
          doi = {10.3847/1538-4357/ac7fee},
archivePrefix = {arXiv},
       eprint = {2207.05106},
 primaryClass = {astro-ph.EP},
       adsurl = {https://ui.adsabs.harvard.edu/abs/2022ApJ...936...16M}
}

@ARTICLE{Krieger2020,
       author = {{Krieger}, A. and {Wolf}, S.},
        title = "{Unbiased Monte Carlo continuum radiative transfer in optically thick regions}",
      journal = {\aap},
         year = 2020,
        month = mar,
       volume = {635},
          eid = {A148},
        pages = {A148},
          doi = {10.1051/0004-6361/201937355},
archivePrefix = {arXiv},
       eprint = {2003.13334},
 primaryClass = {astro-ph.IM},
       adsurl = {https://ui.adsabs.harvard.edu/abs/2020A&A...635A.148K}
}

@ARTICLE{Bae21,
       author = {{Bae}, Jaehan and {Teague}, Richard and {Zhu}, Zhaohuan},
        title = "{Observational Signature of Tightly Wound Spirals Driven by Buoyancy Resonances in Protoplanetary Disks}",
      journal = {\apj},
         year = 2021,
        month = may,
       volume = {912},
       number = {1},
          eid = {56},
        pages = {56},
          doi = {10.3847/1538-4357/abe45e},
archivePrefix = {arXiv},
       eprint = {2102.03899},
 primaryClass = {astro-ph.EP},
       adsurl = {https://ui.adsabs.harvard.edu/abs/2021ApJ...912...56B}
}

@ARTICLE{MelonFuksman2019,
       author = {{Melon Fuksman}, Julio David and {Mignone}, Andrea},
        title = "{A Radiative Transfer Module for Relativistic Magnetohydrodynamics in the PLUTO Code}",
      journal = {\apjs},
         year = 2019,
        month = jun,
       volume = {242},
       number = {2},
          eid = {20},
        pages = {20},
          doi = {10.3847/1538-4365/ab18ff},
archivePrefix = {arXiv},
       eprint = {1903.10456},
 primaryClass = {astro-ph.IM},
       adsurl = {https://ui.adsabs.harvard.edu/abs/2019ApJS..242...20M}
}

@ARTICLE{Zhu2015,
       author = {{Zhu}, Zhaohuan and {Dong}, Ruobing and {Stone}, James M. and {Rafikov}, Roman R.},
        title = "{The Structure of Spiral Shocks Excited by Planetary-mass Companions}",
      journal = {\apj},
         year = 2015,
        month = nov,
       volume = {813},
       number = {2},
          eid = {88},
        pages = {88},
          doi = {10.1088/0004-637X/813/2/88},
archivePrefix = {arXiv},
       eprint = {1507.03599},
 primaryClass = {astro-ph.SR},
       adsurl = {https://ui.adsabs.harvard.edu/abs/2015ApJ...813...88Z}
}

@ARTICLE{Goldreich1978,
       author = {{Goldreich}, P. and {Tremaine}, S.},
        title = "{The excitation and evolution of density waves.}",
      journal = {\apj},
         year = 1978,
        month = jun,
       volume = {222},
        pages = {850-858},
          doi = {10.1086/156203},
       adsurl = {https://ui.adsabs.harvard.edu/abs/1978ApJ...222..850G}
}

@ARTICLE{Goldreich1979,
       author = {{Goldreich}, P. and {Tremaine}, S.},
        title = "{The excitation of density waves at the Lindblad and corotation resonances by an external potential.}",
      journal = {\apj},
         year = 1979,
        month = nov,
       volume = {233},
        pages = {857-871},
          doi = {10.1086/157448},
       adsurl = {https://ui.adsabs.harvard.edu/abs/1979ApJ...233..857G}
}

@ARTICLE{Goldreich1980,
       author = {{Goldreich}, P. and {Tremaine}, S.},
        title = "{Disk-satellite interactions.}",
      journal = {\apj},
         year = 1980,
        month = oct,
       volume = {241},
        pages = {425-441},
          doi = {10.1086/158356},
       adsurl = {https://ui.adsabs.harvard.edu/abs/1980ApJ...241..425G}
}

@ARTICLE{Ren2018,
       author = {{Ren}, Bin and {Dong}, Ruobing and {Esposito}, Thomas M. and {Pueyo}, Laurent and {Debes}, John H. and {Poteet}, Charles A. and {Choquet}, {\'E}lodie and {Benisty}, Myriam and {Chiang}, Eugene and {Grady}, Carol A. and {Hines}, Dean C. and {Schneider}, Glenn and {Soummer}, R{\'e}mi},
        title = "{A Decade of MWC 758 Disk Images: Where Are the Spiral-arm-driving Planets?}",
      journal = {\apjl},
         year = 2018,
        month = apr,
       volume = {857},
       number = {1},
          eid = {L9},
        pages = {L9},
          doi = {10.3847/2041-8213/aab7f5},
archivePrefix = {arXiv},
       eprint = {1803.06776},
 primaryClass = {astro-ph.EP},
       adsurl = {https://ui.adsabs.harvard.edu/abs/2018ApJ...857L...9R}
}

@ARTICLE{Teague2019,
       author = {{Teague}, Richard and {Bae}, Jaehan and {Huang}, Jane and {Bergin}, Edwin A.},
        title = "{Spiral Structure in the Gas Disk of TW Hya}",
      journal = {\apjl},
         year = 2019,
        month = oct,
       volume = {884},
       number = {2},
          eid = {L56},
        pages = {L56},
          doi = {10.3847/2041-8213/ab4a83},
archivePrefix = {arXiv},
       eprint = {1910.01532},
 primaryClass = {astro-ph.EP},
       adsurl = {https://ui.adsabs.harvard.edu/abs/2019ApJ...884L..56T}
}

@ARTICLE{Wagner2019,
       author = {{Wagner}, Kevin and {Stone}, Jordan M. and {Spalding}, Eckhart and {Apai}, Daniel and {Dong}, Ruobing and {Ertel}, Steve and {Leisenring}, Jarron and {Webster}, Ryan},
        title = "{Thermal Infrared Imaging of MWC 758 with the Large Binocular Telescope: Planetary-driven Spiral Arms?}",
      journal = {\apj},
         year = 2019,
        month = sep,
       volume = {882},
       number = {1},
          eid = {20},
        pages = {20},
          doi = {10.3847/1538-4357/ab32ea},
archivePrefix = {arXiv},
       eprint = {1907.06655},
 primaryClass = {astro-ph.SR},
       adsurl = {https://ui.adsabs.harvard.edu/abs/2019ApJ...882...20W}
}

@ARTICLE{Dong2016,
       author = {{Dong}, Ruobing and {Fung}, Jeffrey and {Chiang}, Eugene},
        title = "{How Spirals and Gaps Driven by Companions in Protoplanetary Disks Appear in Scattered Light at Arbitrary Viewing Angles}",
      journal = {\apj},
         year = 2016,
        month = jul,
       volume = {826},
       number = {1},
          eid = {75},
        pages = {75},
          doi = {10.3847/0004-637X/826/1/75},
archivePrefix = {arXiv},
       eprint = {1602.04814},
 primaryClass = {astro-ph.EP},
       adsurl = {https://ui.adsabs.harvard.edu/abs/2016ApJ...826...75D}
}

@ARTICLE{Dong2017,
       author = {{Dong}, Ruobing and {Fung}, Jeffrey},
        title = "{How Bright are Planet-induced Spiral Arms in Scattered Light?}",
      journal = {\apj},
         year = 2017,
        month = jan,
       volume = {835},
       number = {1},
          eid = {38},
        pages = {38},
          doi = {10.3847/1538-4357/835/1/38},
archivePrefix = {arXiv},
       eprint = {1612.00446},
 primaryClass = {astro-ph.EP},
       adsurl = {https://ui.adsabs.harvard.edu/abs/2017ApJ...835...38D}
}

@ARTICLE{Juhasz2018,
       author = {{Juh{\'a}sz}, Attila and {Rosotti}, Giovanni P.},
        title = "{Spiral arms in thermally stratified protoplanetary discs}",
      journal = {\mnras},
         year = 2018,
        month = feb,
       volume = {474},
       number = {1},
        pages = {L32-L36},
          doi = {10.1093/mnrasl/slx182},
archivePrefix = {arXiv},
       eprint = {1711.03559},
 primaryClass = {astro-ph.EP},
       adsurl = {https://ui.adsabs.harvard.edu/abs/2018MNRAS.474L..32J}
}

@ARTICLE{Fung2015,
       author = {{Fung}, Jeffrey and {Dong}, Ruobing},
        title = "{Inferring Planet Mass from Spiral Structures in Protoplanetary Disks}",
      journal = {\apjl},
         year = 2015,
        month = dec,
       volume = {815},
       number = {2},
          eid = {L21},
        pages = {L21},
          doi = {10.1088/2041-8205/815/2/L21},
archivePrefix = {arXiv},
       eprint = {1511.01178},
 primaryClass = {astro-ph.EP},
       adsurl = {https://ui.adsabs.harvard.edu/abs/2015ApJ...815L..21F}
}

@ARTICLE{Huang2022,
       author = {{Huang}, Pinghui and {Bai}, Xue-Ning},
        title = "{A Multifluid Dust Module in Athena++: Algorithms and Numerical Tests}",
      journal = {\apjs},
         year = 2022,
        month = sep,
       volume = {262},
       number = {1},
          eid = {11},
        pages = {11},
          doi = {10.3847/1538-4365/ac76cb},
archivePrefix = {arXiv},
       eprint = {2206.01023},
 primaryClass = {astro-ph.EP},
       adsurl = {https://ui.adsabs.harvard.edu/abs/2022ApJS..262...11H}
}

@ARTICLE{BenitezLlambay2019,
       author = {{Ben{\'\i}tez-Llambay}, Pablo and {Krapp}, Leonardo and {Pessah}, Martin E.},
        title = "{Asymptotically Stable Numerical Method for Multispecies Momentum Transfer: Gas and Multifluid Dust Test Suite and Implementation in FARGO3D}",
      journal = {\apjs},
         year = 2019,
        month = apr,
       volume = {241},
       number = {2},
          eid = {25},
        pages = {25},
          doi = {10.3847/1538-4365/ab0a0e},
archivePrefix = {arXiv},
       eprint = {1811.07925},
 primaryClass = {astro-ph.EP},
       adsurl = {https://ui.adsabs.harvard.edu/abs/2019ApJS..241...25B}
}

@ARTICLE{Mignone2007,
       author = {{Mignone}, A. and {Bodo}, G. and {Massaglia}, S. and {Matsakos}, T. and {Tesileanu}, O. and {Zanni}, C. and {Ferrari}, A.},
        title = "{PLUTO: A Numerical Code for Computational Astrophysics}",
      journal = {\apjs},
         year = 2007,
        month = may,
       volume = {170},
       number = {1},
        pages = {228-242},
          doi = {10.1086/513316},
archivePrefix = {arXiv},
       eprint = {astro-ph/0701854},
 primaryClass = {astro-ph},
       adsurl = {https://ui.adsabs.harvard.edu/abs/2007ApJS..170..228M}
}

@ARTICLE{Pinte2022,
       author = {{Pinte}, Christophe and {Teague}, Richard and {Flaherty}, Kevin and {Hall}, Cassandra and {Facchini}, Stefano and {Casassus}, Simon},
        title = "{Kinematic Structures in Planet-Forming Disks}",
      journal = {arXiv e-prints},
         year = 2022,
        month = mar,
          eid = {arXiv:2203.09528},
        pages = {arXiv:2203.09528},
          doi = {10.48550/arXiv.2203.09528},
archivePrefix = {arXiv},
       eprint = {2203.09528},
 primaryClass = {astro-ph.EP},
       adsurl = {https://ui.adsabs.harvard.edu/abs/2022arXiv220309528P}
}

@ARTICLE{Bae2018b,
       author = {{Bae}, Jaehan and {Zhu}, Zhaohuan},
        title = "{Planet-driven Spiral Arms in Protoplanetary Disks. II. Implications}",
      journal = {\apj},
         year = 2018,
        month = jun,
       volume = {859},
       number = {2},
          eid = {119},
        pages = {119},
          doi = {10.3847/1538-4357/aabf93},
archivePrefix = {arXiv},
       eprint = {1711.08166},
 primaryClass = {astro-ph.EP},
       adsurl = {https://ui.adsabs.harvard.edu/abs/2018ApJ...859..119B}
}

@ARTICLE{Bae2018a,
       author = {{Bae}, Jaehan and {Zhu}, Zhaohuan},
        title = "{Planet-driven Spiral Arms in Protoplanetary Disks. I. Formation Mechanism}",
      journal = {\apj},
         year = 2018,
        month = jun,
       volume = {859},
       number = {2},
          eid = {118},
        pages = {118},
          doi = {10.3847/1538-4357/aabf8c},
archivePrefix = {arXiv},
       eprint = {1711.08161},
 primaryClass = {astro-ph.EP},
       adsurl = {https://ui.adsabs.harvard.edu/abs/2018ApJ...859..118B}
}

@ARTICLE{FSC14,
       author = {{Fung}, Jeffrey and {Shi}, Ji-Ming and {Chiang}, Eugene},
        title = "{How Empty are Disk Gaps Opened by Giant Planets?}",
      journal = {\apj},
         year = 2014,
        month = feb,
       volume = {782},
       number = {2},
          eid = {88},
        pages = {88},
          doi = {10.1088/0004-637X/782/2/88},
archivePrefix = {arXiv},
       eprint = {1310.0156},
 primaryClass = {astro-ph.EP},
       adsurl = {https://ui.adsabs.harvard.edu/abs/2014ApJ...782...88F}
}

@ARTICLE{Fung2016,
       author = {{Fung}, Jeffrey and {Chiang}, Eugene},
        title = "{Gap Opening in 3D: Single-planet Gaps}",
      journal = {\apj},
         year = 2016,
        month = dec,
       volume = {832},
       number = {2},
          eid = {105},
        pages = {105},
          doi = {10.3847/0004-637X/832/2/105},
archivePrefix = {arXiv},
       eprint = {1606.02299},
 primaryClass = {astro-ph.EP},
       adsurl = {https://ui.adsabs.harvard.edu/abs/2016ApJ...832..105F}
}

@ARTICLE{Dong2019,
       author = {{Dong}, Ruobing and {Liu}, Sheng-Yuan and {Fung}, Jeffrey},
        title = "{Observational Signatures of Planets in Protoplanetary Disks: Planet-induced Line Broadening in Gaps}",
      journal = {\apj},
         year = 2019,
        month = jan,
       volume = {870},
       number = {2},
          eid = {72},
        pages = {72},
          doi = {10.3847/1538-4357/aaf38e},
archivePrefix = {arXiv},
       eprint = {1811.09629},
 primaryClass = {astro-ph.EP},
       adsurl = {https://ui.adsabs.harvard.edu/abs/2019ApJ...870...72D}
}

@ARTICLE{Lubow2014,
       author = {{Lubow}, Stephen H. and {Zhu}, Zhaohuan},
        title = "{An Analytic Model for Buoyancy Resonances in Protoplanetary Disks}",
      journal = {\apj},
         year = 2014,
        month = apr,
       volume = {785},
       number = {1},
          eid = {32},
        pages = {32},
          doi = {10.1088/0004-637X/785/1/32},
archivePrefix = {arXiv},
       eprint = {1402.4162},
 primaryClass = {astro-ph.EP},
       adsurl = {https://ui.adsabs.harvard.edu/abs/2014ApJ...785...32L}
}

@ARTICLE{Muley2023,
       author = {{Muley}, Dhruv and {Melon Fuksman}, Julio David and {Klahr}, Hubert},
        title = "{Three-temperature radiation hydrodynamics with PLUTO. Tests and applications in the context of protoplanetary disks}",
      journal = {\aap},
         year = 2023,
        month = oct,
       volume = {678},
          eid = {A162},
        pages = {A162},
          doi = {10.1051/0004-6361/202347101},
archivePrefix = {arXiv},
       eprint = {2308.03504},
 primaryClass = {astro-ph.IM},
       adsurl = {https://ui.adsabs.harvard.edu/abs/2023A&A...678A.162M}
}

@ARTICLE{Pinte2019,
       author = {{Pinte}, C. and {van der Plas}, G. and {M{\'e}nard}, F. and {Price}, D.~J. and {Christiaens}, V. and {Hill}, T. and {Mentiplay}, D. and {Ginski}, C. and {Choquet}, E. and {Boehler}, Y. and {Duch{\^e}ne}, G. and {Perez}, S. and {Casassus}, S.},
        title = "{Kinematic detection of a planet carving a gap in a protoplanetary disk}",
      journal = {Nature Astronomy},
         year = 2019,
        month = aug,
       volume = {3},
        pages = {1109-1114},
          doi = {10.1038/s41550-019-0852-6},
archivePrefix = {arXiv},
       eprint = {1907.02538},
 primaryClass = {astro-ph.SR},
       adsurl = {https://ui.adsabs.harvard.edu/abs/2019NatAs...3.1109P}
}

@ARTICLE{Fung2015b,
       author = {{Fung}, Jeffrey and {Artymowicz}, Pawel and {Wu}, Yanqin},
        title = "{The 3D Flow Field Around an Embedded Planet}",
      journal = {\apj},
         year = 2015,
        month = oct,
       volume = {811},
       number = {2},
          eid = {101},
        pages = {101},
          doi = {10.1088/0004-637X/811/2/101},
archivePrefix = {arXiv},
       eprint = {1505.03152},
 primaryClass = {astro-ph.EP},
       adsurl = {https://ui.adsabs.harvard.edu/abs/2015ApJ...811..101F}
}

@ARTICLE{BarrazaAlfaro2023,
       author = {{Barraza-Alfaro}, Marcelo and {Flock}, Mario and {Henning}, Thomas},
        title = "{Kinematic signatures of planet-disk interactions in VSI-turbulent protoplanetary disks}",
      journal = {arXiv e-prints},
         year = 2023,
        month = oct,
          eid = {arXiv:2310.18484},
        pages = {arXiv:2310.18484},
          doi = {10.48550/arXiv.2310.18484},
archivePrefix = {arXiv},
       eprint = {2310.18484},
 primaryClass = {astro-ph.EP},
       adsurl = {https://ui.adsabs.harvard.edu/abs/2023arXiv231018484B}
}

@ARTICLE{Zhu2012,
       author = {{Zhu}, Zhaohuan and {Stone}, James M. and {Rafikov}, Roman R.},
        title = "{Planet-Disk Interaction in Three Dimensions: The Importance of Buoyancy Waves}",
      journal = {\apjl},
         year = 2012,
        month = oct,
       volume = {758},
       number = {2},
          eid = {L42},
        pages = {L42},
          doi = {10.1088/2041-8205/758/2/L42},
archivePrefix = {arXiv},
       eprint = {1209.4358},
 primaryClass = {astro-ph.EP},
       adsurl = {https://ui.adsabs.harvard.edu/abs/2012ApJ...758L..42Z}
}

@ARTICLE{Teague2022,
       author = {{Teague}, Richard and {Bae}, Jaehan and {Andrews}, Sean M. and {Benisty}, Myriam and {Bergin}, Edwin A. and {Facchini}, Stefano and {Huang}, Jane and {Longarini}, Cristiano and {Wilner}, David},
        title = "{Mapping the Complex Kinematic Substructure in the TW Hya Disk}",
      journal = {\apj},
         year = 2022,
        month = sep,
       volume = {936},
       number = {2},
          eid = {163},
        pages = {163},
          doi = {10.3847/1538-4357/ac88ca},
archivePrefix = {arXiv},
       eprint = {2208.04837},
 primaryClass = {astro-ph.EP},
       adsurl = {https://ui.adsabs.harvard.edu/abs/2022ApJ...936..163T}
}

@article{Pinte2018,
	author = {C. Pinte and D. J. Price and F. M{\'e}nard and G. Duch{\^e}ne and W. R. F. Dent and T. Hill and I. de Gregorio-Monsalvo and A. Hales and D. Mentiplay},
	doi = {10.3847/2041-8213/aac6dc},
	journal = {The Astrophysical Journal Letters},
	month = {jun},
	number = {1},
	pages = {L13},
	publisher = {The American Astronomical Society},
	title = {Kinematic Evidence for an Embedded Protoplanet in a Circumstellar Disk},
	url = {https://dx.doi.org/10.3847/2041-8213/aac6dc},
	volume = {860},
	year = {2018}}

@ARTICLE{Davis2012,
       author = {{Davis}, Shane W. and {Stone}, James M. and {Jiang}, Yan-Fei},
        title = "{A Radiation Transfer Solver for Athena Using Short Characteristics}",
      journal = {\apjs},
         year = 2012,
        month = mar,
       volume = {199},
       number = {1},
          eid = {9},
        pages = {9},
          doi = {10.1088/0067-0049/199/1/9},
archivePrefix = {arXiv},
       eprint = {1201.2222},
 primaryClass = {astro-ph.IM},
       adsurl = {https://ui.adsabs.harvard.edu/abs/2012ApJS..199....9D}
}

@ARTICLE{Bi2023,
       author = {{Bi}, Jiaqing and {Lin}, Min-Kai and {Dong}, Ruobing},
        title = "{Gap-opening Planets Make Dust Rings Wider}",
      journal = {\apj},
         year = 2023,
        month = jan,
       volume = {942},
       number = {2},
          eid = {80},
        pages = {80},
          doi = {10.3847/1538-4357/aca1b1},
archivePrefix = {arXiv},
       eprint = {2210.11488},
 primaryClass = {astro-ph.EP},
       adsurl = {https://ui.adsabs.harvard.edu/abs/2023ApJ...942...80B}
}

@ARTICLE{Montesinos2018,
       author = {{Montesinos}, Mat{\'\i}as and {Cuello}, Nicol{\'a}s},
        title = "{Planetary-like spirals caused by moving shadows in transition discs}",
      journal = {\mnras},
         year = 2018,
        month = mar,
       volume = {475},
       number = {1},
        pages = {L35-L39},
          doi = {10.1093/mnrasl/sly001},
archivePrefix = {arXiv},
       eprint = {1712.09157},
 primaryClass = {astro-ph.EP},
       adsurl = {https://ui.adsabs.harvard.edu/abs/2018MNRAS.475L..35M}
}

@ARTICLE{Shuai2022,
       author = {{Shuai}, Linling and {Ren}, Bin B. and {Dong}, Ruobing and {Zhou}, Xingyu and {Pueyo}, Laurent and {De Rosa}, Robert J. and {Fang}, Taotao and {Mawet}, Dimitri},
        title = "{Stellar Flyby Analysis for Spiral Arm Hosts with Gaia DR3}",
      journal = {\apjs},
         year = 2022,
        month = dec,
       volume = {263},
       number = {2},
          eid = {31},
        pages = {31},
          doi = {10.3847/1538-4365/ac98fd},
archivePrefix = {arXiv},
       eprint = {2210.03725},
 primaryClass = {astro-ph.EP},
       adsurl = {https://ui.adsabs.harvard.edu/abs/2022ApJS..263...31S}
}

@ARTICLE{AsensioTorres2021,
       author = {{Asensio-Torres}, R. and {Henning}, Th. and {Cantalloube}, F. and {Pinilla}, P. and {Mesa}, D. and {Garufi}, A. and {Jorquera}, S. and {Gratton}, R. and {Chauvin}, G. and {Szul{\'a}gyi}, J. and {van Boekel}, R. and {Dong}, R. and {Marleau}, G. -D. and {Benisty}, M. and {Villenave}, M. and {Bergez-Casalou}, C. and {Desgrange}, C. and {Janson}, M. and {Keppler}, M. and {Langlois}, M. and {M{\'e}nard}, F. and {Rickman}, E. and {Stolker}, T. and {Feldt}, M. and {Fusco}, T. and {Gluck}, L. and {Pavlov}, A. and {Ramos}, J.},
        title = "{Perturbers: SPHERE detection limits to planetary-mass companions in protoplanetary disks}",
      journal = {\aap},
         year = 2021,
        month = aug,
       volume = {652},
          eid = {A101},
        pages = {A101},
          doi = {10.1051/0004-6361/202140325},
archivePrefix = {arXiv},
       eprint = {2103.05377},
 primaryClass = {astro-ph.EP},
       adsurl = {https://ui.adsabs.harvard.edu/abs/2021A&A...652A.101A}
}

@ARTICLE{Montesinos2016,
       author = {{Montesinos}, Mat{\'\i}as and {Perez}, Sebastian and {Casassus}, Simon and {Marino}, Sebastian and {Cuadra}, Jorge and {Christiaens}, Valentin},
        title = "{Spiral Waves Triggered by Shadows in Transition Disks}",
      journal = {\apjl},
         year = 2016,
        month = may,
       volume = {823},
       number = {1},
          eid = {L8},
        pages = {L8},
          doi = {10.3847/2041-8205/823/1/L8},
archivePrefix = {arXiv},
       eprint = {1601.07912},
 primaryClass = {astro-ph.EP},
       adsurl = {https://ui.adsabs.harvard.edu/abs/2016ApJ...823L...8M}
}

@ARTICLE{Muley2024,
       author = {{Muley}, D. and {Melon Fuksman}, Julio David and {Klahr}, Hubert},
        title = "{Three-temperature radiation hydrodynamics with PLUTO: Thermal and kinematic signatures of planet-driven spirals}",
      journal = {\aap},
         year = 2024,
        month = feb,
        volume = {submitted}
}

@ARTICLE{Cugno2024,
       author = {{Cugno}, Gabriele and {Leisenring}, Jarron and {Wagner}, Kevin R. and {Mullin}, Camryn and {Dong}, Roubing and {Greene}, Thomas and {Johnstone}, Doug and {Meyer}, Michael R. and {Wolff}, Schuyler G. and {Beichman}, Charles and {Boyer}, Martha and {Horner}, Scott and {Hodapp}, Klaus and {Kelly}, Doug and {McCarthy}, Don and {Roellig}, Thomas and {Rieke}, George and {Rieke}, Marcia and {Stansberry}, John and {Young}, Erick},
        title = "{JWST/NIRCam Imaging of Young Stellar Objects. II. Deep Constraints on Giant Planets and a Planet Candidate Outside of the Spiral Disk Around SAO 206462}",
      journal = {\aj},
         year = 2024,
        month = apr,
       volume = {167},
       number = {4},
          eid = {182},
        pages = {182},
          doi = {10.3847/1538-3881/ad1ffc},
archivePrefix = {arXiv},
       eprint = {2401.02834},
 primaryClass = {astro-ph.EP},
       adsurl = {https://ui.adsabs.harvard.edu/abs/2024AJ....167..182C}
}

@ARTICLE{Wagner2024,
       author = {{Wagner}, Kevin and {Leisenring}, Jarron and {Cugno}, Gabriele and {Mullin}, Camryn and {Dong}, Ruobing and {Wolff}, Schuyler G. and {Greene}, Thomas and {Johnstone}, Doug and {Meyer}, Michael R. and {Beichman}, Charles and {Boyer}, Martha and {Horner}, Scott and {Hodapp}, Klaus and {Kelly}, Doug and {McCarthy}, Don and {Roellig}, Tom and {Rieke}, George and {Rieke}, Marcia and {Sitko}, Michael and {Stansberry}, John and {Young}, Erick},
        title = "{JWST/NIRCam Imaging of Young Stellar Objects. I. Constraints on Planets Exterior to the Spiral Disk Around MWC 758}",
      journal = {\aj},
         year = 2024,
        month = apr,
       volume = {167},
       number = {4},
          eid = {181},
        pages = {181},
          doi = {10.3847/1538-3881/ad11d5},
archivePrefix = {arXiv},
       eprint = {2401.02830},
 primaryClass = {astro-ph.EP},
       adsurl = {https://ui.adsabs.harvard.edu/abs/2024AJ....167..181W}
}

@ARTICLE{Follette2023,
       author = {{Follette}, Katherine B. and {Close}, Laird M. and {Males}, Jared R. and {Ward-Duong}, Kimberly and {Balmer}, William O. and {Redai}, J{\'e}a Adams and {Morales}, Julio and {Sarosi}, Catherine and {Dacus}, Beck and {De Rosa}, Robert J. and {Garcia Toro}, Fernando and {Leonard}, Clare and {Macintosh}, Bruce and {Morzinski}, Katie M. and {Mullen}, Wyatt and {Palmo}, Joseph and {Saitoti}, Raymond Nzaba and {Spiro}, Elijah and {Treiber}, Helena and {Wagner}, Kevin and {Wang}, Jason and {Wang}, David and {Watson}, Alex and {Weinberger}, Alycia J.},
        title = "{The Giant Accreting Protoplanet Survey (GAPlanetS)-Results from a 6 yr Campaign to Image Accreting Protoplanets}",
      journal = {\aj},
         year = 2023,
        month = jun,
       volume = {165},
       number = {6},
          eid = {225},
        pages = {225},
          doi = {10.3847/1538-3881/acc183},
archivePrefix = {arXiv},
       eprint = {2211.02109},
 primaryClass = {astro-ph.EP},
       adsurl = {https://ui.adsabs.harvard.edu/abs/2023AJ....165..225F}
}

@ARTICLE{Desidera2021,
       author = {{Desidera}, S. and {Chauvin}, G. and {Bonavita}, M. and {Messina}, S. and {LeCoroller}, H. and {Schmidt}, T. and {Gratton}, R. and {Lazzoni}, C. and {Meyer}, M. and {Schlieder}, J. and {Cheetham}, A. and {Hagelberg}, J. and {Bonnefoy}, M. and {Feldt}, M. and {Lagrange}, A. -M. and {Langlois}, M. and {Vigan}, A. and {Tan}, T.~G. and {Hambsch}, F. -J. and {Millward}, M. and {Alcal{\'a}}, J. and {Benatti}, S. and {Brandner}, W. and {Carson}, J. and {Covino}, E. and {Delorme}, P. and {D'Orazi}, V. and {Janson}, M. and {Rigliaco}, E. and {Beuzit}, J. -L. and {Biller}, B. and {Boccaletti}, A. and {Dominik}, C. and {Cantalloube}, F. and {Fontanive}, C. and {Galicher}, R. and {Henning}, Th. and {Lagadec}, E. and {Ligi}, R. and {Maire}, A. -L. and {Menard}, F. and {Mesa}, D. and {M{\"u}ller}, A. and {Samland}, M. and {Schmid}, H.~M. and {Sissa}, E. and {Turatto}, M. and {Udry}, S. and {Zurlo}, A. and {Asensio-Torres}, R. and {Kopytova}, T. and {Rickman}, E. and {Abe}, L. and {Antichi}, J. and {Baruffolo}, A. and {Baudoz}, P. and {Baudrand}, J. and {Blanchard}, P. and {Bazzon}, A. and {Buey}, T. and {Carbillet}, M. and {Carle}, M. and {Charton}, J. and {Cascone}, E. and {Claudi}, R. and {Costille}, A. and {Deboulb{\'e}}, A. and {De Caprio}, V. and {Dohlen}, K. and {Fantinel}, D. and {Feautrier}, P. and {Fusco}, T. and {Gigan}, P. and {Giro}, E. and {Gisler}, D. and {Gluck}, L. and {Hubin}, N. and {Hugot}, E. and {Jaquet}, M. and {Kasper}, M. and {Madec}, F. and {Magnard}, Y. and {Martinez}, P. and {Maurel}, D. and {Le Mignant}, D. and {M{\"o}ller-Nilsson}, O. and {Llored}, M. and {Moulin}, T. and {Orign{\'e}}, A. and {Pavlov}, A. and {Perret}, D. and {Petit}, C. and {Pragt}, J. and {Puget}, P. and {Rabou}, P. and {Ramos}, J. and {Rigal}, F. and {Rochat}, S. and {Roelfsema}, R. and {Rousset}, G. and {Roux}, A. and {Salasnich}, B. and {Sauvage}, J. -F. and {Sevin}, A. and {Soenke}, C. and {Stadler}, E. and {Suarez}, M. and {Weber}, L. and {Wildi}, F.},
        title = "{The SPHERE infrared survey for exoplanets (SHINE). I. Sample definition and target characterization}",
      journal = {\aap},
         year = 2021,
        month = jul,
       volume = {651},
          eid = {A70},
        pages = {A70},
          doi = {10.1051/0004-6361/202038806},
archivePrefix = {arXiv},
       eprint = {2103.04366},
 primaryClass = {astro-ph.EP},
       adsurl = {https://ui.adsabs.harvard.edu/abs/2021A&A...651A..70D}
}

@ARTICLE{Biddle2024,
       author = {{Biddle}, Lauren I. and {Bowler}, Brendan P. and {Zhou}, Yifan and {Franson}, Kyle and {Zhang}, Zhoujian},
        title = "{Deep Pa$\beta$ Imaging of the Candidate Accreting Protoplanet AB Aur b}",
      journal = {arXiv e-prints},
         year = 2024,
        month = feb,
          eid = {arXiv:2402.12601},
        pages = {arXiv:2402.12601},
          doi = {10.48550/arXiv.2402.12601},
archivePrefix = {arXiv},
       eprint = {2402.12601},
 primaryClass = {astro-ph.EP},
       adsurl = {https://ui.adsabs.harvard.edu/abs/2024arXiv240212601B}
}

@ARTICLE{Zhang2024,
       author = {{Zhang}, Minghao and {Huang}, Pinghui and {Dong}, Ruobing},
        title = "{The Dependence of the Structure of Planet-opened Gaps in Protoplanetary Disks on Radiative Cooling}",
      journal = {\apj},
         year = 2024,
        month = jan,
       volume = {961},
       number = {1},
          eid = {86},
        pages = {86},
          doi = {10.3847/1538-4357/ad055c},
archivePrefix = {arXiv},
       eprint = {2310.11757},
 primaryClass = {astro-ph.EP},
       adsurl = {https://ui.adsabs.harvard.edu/abs/2024ApJ...961...86Z}
}

@ARTICLE{Muto2012,
       author = {{Muto}, T. and {Grady}, C.~A. and {Hashimoto}, J. and {Fukagawa}, M. and {Hornbeck}, J.~B. and {Sitko}, M. and {Russell}, R. and {Werren}, C. and {Cur{\'e}}, M. and {Currie}, T. and {Ohashi}, N. and {Okamoto}, Y. and {Momose}, M. and {Honda}, M. and {Inutsuka}, S. and {Takeuchi}, T. and {Dong}, R. and {Abe}, L. and {Brandner}, W. and {Brandt}, T. and {Carson}, J. and {Egner}, S. and {Feldt}, M. and {Fukue}, T. and {Goto}, M. and {Guyon}, O. and {Hayano}, Y. and {Hayashi}, M. and {Hayashi}, S. and {Henning}, T. and {Hodapp}, K.~W. and {Ishii}, M. and {Iye}, M. and {Janson}, M. and {Kandori}, R. and {Knapp}, G.~R. and {Kudo}, T. and {Kusakabe}, N. and {Kuzuhara}, M. and {Matsuo}, T. and {Mayama}, S. and {McElwain}, M.~W. and {Miyama}, S. and {Morino}, J. -I. and {Moro-Martin}, A. and {Nishimura}, T. and {Pyo}, T. -S. and {Serabyn}, E. and {Suto}, H. and {Suzuki}, R. and {Takami}, M. and {Takato}, N. and {Terada}, H. and {Thalmann}, C. and {Tomono}, D. and {Turner}, E.~L. and {Watanabe}, M. and {Wisniewski}, J.~P. and {Yamada}, T. and {Takami}, H. and {Usuda}, T. and {Tamura}, M.},
        title = "{Discovery of Small-scale Spiral Structures in the Disk of SAO 206462 (HD 135344B): Implications for the Physical State of the Disk from Spiral Density Wave Theory}",
      journal = {\apjl},
         year = 2012,
        month = apr,
       volume = {748},
       number = {2},
          eid = {L22},
        pages = {L22},
          doi = {10.1088/2041-8205/748/2/L22},
archivePrefix = {arXiv},
       eprint = {1202.6139},
 primaryClass = {astro-ph.EP},
       adsurl = {https://ui.adsabs.harvard.edu/abs/2012ApJ...748L..22M}
}

@ARTICLE{Grady2013,
       author = {{Grady}, C.~A. and {Muto}, T. and {Hashimoto}, J. and {Fukagawa}, M. and {Currie}, T. and {Biller}, B. and {Thalmann}, C. and {Sitko}, M.~L. and {Russell}, R. and {Wisniewski}, J. and {Dong}, R. and {Kwon}, J. and {Sai}, S. and {Hornbeck}, J. and {Schneider}, G. and {Hines}, D. and {Moro Mart{\'\i}n}, A. and {Feldt}, M. and {Henning}, Th. and {Pott}, J. -U. and {Bonnefoy}, M. and {Bouwman}, J. and {Lacour}, S. and {Mueller}, A. and {Juh{\'a}sz}, A. and {Crida}, A. and {Chauvin}, G. and {Andrews}, S. and {Wilner}, D. and {Kraus}, A. and {Dahm}, S. and {Robitaille}, T. and {Jang-Condell}, H. and {Abe}, L. and {Akiyama}, E. and {Brandner}, W. and {Brandt}, T. and {Carson}, J. and {Egner}, S. and {Follette}, K.~B. and {Goto}, M. and {Guyon}, O. and {Hayano}, Y. and {Hayashi}, M. and {Hayashi}, S. and {Hodapp}, K. and {Ishii}, M. and {Iye}, M. and {Janson}, M. and {Kandori}, R. and {Knapp}, G. and {Kudo}, T. and {Kusakabe}, N. and {Kuzuhara}, M. and {Mayama}, S. and {McElwain}, M. and {Matsuo}, T. and {Miyama}, S. and {Morino}, J. -I. and {Nishimura}, T. and {Pyo}, T. -S. and {Serabyn}, G. and {Suto}, H. and {Suzuki}, R. and {Takami}, M. and {Takato}, N. and {Terada}, H. and {Tomono}, D. and {Turner}, E. and {Watanabe}, M. and {Yamada}, T. and {Takami}, H. and {Usuda}, T. and {Tamura}, M.},
        title = "{Spiral Arms in the Asymmetrically Illuminated Disk of MWC 758 and Constraints on Giant Planets}",
      journal = {\apj},
         year = 2013,
        month = jan,
       volume = {762},
       number = {1},
          eid = {48},
        pages = {48},
          doi = {10.1088/0004-637X/762/1/48},
archivePrefix = {arXiv},
       eprint = {1212.1466},
 primaryClass = {astro-ph.SR},
       adsurl = {https://ui.adsabs.harvard.edu/abs/2013ApJ...762...48G}
}

@ARTICLE{Ren2024,
       author = {{Ren}, Bin B. and {Xie}, Chen and {Benisty}, Myriam and {Dong}, Ruobing and {Bae}, Jaehan and {Stolker}, Tomas and {van Holstein}, Rob G. and {Debes}, John H. and {Garufi}, Antonio and {Ginski}, Christian and {Kraus}, Stefan},
        title = "{A companion in V1247 Ori supported by motion in the pattern of the spiral arm}",
      journal = {\aap},
         year = 2024,
        month = jan,
       volume = {681},
          eid = {L2},
        pages = {L2},
          doi = {10.1051/0004-6361/202348114},
archivePrefix = {arXiv},
       eprint = {2310.15430},
 primaryClass = {astro-ph.EP},
       adsurl = {https://ui.adsabs.harvard.edu/abs/2024A&A...681L...2R}
}

@ARTICLE{Krapp2024,
       author = {{Krapp}, Leonardo and {Garrido-Deutelmoser}, Juan and {Ben{\'\i}tez-Llambay}, Pablo and {Kratter}, Kaitlin M.},
        title = "{A Fast Second-order Solver for Stiff Multifluid Dust and Gas Hydrodynamics}",
      journal = {\apjs},
         year = 2024,
        month = mar,
       volume = {271},
       number = {1},
          eid = {7},
        pages = {7},
          doi = {10.3847/1538-4365/ad14f9},
archivePrefix = {arXiv},
       eprint = {2310.04435},
 primaryClass = {physics.comp-ph},
       adsurl = {https://ui.adsabs.harvard.edu/abs/2024ApJS..271....7K}
}

@ARTICLE{Jiang2021,
       author = {{Jiang}, Yan-Fei},
        title = "{An Implicit Finite Volume Scheme to Solve the Time-dependent Radiation Transport Equation Based on Discrete Ordinates}",
      journal = {\apjs},
         year = 2021,
        month = apr,
       volume = {253},
       number = {2},
          eid = {49},
        pages = {49},
          doi = {10.3847/1538-4365/abe303},
archivePrefix = {arXiv},
       eprint = {2102.02212},
 primaryClass = {astro-ph.IM},
       adsurl = {https://ui.adsabs.harvard.edu/abs/2021ApJS..253...49J}
}

@ARTICLE{Su2024,
       author = {{Su}, Zehao and {Bai}, Xue-Ning},
        title = "{Dynamical Consequence of Shadows Cast to the Outer Protoplanetary Disks. I. Two-dimensional Simulations}",
      journal = {\apj},
         year = 2024,
        month = nov,
       volume = {975},
       number = {1},
          eid = {126},
        pages = {126},
          doi = {10.3847/1538-4357/ad7581},
archivePrefix = {arXiv},
       eprint = {2407.12659},
 primaryClass = {astro-ph.EP},
       adsurl = {https://ui.adsabs.harvard.edu/abs/2024ApJ...975..126S}
}

@ARTICLE{Brown2021,
       author = {{Brown-Sevilla}, S.~B. and {Keppler}, M. and {Barraza-Alfaro}, M. and {Melon Fuksman}, J.~D. and {Kurtovic}, N. and {Pinilla}, P. and {Feldt}, M. and {Brandner}, W. and {Ginski}, C. and {Henning}, Th. and {Klahr}, H. and {Asensio-Torres}, R. and {Cantalloube}, F. and {Garufi}, A. and {van Holstein}, R.~G. and {Langlois}, M. and {M{\'e}nard}, F. and {Rickman}, E. and {Benisty}, M. and {Chauvin}, G. and {Zurlo}, A. and {Weber}, P. and {Pavlov}, A. and {Ramos}, J. and {Rochat}, S. and {Roelfsema}, R.},
        title = "{A multiwavelength analysis of the spiral arms in the protoplanetary disk around WaOph 6}",
      journal = {\aap},
         year = 2021,
        month = oct,
       volume = {654},
          eid = {A35},
        pages = {A35},
          doi = {10.1051/0004-6361/202140783},
archivePrefix = {arXiv},
       eprint = {2107.13560},
 primaryClass = {astro-ph.EP},
       adsurl = {https://ui.adsabs.harvard.edu/abs/2021A&A...654A..35B}
}

@ARTICLE{Zhang2024_1,
       author = {{Zhang}, Shangjia and {Zhu}, Zhaohuan},
        title = "{3D Radiation-hydrodynamical Simulations of Shadows on Transition Disks}",
      journal = {\apjl},
         year = 2024,
        month = oct,
       volume = {974},
       number = {2},
          eid = {L38},
        pages = {L38},
          doi = {10.3847/2041-8213/ad815f},
archivePrefix = {arXiv},
       eprint = {2409.08373},
 primaryClass = {astro-ph.EP},
       adsurl = {https://ui.adsabs.harvard.edu/abs/2024ApJ...974L..38Z}
}

@ARTICLE{Muley2024b,
       author = {{Muley}, Dhruv and {Melon Fuksman}, Julio David and {Klahr}, Hubert},
    shorthand = {Paper I},
        title = "{Spiral excitation in protoplanetary disks through gap-edge illumination: Three-temperature radiation hydrodynamics and NIR image modeling}",
      journal = {\aap},
         year = 2024,
        month = oct,
       volume = {690},
          eid = {A355},
        pages = {A355},
          doi = {10.1051/0004-6361/202451554},
archivePrefix = {arXiv},
       eprint = {2408.16461},
 primaryClass = {astro-ph.EP},
       adsurl = {https://ui.adsabs.harvard.edu/abs/2024A&A...690A.355M}
}

@ARTICLE{Schulik2019,
       author = {{Schulik}, M. and {Johansen}, A. and {Bitsch}, B. and {Lega}, E.},
        title = "{Global 3D radiation-hydrodynamic simulations of gas accretion: Opacity-dependent growth of Saturn-mass planets}",
      journal = {\aap},
         year = 2019,
        month = dec,
       volume = {632},
          eid = {A118},
        pages = {A118},
          doi = {10.1051/0004-6361/201935473},
archivePrefix = {arXiv},
       eprint = {1909.08359},
 primaryClass = {astro-ph.EP},
       adsurl = {https://ui.adsabs.harvard.edu/abs/2019A&A...632A.118S}
}

@ARTICLE{Cassasus2019,
       author = {{Casassus}, Simon and {P{\'e}rez}, Sebasti{\'a}n},
        title = "{Kinematic Detections of Protoplanets: A Doppler Flip in the Disk of HD 100546}",
      journal = {\apjl},
         year = 2019,
        month = oct,
       volume = {883},
       number = {2},
          eid = {L41},
        pages = {L41},
          doi = {10.3847/2041-8213/ab4425},
archivePrefix = {arXiv},
       eprint = {1906.06302},
 primaryClass = {astro-ph.EP},
       adsurl = {https://ui.adsabs.harvard.edu/abs/2019ApJ...883L..41C}
}

@article{Teague2018,
doi = {10.3847/2515-5172/aae265},
url = {https://doi.org/10.3847/2515-5172/aae265},
year = {2018},
month = {sep},
publisher = {The American Astronomical Society},
volume = {2},
number = {3},
pages = {173},
author = {Teague, Richard and Foreman-Mackey, Daniel},
title = {A Robust Method to Measure Centroids of Spectral Lines},
journal = {Research Notes of the AAS}
}

@ARTICLE{Hashimoto2015,
       author = {{Hashimoto}, J. and {Tsukagoshi}, T. and {Brown}, J.~M. and {Dong}, R. and {Muto}, T. and {Zhu}, Z. and {Wisniewski}, J. and {Ohashi}, N. and {kudo}, T. and {Kusakabe}, N. and {Abe}, L. and {Akiyama}, E. and {Brandner}, W. and {Brandt}, T. and {Carson}, J. and {Currie}, T. and {Egner}, S. and {Feldt}, M. and {Grady}, C.~A. and {Guyon}, O. and {Hayano}, Y. and {Hayashi}, M. and {Hayashi}, S. and {Henning}, T. and {Hodapp}, K. and {Ishii}, M. and {Iye}, M. and {Janson}, M. and {Kandori}, R. and {Knapp}, G. and {Kuzuhara}, M. and {Kwon}, J. and {Matsuo}, T. and {McElwain}, M.~W. and {Mayama}, S. and {Mede}, K. and {Miyama}, S. and {Morino}, J. -I. and {Moro-Martin}, A. and {Nishimura}, T. and {Pyo}, T. -S. and {Serabyn}, G. and {Suenaga}, T. and {Suto}, H. and {Suzuki}, R. and {Takahashi}, Y. and {Takami}, M. and {Takato}, N. and {Terada}, H. and {Thalmann}, C. and {Tomono}, D. and {Turner}, E.~L. and {Watanabe}, M. and {Yamada}, T. and {Takami}, H. and {Usuda}, T. and {Tamura}, M.},
        title = "{The Structure of Pre-transitional Protoplanetary Disks. II. Azimuthal Asymmetries, Different Radial Distributions of Large and Small Dust Grains in PDS 70}",
      journal = {\apj},
         year = 2015,
        month = jan,
       volume = {799},
       number = {1},
          eid = {43},
        pages = {43},
          doi = {10.1088/0004-637X/799/1/43},
archivePrefix = {arXiv},
       eprint = {1411.2587},
 primaryClass = {astro-ph.SR},
       adsurl = {https://ui.adsabs.harvard.edu/abs/2015ApJ...799...43H}
}

@ARTICLE{Muley2019,
       author = {{Muley}, Dhruv and {Fung}, Jeffrey and {van der Marel}, Nienke},
        title = "{PDS 70: A Transition Disk Sculpted by a Single Planet}",
      journal = {\apjl},
         year = 2019,
        month = jul,
       volume = {879},
       number = {1},
          eid = {L2},
        pages = {L2},
          doi = {10.3847/2041-8213/ab24d0},
archivePrefix = {arXiv},
       eprint = {1902.07191},
 primaryClass = {astro-ph.EP},
       adsurl = {https://ui.adsabs.harvard.edu/abs/2019ApJ...879L...2M}
}

@ARTICLE{Dullemond2018,
       author = {{Dullemond}, Cornelis P. and {Birnstiel}, Tilman and {Huang}, Jane and {Kurtovic}, Nicol{\'a}s T. and {Andrews}, Sean M. and {Guzm{\'a}n}, Viviana V. and {P{\'e}rez}, Laura M. and {Isella}, Andrea and {Zhu}, Zhaohuan and {Benisty}, Myriam and {Wilner}, David J. and {Bai}, Xue-Ning and {Carpenter}, John M. and {Zhang}, Shangjia and {Ricci}, Luca},
        title = "{The Disk Substructures at High Angular Resolution Project (DSHARP). VI. Dust Trapping in Thin-ringed Protoplanetary Disks}",
      journal = {\apjl},
         year = 2018,
        month = dec,
       volume = {869},
       number = {2},
          eid = {L46},
        pages = {L46},
          doi = {10.3847/2041-8213/aaf742},
archivePrefix = {arXiv},
       eprint = {1812.04044},
 primaryClass = {astro-ph.EP},
       adsurl = {https://ui.adsabs.harvard.edu/abs/2018ApJ...869L..46D}
}

@ARTICLE{Andrews2018,
       author = {{Andrews}, Sean M. and {Huang}, Jane and {P{\'e}rez}, Laura M. and {Isella}, Andrea and {Dullemond}, Cornelis P. and {Kurtovic}, Nicol{\'a}s T. and {Guzm{\'a}n}, Viviana V. and {Carpenter}, John M. and {Wilner}, David J. and {Zhang}, Shangjia and {Zhu}, Zhaohuan and {Birnstiel}, Tilman and {Bai}, Xue-Ning and {Benisty}, Myriam and {Hughes}, A. Meredith and {{\"O}berg}, Karin I. and {Ricci}, Luca},
        title = "{The Disk Substructures at High Angular Resolution Project (DSHARP). I. Motivation, Sample, Calibration, and Overview}",
      journal = {\apjl},
         year = 2018,
        month = dec,
       volume = {869},
       number = {2},
          eid = {L41},
        pages = {L41},
          doi = {10.3847/2041-8213/aaf741},
archivePrefix = {arXiv},
       eprint = {1812.04040},
 primaryClass = {astro-ph.SR},
       adsurl = {https://ui.adsabs.harvard.edu/abs/2018ApJ...869L..41A}
}

@ARTICLE{MelonFuksman2025,
       author = {{Melon Fuksman}, David and {Flock}, Mario and {Klahr}, Hubert and {Mattia}, Giancarlo and {Muley}, Dhruv},
        title = "{Multidimensional half-moment multigroup radiative transfer: Improving moment-based thermal models of circumstellar disks}",
      journal = {\aap},
         year = 2025,
        month = sep,
       volume = {701},
          eid = {A97},
        pages = {A97},
          doi = {10.1051/0004-6361/202554994},
archivePrefix = {arXiv},
       eprint = {2504.13999},
 primaryClass = {astro-ph.IM},
       adsurl = {https://ui.adsabs.harvard.edu/abs/2025A&A...701A..97M}
}

@ARTICLE{Matsumoto2023,
       author = {{Matsumoto}, Kosei and {Camps}, Peter and {Baes}, Maarten and {De Ceuster}, Frederik and {Wada}, Keiichi and {Nakagawa}, Takao and {Nagamine}, Kentaro},
        title = "{Self-consistent dust and non-LTE line radiative transfer with SKIRT}",
      journal = {\aap},
         year = 2023,
        month = oct,
       volume = {678},
          eid = {A175},
        pages = {A175},
          doi = {10.1051/0004-6361/202347376},
archivePrefix = {arXiv},
       eprint = {2309.02628},
 primaryClass = {astro-ph.GA},
       adsurl = {https://ui.adsabs.harvard.edu/abs/2023A&A...678A.175M}
}

@ARTICLE{Huehn2025,
       author = {{H{\"u}hn}, L. -A. and {Jiang}, H. -C. and {Dullemond}, C.~P.},
        title = "{Late accretion offers pathway to misaligned disk around the planet-hosting IRAS 04125+2902}",
      journal = {\aap},
         year = 2025,
        month = sep,
       volume = {701},
          eid = {L15},
        pages = {L15},
          doi = {10.1051/0004-6361/202555391},
archivePrefix = {arXiv},
       eprint = {2509.06564},
 primaryClass = {astro-ph.EP},
       adsurl = {https://ui.adsabs.harvard.edu/abs/2025A&A...701L..15H}
}

@ARTICLE{Kuznetsova2022,
       author = {{Kuznetsova}, Aleksandra and {Bae}, Jaehan and {Hartmann}, Lee and {Mac Low}, Mordecai-Mark},
        title = "{Anisotropic Infall and Substructure Formation in Embedded Disks}",
      journal = {\apj},
         year = 2022,
        month = mar,
       volume = {928},
       number = {1},
          eid = {92},
        pages = {92},
          doi = {10.3847/1538-4357/ac54a8},
archivePrefix = {arXiv},
       eprint = {2202.05301},
 primaryClass = {astro-ph.EP},
       adsurl = {https://ui.adsabs.harvard.edu/abs/2022ApJ...928...92K}
}

@ARTICLE{Smallwood2023,
       author = {{Smallwood}, Jeremy L. and {Yang}, Chao-Chin and {Zhu}, Zhaohuan and {Martin}, Rebecca G. and {Dong}, Ruobing and {Cuello}, Nicol{\'a}s and {Isella}, Andrea},
        title = "{Exciting spiral arms in protoplanetary discs from flybys}",
      journal = {\mnras},
         year = 2023,
        month = may,
       volume = {521},
       number = {3},
        pages = {3500-3516},
          doi = {10.1093/mnras/stad742},
archivePrefix = {arXiv},
       eprint = {2303.05753},
 primaryClass = {astro-ph.EP},
       adsurl = {https://ui.adsabs.harvard.edu/abs/2023MNRAS.521.3500S}
}

@ARTICLE{Zhu2025,
       author = {{Zhu}, Zhaohuan and {Zhang}, Shangjia and {Johnson}, Ted M.},
        title = "{Asymmetric Temperature Variations In Protoplanetary Disks. I. Linear Theory, Corotating Spirals, and Ring Formation}",
      journal = {\apj},
         year = 2025,
        month = feb,
       volume = {980},
       number = {2},
          eid = {259},
        pages = {259},
          doi = {10.3847/1538-4357/adae0d},
archivePrefix = {arXiv},
       eprint = {2412.09571},
 primaryClass = {astro-ph.EP},
       adsurl = {https://ui.adsabs.harvard.edu/abs/2025ApJ...980..259Z}
}

@ARTICLE{Ziampras2025,
       author = {{Ziampras}, Alexandros and {Dullemond}, Cornelis P. and {Birnstiel}, Tilman and {Benisty}, Myriam and {Nelson}, Richard P.},
        title = "{Spirals, rings, and vortices shaped by shadows in protoplanetary discs: from radiative hydrodynamical simulations to observable signatures}",
      journal = {\mnras},
         year = 2025,
        month = jun,
       volume = {540},
       number = {1},
        pages = {1185-1201},
          doi = {10.1093/mnras/staf785},
archivePrefix = {arXiv},
       eprint = {2410.13932},
 primaryClass = {astro-ph.EP},
       adsurl = {https://ui.adsabs.harvard.edu/abs/2025MNRAS.540.1185Z}
}

@ARTICLE{Cugno2025,
       author = {{Cugno}, G. and {Facchini}, S. and {Alarcon}, F. and {Bae}, J. and {Benisty}, M. and {Eilers}, A. -C. and {Leung}, G.~C.~K. and {Meyer}, M. and {Pueyo}, L. and {Teague}, R. and {Bergin}, E. and {Girard}, J. and {Helled}, R. and {Huang}, J. and {Leisenring}, J.},
        title = "{Direct Measurement of Extinction in a Planet-Hosting Gap}",
      journal = {arXiv e-prints},
         year = 2025,
        month = sep,
          eid = {arXiv:2509.26617},
        pages = {arXiv:2509.26617},
          doi = {10.48550/arXiv.2509.26617},
archivePrefix = {arXiv},
       eprint = {2509.26617},
 primaryClass = {astro-ph.EP},
       adsurl = {https://ui.adsabs.harvard.edu/abs/2025arXiv250926617C}
}

@ARTICLE{Teague2021,
       author = {{Teague}, Richard and {Bae}, Jaehan and {Aikawa}, Yuri and {Andrews}, Sean M. and {Bergin}, Edwin A. and {Bergner}, Jennifer B. and {Boehler}, Yann and {Booth}, Alice S. and {Bosman}, Arthur D. and {Cataldi}, Gianni and {Czekala}, Ian and {Guzm{\'a}n}, Viviana V. and {Huang}, Jane and {Ilee}, John D. and {Law}, Charles J. and {Le Gal}, Romane and {Long}, Feng and {Loomis}, Ryan A. and {M{\'e}nard}, Fran{\c{c}}ois and {{\"O}berg}, Karin I. and {P{\'e}rez}, Laura M. and {Schwarz}, Kamber R. and {Sierra}, Anibal and {Walsh}, Catherine and {Wilner}, David J. and {Yamato}, Yoshihide and {Zhang}, Ke},
        title = "{Molecules with ALMA at Planet-forming Scales (MAPS). XVIII. Kinematic Substructures in the Disks of HD 163296 and MWC 480}",
      journal = {\apjs},
         year = 2021,
        month = nov,
       volume = {257},
       number = {1},
          eid = {18},
        pages = {18},
          doi = {10.3847/1538-4365/ac1438},
archivePrefix = {arXiv},
       eprint = {2109.06218},
 primaryClass = {astro-ph.EP},
       adsurl = {https://ui.adsabs.harvard.edu/abs/2021ApJS..257...18T}
}

@ARTICLE{Flock2015,
       author = {{Flock}, M. and {Ruge}, J.~P. and {Dzyurkevich}, N. and {Henning}, Th. and {Klahr}, H. and {Wolf}, S.},
        title = "{Gaps, rings, and non-axisymmetric structures in protoplanetary disks. From simulations to ALMA observations}",
      journal = {\aap},
         year = 2015,
        month = feb,
       volume = {574},
          eid = {A68},
        pages = {A68},
          doi = {10.1051/0004-6361/201424693},
archivePrefix = {arXiv},
       eprint = {1411.2736},
 primaryClass = {astro-ph.EP},
       adsurl = {https://ui.adsabs.harvard.edu/abs/2015A&A...574A..68F}
}

@ARTICLE{BarrazaAlfaro2021,
       author = {{Barraza-Alfaro}, Marcelo and {Flock}, Mario and {Marino}, Sebastian and {P{\'e}rez}, Sebasti{\'a}n},
        title = "{Observability of the vertical shear instability in protoplanetary disk CO kinematics}",
      journal = {\aap},
         year = 2021,
        month = sep,
       volume = {653},
          eid = {A113},
        pages = {A113},
          doi = {10.1051/0004-6361/202140535},
archivePrefix = {arXiv},
       eprint = {2106.01159},
 primaryClass = {astro-ph.EP},
       adsurl = {https://ui.adsabs.harvard.edu/abs/2021A&A...653A.113B}
}

@ARTICLE{Ercolano2022,
       author = {{Ercolano}, Barbara and {Picogna}, Giovanni},
        title = "{Modelling photoevaporation in planet forming discs}",
      journal = {European Physical Journal Plus},
         year = 2022,
        month = dec,
       volume = {137},
       number = {12},
          eid = {1357},
        pages = {1357},
          doi = {10.1140/epjp/s13360-022-03515-8},
archivePrefix = {arXiv},
       eprint = {2211.10130},
 primaryClass = {astro-ph.EP},
       adsurl = {https://ui.adsabs.harvard.edu/abs/2022EPJP..137.1357E}
}

@ARTICLE{Calcino2025,
       author = {{Calcino}, Josh and {Price}, Daniel J. and {Hilder}, Thomas and {Christiaens}, Valentin and {Speedie}, Jessica and {Ormel}, Chris W.},
        title = "{Anatomy of a fall: stationary and super-Keplerian spiral arms generated by accretion streamers in protostellar discs}",
      journal = {\mnras},
         year = 2025,
        month = mar,
       volume = {537},
       number = {3},
        pages = {2695-2707},
          doi = {10.1093/mnras/staf135},
archivePrefix = {arXiv},
       eprint = {2410.18521},
 primaryClass = {astro-ph.EP},
       adsurl = {https://ui.adsabs.harvard.edu/abs/2025MNRAS.537.2695C}
}

@ARTICLE{Hu2022,
       author = {{Hu}, Xiao and {Li}, Zhi-Yun and {Zhu}, Zhaohuan and {Yang}, Chao-Chin},
        title = "{Formation of dust rings and gaps in non-ideal MHD discs through meridional gas flows}",
      journal = {\mnras},
         year = 2022,
        month = oct,
       volume = {516},
       number = {2},
        pages = {2006-2022},
          doi = {10.1093/mnras/stac1799},
archivePrefix = {arXiv},
       eprint = {2203.05629},
 primaryClass = {astro-ph.EP},
       adsurl = {https://ui.adsabs.harvard.edu/abs/2022MNRAS.516.2006H}
}

@ARTICLE{Paneque2025,
       author = {{Paneque-Carre{\~n}o}, T. and {Miotello}, A. and {van Dishoeck}, E.~F. and {Rosotti}, G. and {Tabone}, B.},
        title = "{Vertical CO surfaces as a probe for protoplanetary disk mass and carbon depletion}",
      journal = {\aap},
         year = 2025,
        month = jun,
       volume = {698},
          eid = {A231},
        pages = {A231},
          doi = {10.1051/0004-6361/202451862},
archivePrefix = {arXiv},
       eprint = {2501.08294},
 primaryClass = {astro-ph.EP},
       adsurl = {https://ui.adsabs.harvard.edu/abs/2025A&A...698A.231P}
}

@ARTICLE{Isella2016,
       author = {{Isella}, Andrea and {Guidi}, Greta and {Testi}, Leonardo and {Liu}, Shangfei and {Li}, Hui and {Li}, Shengtai and {Weaver}, Erik and {Boehler}, Yann and {Carperter}, John M. and {De Gregorio-Monsalvo}, Itziar and {Manara}, Carlo F. and {Natta}, Antonella and {P{\'e}rez}, Laura M. and {Ricci}, Luca and {Sargent}, Anneila and {Tazzari}, Marco and {Turner}, Neal},
        title = "{Ringed Structures of the HD 163296 Protoplanetary Disk Revealed by ALMA}",
      journal = {\prl},
         year = 2016,
        month = dec,
       volume = {117},
       number = {25},
          eid = {251101},
        pages = {251101},
          doi = {10.1103/PhysRevLett.117.251101},
       adsurl = {https://ui.adsabs.harvard.edu/abs/2016PhRvL.117y1101I}
}

@ARTICLE{Bethune2021,
       author = {{B{\'e}thune}, William and {Latter}, Henrik and {Kley}, Wilhelm},
        title = "{Spiral structures in gravito-turbulent gaseous disks}",
      journal = {\aap},
         year = 2021,
        month = jun,
       volume = {650},
          eid = {A49},
        pages = {A49},
          doi = {10.1051/0004-6361/202040094},
archivePrefix = {arXiv},
       eprint = {2102.00775},
 primaryClass = {astro-ph.EP},
       adsurl = {https://ui.adsabs.harvard.edu/abs/2021A&A...650A..49B}
}

@ARTICLE{Dong2015,
       author = {{Dong}, Ruobing and {Hall}, Cassandra and {Rice}, Ken and {Chiang}, Eugene},
        title = "{Spiral Arms in Gravitationally Unstable Protoplanetary Disks as Imaged in Scattered Light}",
      journal = {\apjl},
         year = 2015,
        month = oct,
       volume = {812},
       number = {2},
          eid = {L32},
        pages = {L32},
          doi = {10.1088/2041-8205/812/2/L32},
archivePrefix = {arXiv},
       eprint = {1510.00396},
 primaryClass = {astro-ph.SR},
       adsurl = {https://ui.adsabs.harvard.edu/abs/2015ApJ...812L..32D}
}

@article{Artymowicz1994,
	author = {Artymowicz, Pawel and Lubow, Stephen H.},
	doi = {10.1086/173679},
	issn = {0004-637X},
	journal = {The Astrophysical Journal},
	language = {en},
	month = feb,
	pages = {651},
	title = {Dynamics of {Binary}-{Disk} {Interaction}. {I}. {Resonances} and {Disk} {Gap} {Sizes}},
	url = {https://ui.adsabs.harvard.edu/abs/1994ApJ...421..651A/abstract},
	urldate = {2025-11-17},
	volume = {421},
	year = {1994}}

@ARTICLE{Xie2024,
       author = {{Xie}, Chen and {Xie}, Chengyan and {Ren}, Bin B. and {Benisty}, Myriam and {Ginski}, Christian and {Fang}, Taotao and {Casassus}, Simon and {Bae}, Jaehan and {Facchini}, Stefano and {M{\'e}nard}, Fran{\c{c}}ois and {van Holstein}, Rob G.},
        title = "{Disk Evolution Study Through Imaging of Nearby Young Stars (DESTINYS): Dynamical Evidence of a Spiral-Arm-Driving and Gap-Opening Protoplanet from SAO 206462 Spiral Motion}",
      journal = {Universe},
         year = 2024,
        month = dec,
       volume = {10},
       number = {12},
          eid = {465},
        pages = {465},
          doi = {10.3390/universe10120465},
archivePrefix = {arXiv},
       eprint = {2412.14402},
 primaryClass = {astro-ph.EP},
       adsurl = {https://ui.adsabs.harvard.edu/abs/2024Univ...10..465X}
}

@ARTICLE{Wagner2023,
       author = {{Wagner}, Kevin and {Stone}, Jordan and {Skemer}, Andrew and {Ertel}, Steve and {Dong}, Ruobing and {Apai}, D{\'a}niel and {Spalding}, Eckhart and {Leisenring}, Jarron and {Sitko}, Michael and {Kratter}, Kaitlin and {Barman}, Travis and {Marley}, Mark and {Miles}, Brittany and {Boccaletti}, Anthony and {Assani}, Korash and {Bayyari}, Ammar and {Uyama}, Taichi and {Woodward}, Charles E. and {Hinz}, Phil and {Briesemeister}, Zackery and {Lawson}, Kellen and {M{\'e}nard}, Fran{\c{c}}ois and {Pantin}, Eric and {Russell}, Ray W. and {Skrutskie}, Michael and {Wisniewski}, John},
        title = "{Direct images and spectroscopy of a giant protoplanet driving spiral arms in MWC 758.}",
      journal = {Nature Astronomy},
         year = 2023,
        month = oct,
       volume = {7},
        pages = {1208-1217},
          doi = {10.1038/s41550-023-02028-3},
archivePrefix = {arXiv},
       eprint = {2307.04021},
 primaryClass = {astro-ph.EP},
       adsurl = {https://ui.adsabs.harvard.edu/abs/2023NatAs...7.1208W}
}

@ARTICLE{Maio2025,
       author = {{Maio}, F. and {Fedele}, D. and {Roccatagliata}, V. and {Facchini}, S. and {Lodato}, G. and {Desidera}, S. and {Garufi}, A. and {Mesa}, D. and {Ruzza}, A. and {Toci}, C. and {Testi}, L. and {Zurlo}, A. and {Rosotti}, G.},
        title = "{Unveiling a protoplanet candidate embedded in the HD 135344B disk with VLT/ERIS}",
      journal = {\aap},
         year = 2025,
        month = jul,
       volume = {699},
          eid = {L10},
        pages = {L10},
          doi = {10.1051/0004-6361/202554472},
       adsurl = {https://ui.adsabs.harvard.edu/abs/2025A&A...699L..10M}
}

@article{disksurf,
  doi = {10.21105/joss.03827},
  url = {https://doi.org/10.21105/joss.03827},
  year = {2021},
  publisher = {The Open Journal},
  volume = {6},
  number = {67},
  pages = {3827},
  author = {Richard Teague and Charles J. Law and Jane Huang and Feilong Meng},
  title = {disksurf: Extracting the 3D Structure of Protoplanetary Disks},
  journal = {Journal of Open Source Software}
}
\bibliographystyle{aa}
\appendix
\section{Total versus differential moment-1 maps}
\begin{figure*}
    \centering
    \begin{overpic}[width=0.9174364896\textwidth,origin=c]{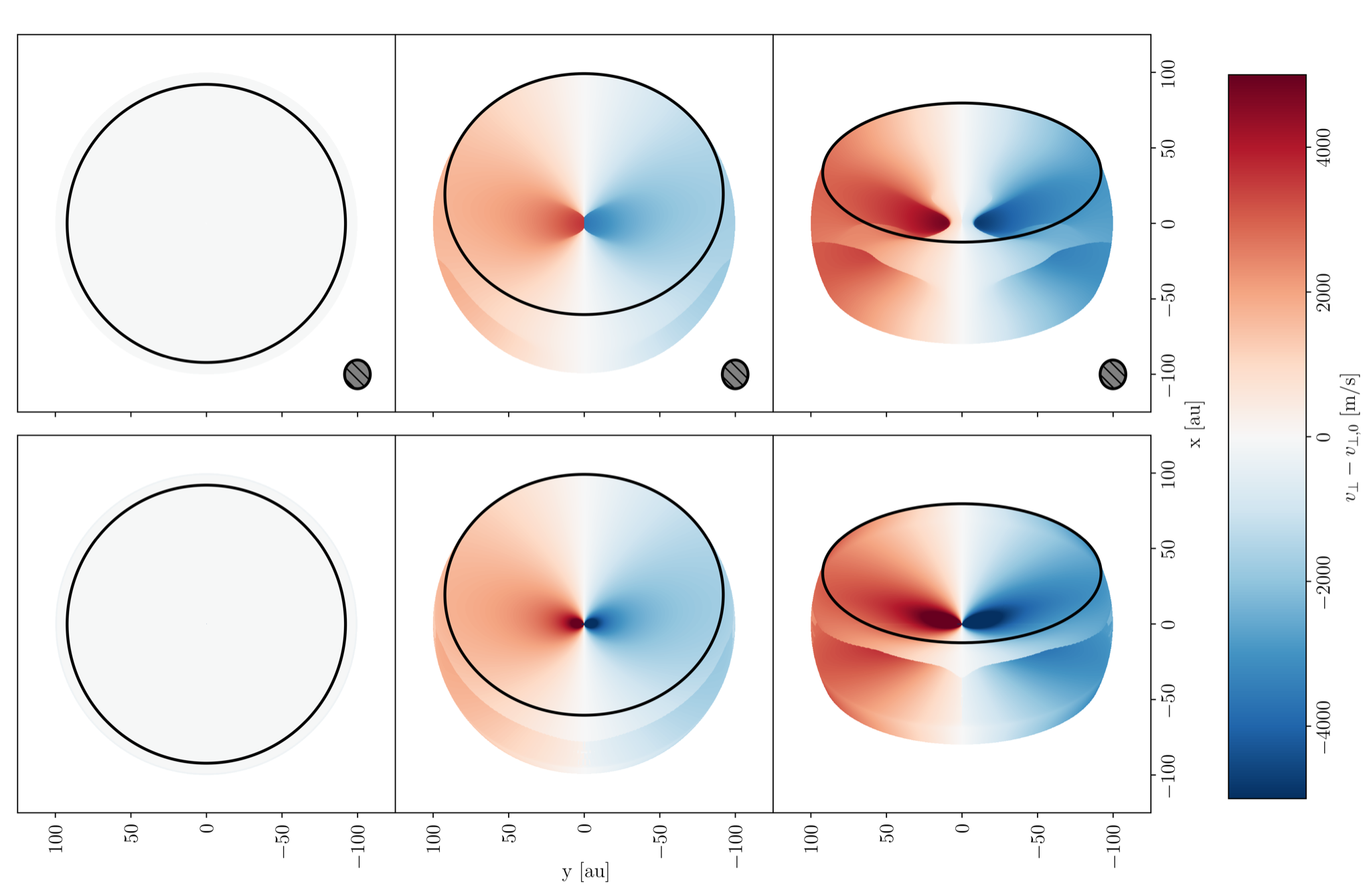}
    \put (3,59) {\Large Initial condition}
    \put (3,37) {\large $0^{\circ}$}
    \put (30.25,37) {\large $30^{\circ}$}
    \put (57.5,37) {\large $60^{\circ}$}
    \end{overpic}
    \begin{overpic}[width=0.9174364896\textwidth,origin=c]{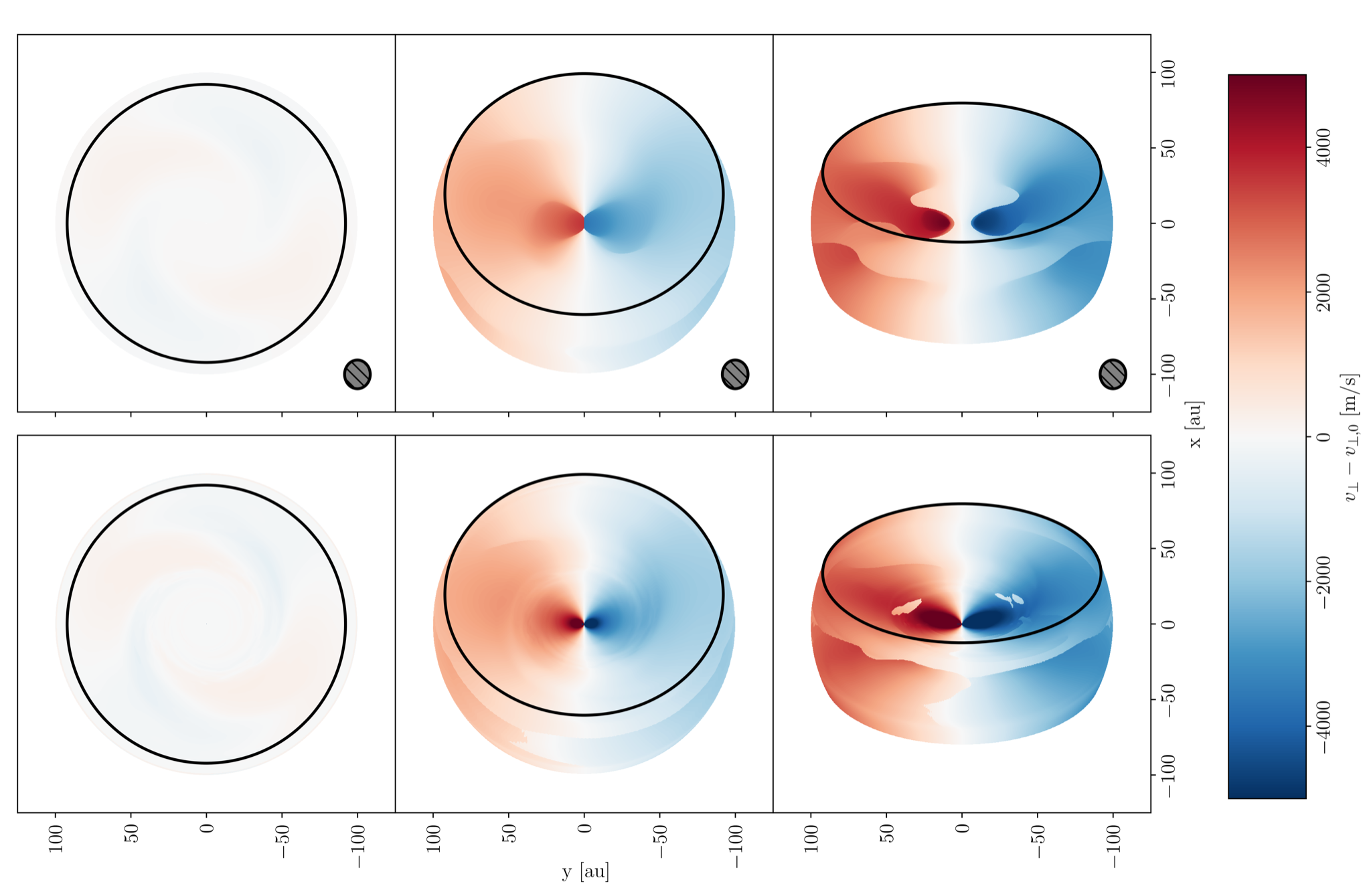}
    \put (3,59) {\Large Final condition}
    \put (3,37) {\large $0^{\circ}$}
    \put (30.25,37) {\large $30^{\circ}$}
    \put (57.5,37) {\large $60^{\circ}$}
    \end{overpic}
    \caption{\revision{Kinematic moment-1 maps for the $J = 2-1$ transition of \moltwelveco \ in the shadow-driven spiral simulations. Here, the initial and final maps are plotted separately, unlike in other figures where the former is used as a background model and subtracted from the latter. Accordingly, we expand the range of the color scale ($\SI{-5}{\km\per\second} \leq v_{\perp} \leq \SI{5}{\km\per\second}$ ) to accommodate the Keplerian flow of the disk. } }
    \label{fig:ic_fc_moment_map}
\end{figure*}
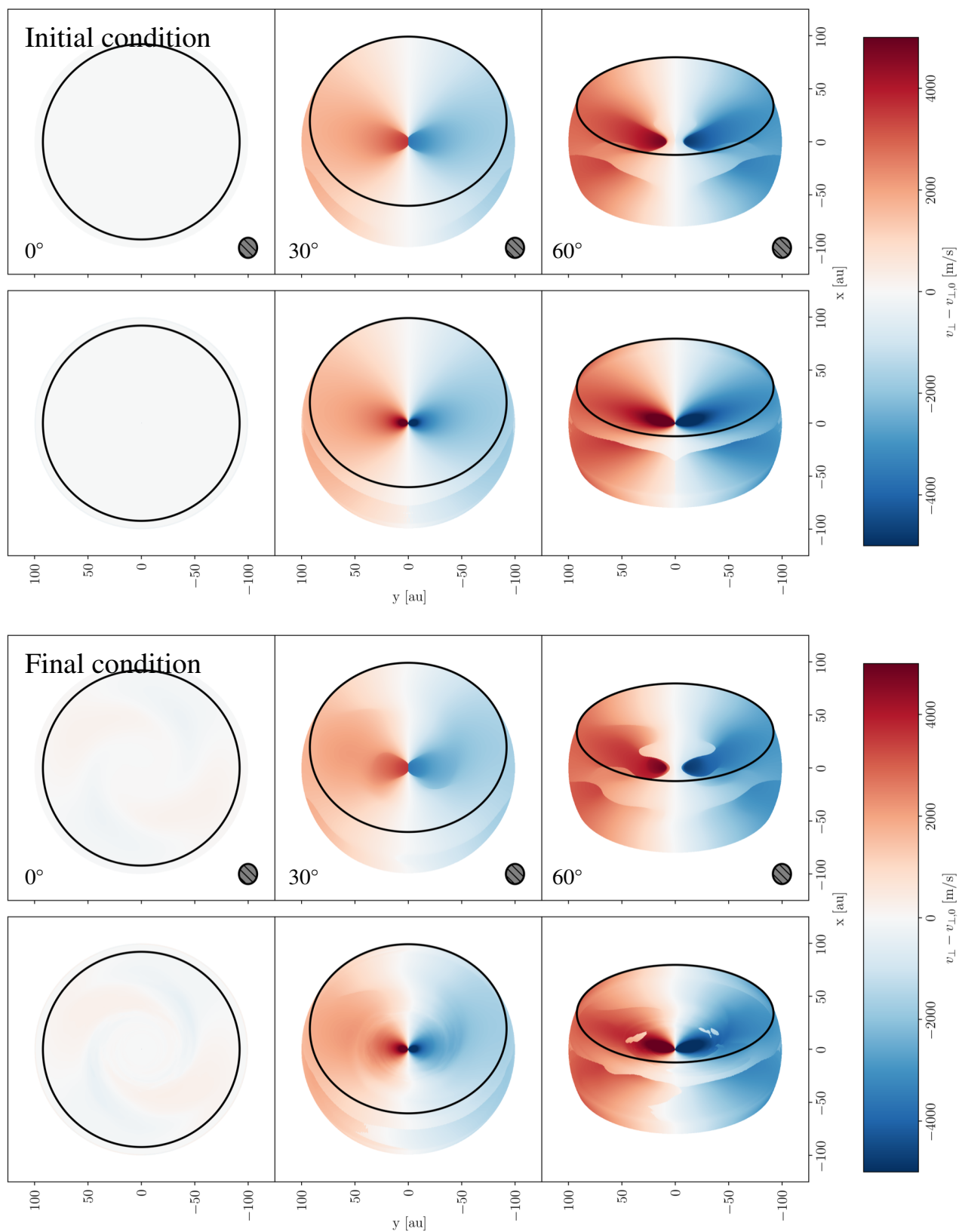

\revision{The figures presented in the main text show differential/residual moment maps, in which the line-of-sight velocities from the final simulations (computed using \texttt{bettermoments}) are regularized by subtracting off the line-of-sight velocities obtained from a background disk model. In this work, we choose for our background model a moment-1 map computed with \texttt{RADMC3D} from the simulation's initial condition, which naturally reproduces the location of the optical surface for each transition, as well as the background velocities (giving the line shift) and temperatures (giving the width) on this surface. In Figure \ref{fig:ic_fc_moment_map}, we plot both the final and initial (background) moment-1 maps, computed from both raw and beam-convolved mock $J=2-1$ \ \moltwelveco \ observations. In the face-on case, the background line-of-sight velocity is zero because the disk is in vertical hydrostatic equilibrium, but in inclined disks it becomes substantial, as the projection of the disk's quasi-Keplerian velocity onto the line-of-sight becomes nonzero. In the inner regions of inclined disks, where approaching and receding quasi-Keplerian flows occur in close angular proximity to one another, beam convolution averages these together to yield no net apparent line-of-sight velocity.

Although our choice of an initial condition-based background model is logical and advantageous in many respects, limitations exist on its applicability to actual observations. Unlike in simulations, there is no way to determine \textit{a priori} the unperturbed state of a real disk while holding all other factors constant. Obtaining a physically-informed, best-fit initial model would not only be highly expensive---due to the need to iterate MCRT and hydrostatic calculations \citep{MelonFuksman2022} at each sampled point in a multidimensional parameter space---but also degenerate in parameters such as disk mass, grain size distribution, opacity/mineralogy, and isotopologue mass fraction, among others. In practice, the emission surfaces of observed disks are often computed directly from channel maps \citep{Pinte2018b,disksurf}, with background velocities and temperatures found by fits to the azimuthally-averaged measurements rather than detailed physical modeling.
}

\section{Full library of differential moment-1 maps}
\label{sec:app_library_desc}
In what follows, we present mock \texttt{bettermoments} moment-1 maps of the disk with the shadow-driven spiral, for all of the transitions and isotopologues under consideration for our study. These are, namely, the $J = 2-1$ and $3-2$ transitions for \moltwelveco, \molthirteenco, and \molceighteeno. As in the main text, we regularize by subtracting off the moment-1 map produced from the initial condition, in order to reveal the spiral structure. 

In these Appendix plots, however, we include moment maps computed both from the beam-convolved images, as well as those from non-convolved images. The latter reveal additional structure, especially in the inner disk and at higher inclination, which could potentially be probed by future, higher-resolution gas observations. However, they also show much stronger impact from artifacts generated by the \texttt{bettermoments} method itself, which the dissipative effect of beam convolution acts to mitigate.

In general, features are very similar in the $J = 2-1$ and $J = 3-2$ transitions of any given isotopologue, an unsurprising result given that both trace similar regions of the disk (Figure \ref{fig:tau1_surface}). Shadow-driven features in \molthirteenco \ and \molceighteeno \ are not as strong as those in \moltwelveco---as demonstrated for face-on disks in the main text---and become weaker still when the system is viewed at an inclination $i_{d} = 30^{\circ}$, particularly when beam convolution is applied.

\begin{figure*}[h]
    \centering
    \begin{overpic}[width=0.9174364896\textwidth]{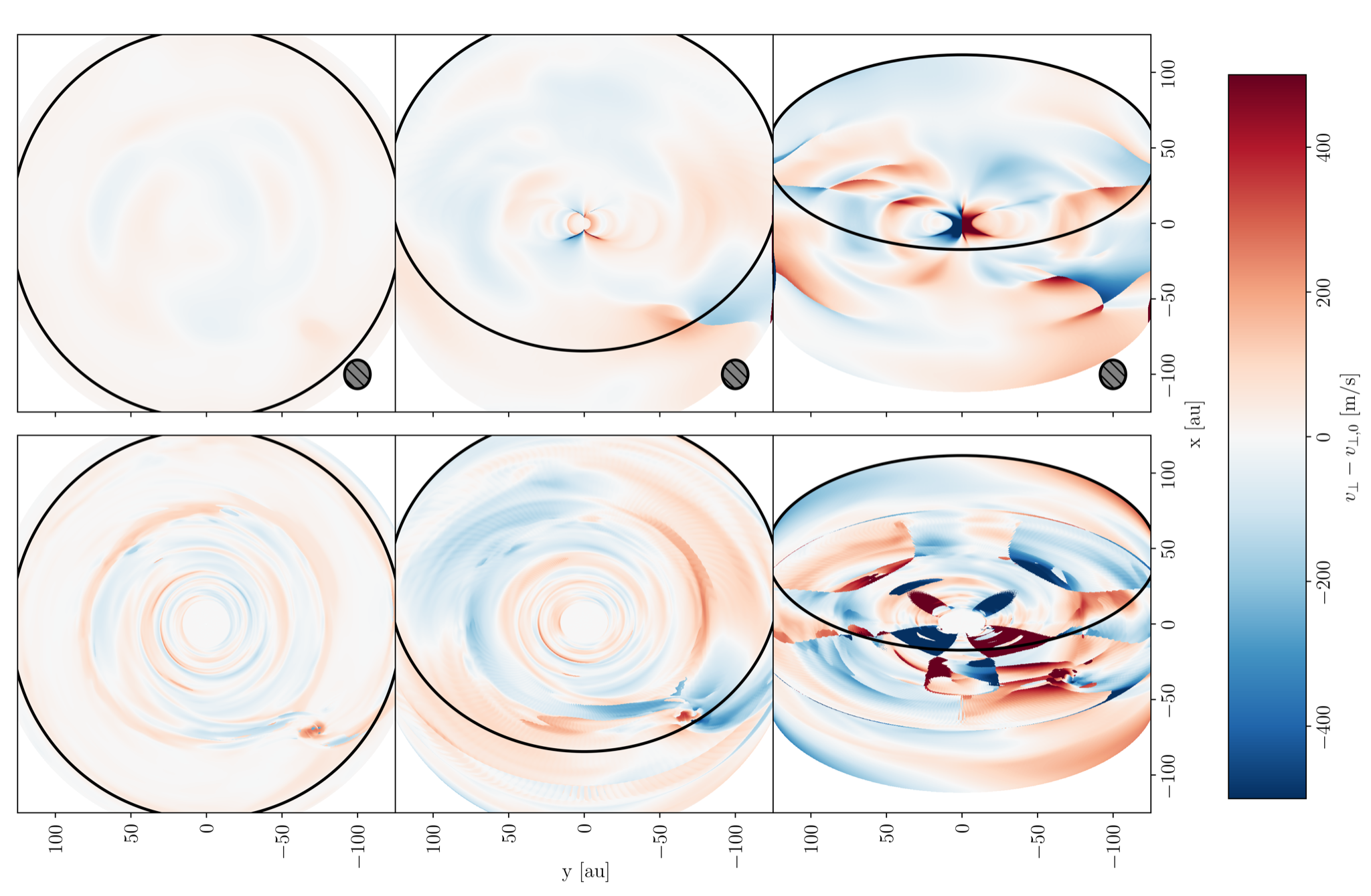}
    \put (3, 59) {\Large \molthirteenco \ $J = 2-1$, planet}
    \end{overpic}
    \begin{overpic}[width=0.9174364896\textwidth]{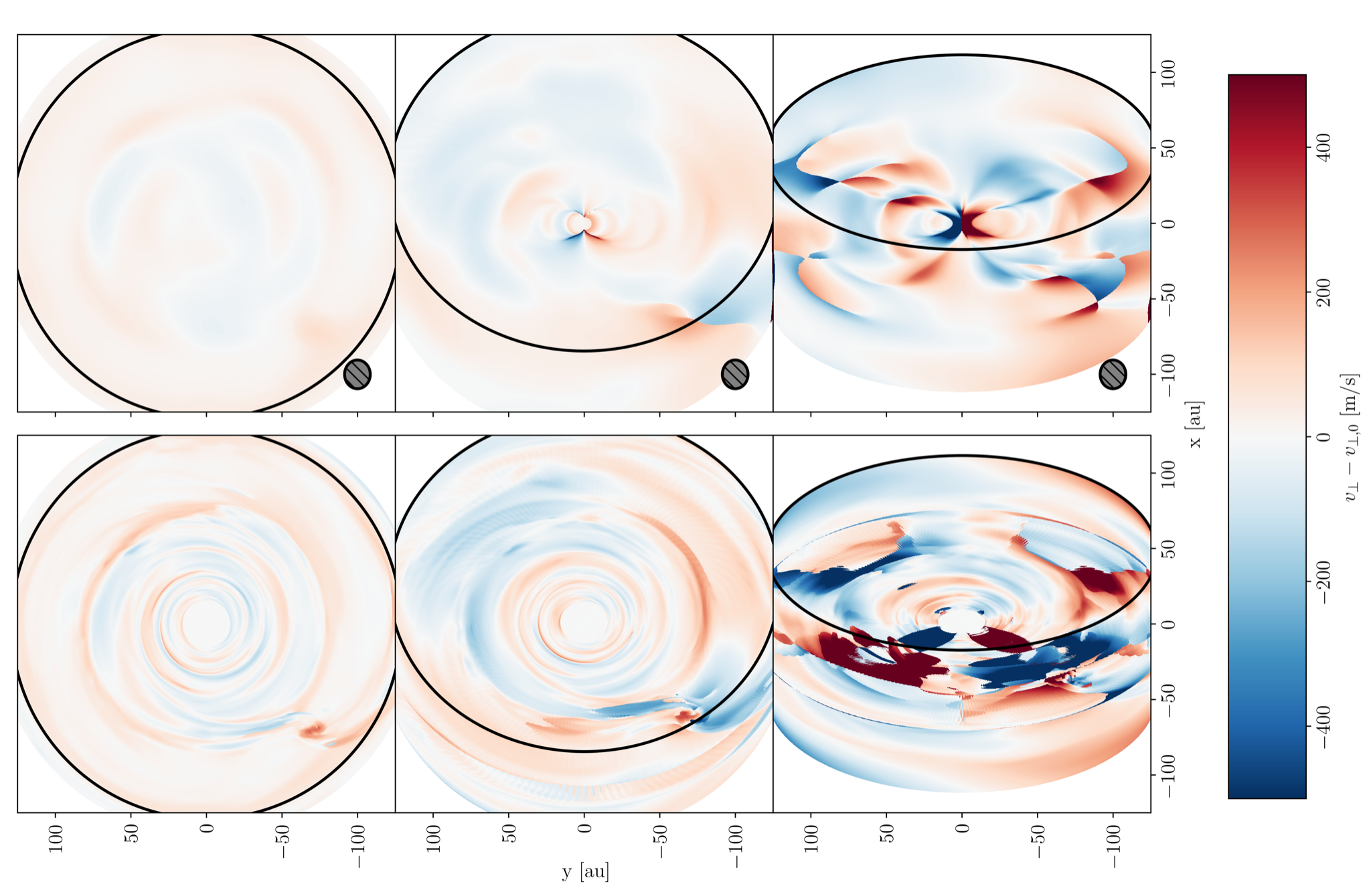}
    \put (3, 59) {\Large \molthirteenco \ $J = 3-2$, planet}
    \end{overpic}
    \vspace*{-1ex}
    \caption{Kinematic maps of planet-driven spirals in \molthirteenco. From left to right, we plot disk inclinations of 0$^\circ$, 30$^\circ$, and 60$^\circ$. Images at 30$^\circ$ in particular reveal the planet-driven spiral structure, as well as the Doppler flip at the planet's location.}
    \label{fig:_13co_.planet}
\end{figure*}

\begin{figure*}[h]
    \centering
    \begin{overpic}[width=0.9174364896\textwidth]{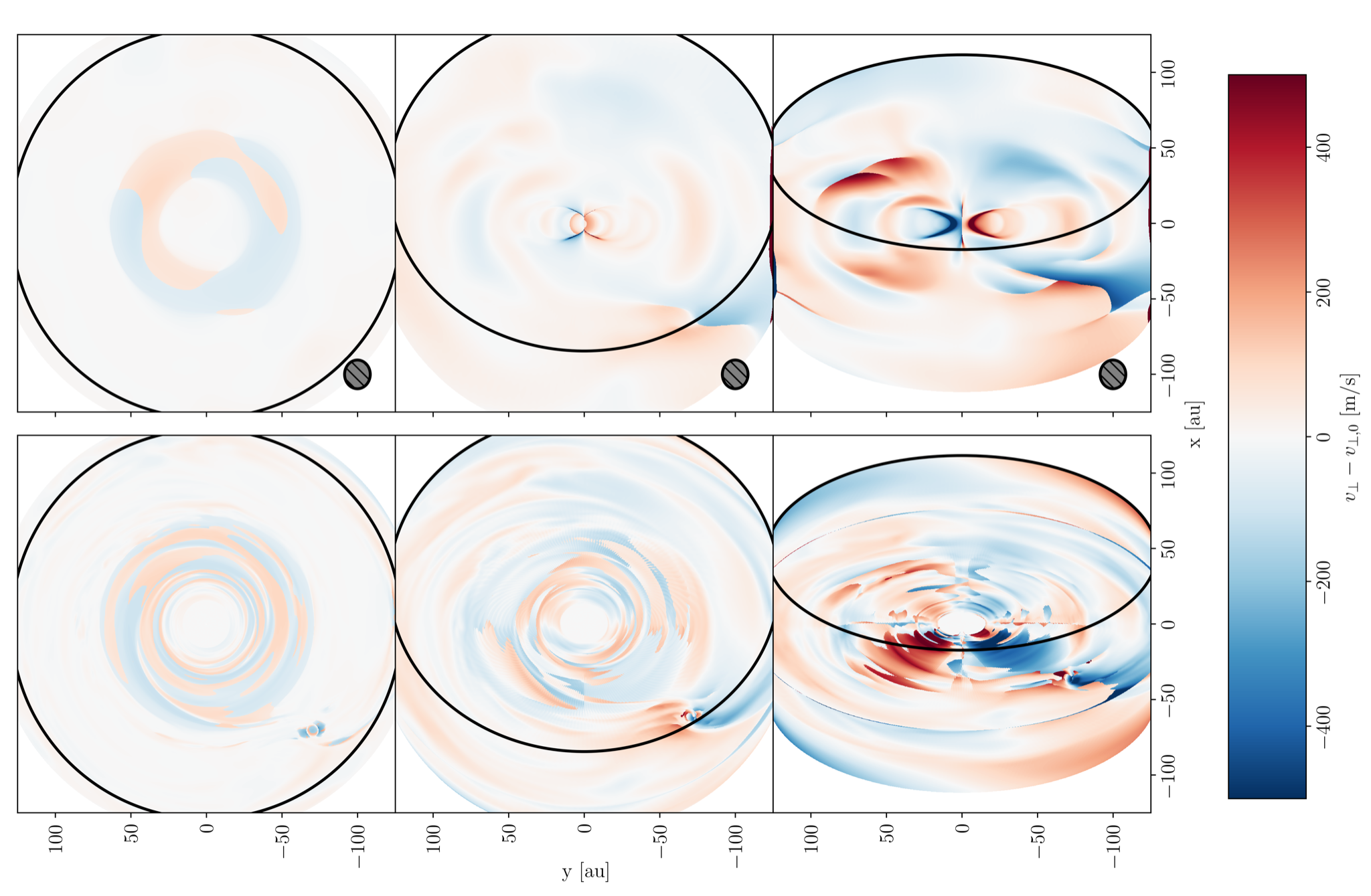}
    \put (3, 59) {\Large \molceighteeno \ $J = 2-1$, planet}
    \end{overpic}
    \begin{overpic}[width=0.9174364896\textwidth]{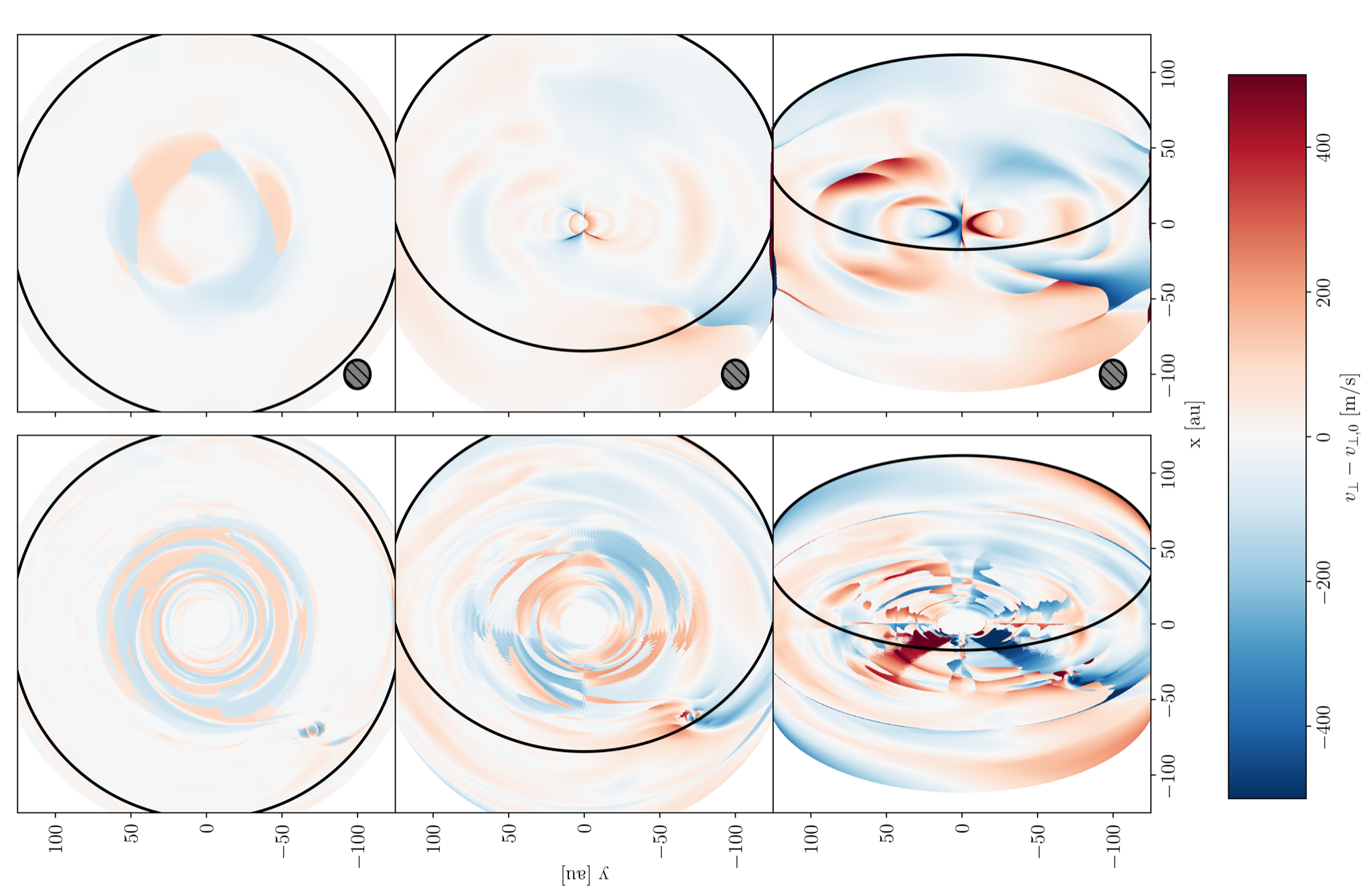}
    \put (3, 59) {\Large \molceighteeno \ $J = 3-2$, planet}
    \end{overpic}    
    \vspace*{-1ex}
    \caption{Kinematic moment-1 (line-of-sight velocity) maps of planet-driven spirals in \molceighteeno.  From left to right, we plot disk inclinations of 0$^\circ$, 30$^\circ$, and 60$^\circ$. The Doppler flip remains prominent in the moderate-inclination case, although the spiral signature itself is degraded by \texttt{bettermoments} artifacts.}
    \label{fig:_c18o_.planet}
\end{figure*}

For the planet-driven case, we see that $i_d = 30^{\circ}$ shows the spiral structure much more clearly than does a face-on orientation ($i_d = 0^{\circ}$). This follows from the fact that in-plane velocity perturbations ($v_r$, $v_{\phi}$) dominate over $v_{\theta}$ in classical disk-planet interaction, and that a nonzero inclination projects the in-plane components onto the line of sight. Indeed, in \molthirteenco \ (Figure \ref{fig:_13co_.planet}) and to a lesser extent \molceighteeno \ (Figure \ref{fig:_c18o_.planet}), the images at $i_d = 30^{\circ}$ emphasize the double-armed nature of the spiral. These images also clearly illustrate the Doppler-flip phenomenon associated with the planet's location, which remains prominent even when beam convolution is considered.

In nearly all cases, save for the \moltwelveco \ maps for the shadow-driven case, and the unconvolved \moltwelveco \ maps for the planet-driven case, an inclination $i_d = 60^{\circ}$ makes it difficult to obtain useful information about the spirals therein from the observed gas kinematics. Nevertheless, alternative features, such as gaps and rings in ALMA dust continuum, may provide evidence for planet formation even in relatively high-inclination disks \citep{Isella2016}.

\onecolumn
\section{Shadow-driven spirals}
\label{sec:app_b}
\subsection{\moltwelveco}
\begin{figure*}[h]
    \centering
    \begin{overpic}[width=0.9174364896\textwidth]{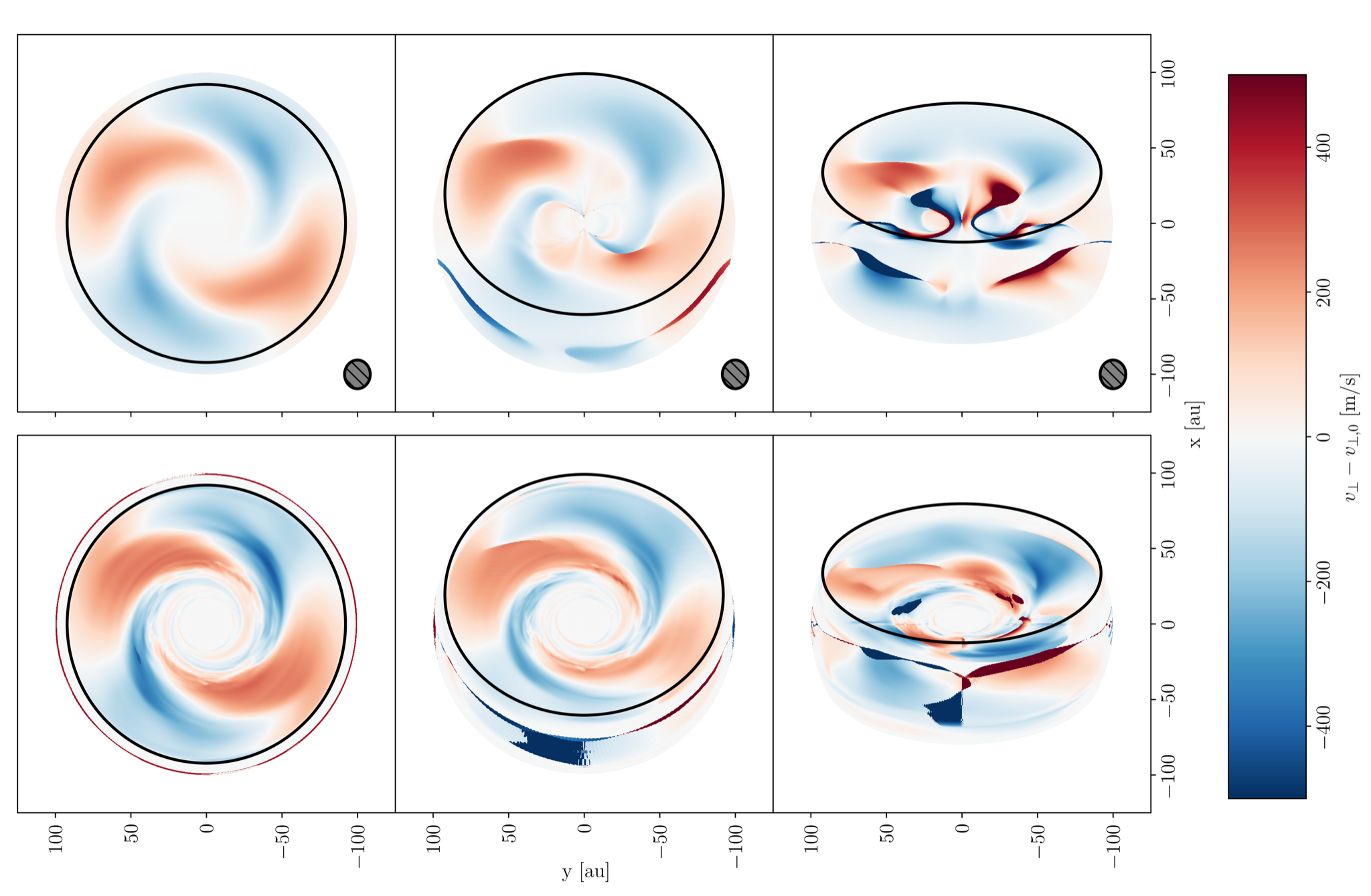}
    \put (3, 59) {\Large \moltwelveco \ $J = 2-1$, shadow}
    \end{overpic}
    
    \begin{overpic}[width=0.9174364896\textwidth]{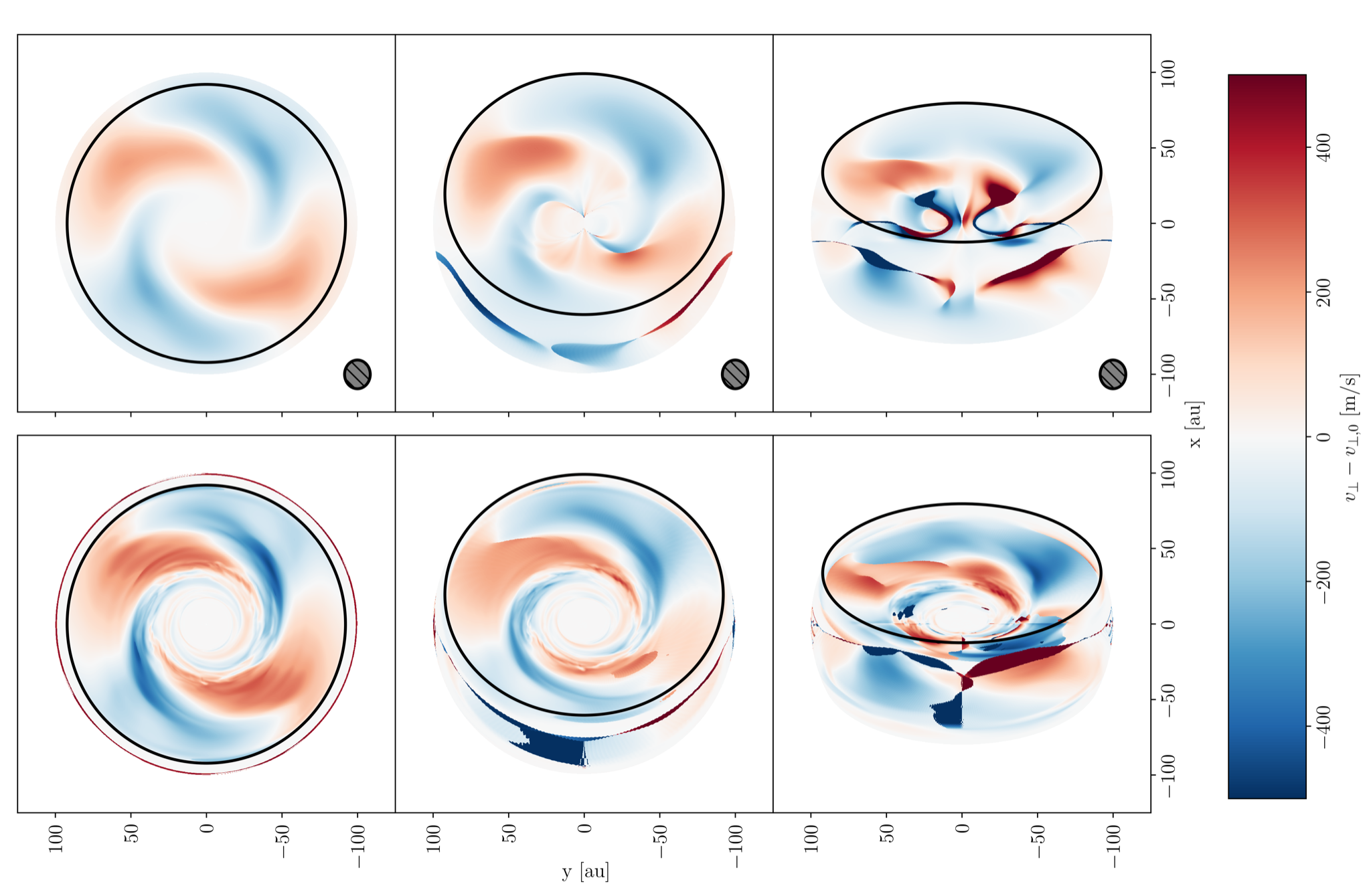}
    \put (3, 59) {\Large \moltwelveco \ $J = 3-2$, shadow}
    \end{overpic}    
    \caption{Kinematic moment-1 (line-of-sight velocity) maps of shadow-driven spirals in \moltwelveco. Upper two rows are the $J = 2-1$ transition, while the lower two are $J = 3-2$. From left to right, we plot disk inclinations of 0$^\circ$, 30$^\circ$, and 60$^\circ$. }
    \label{fig:_12co_.shadow}
\end{figure*}

\subsection{\molthirteenco}
\begin{figure*}[h]
    \centering
    \begin{overpic}[width=0.9174364896\textwidth]{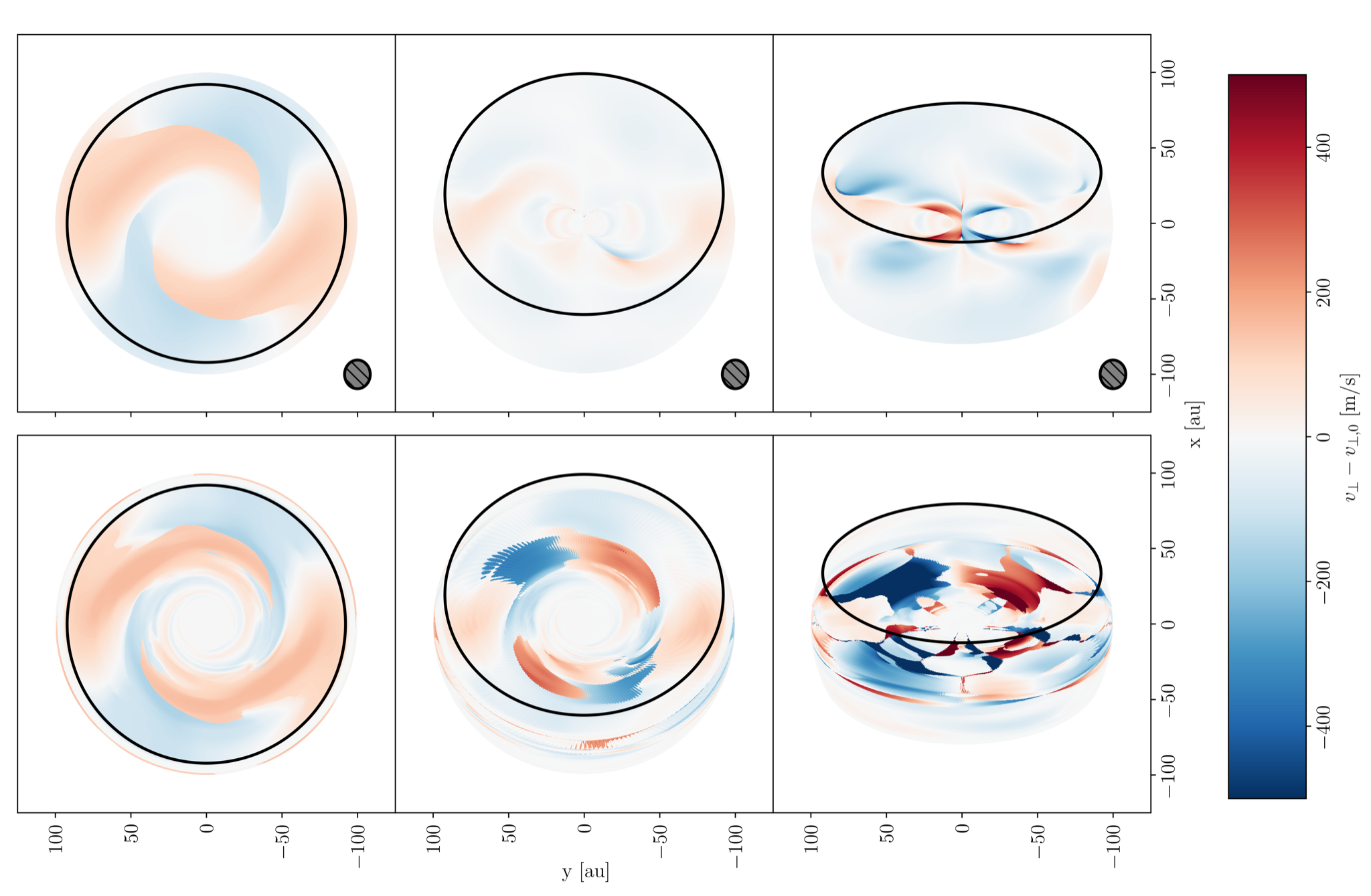}
     \put (3, 59) {\Large \molthirteenco \ $J = 2-1$, shadow}
    \end{overpic}
    \begin{overpic}[width=0.9174364896\textwidth]{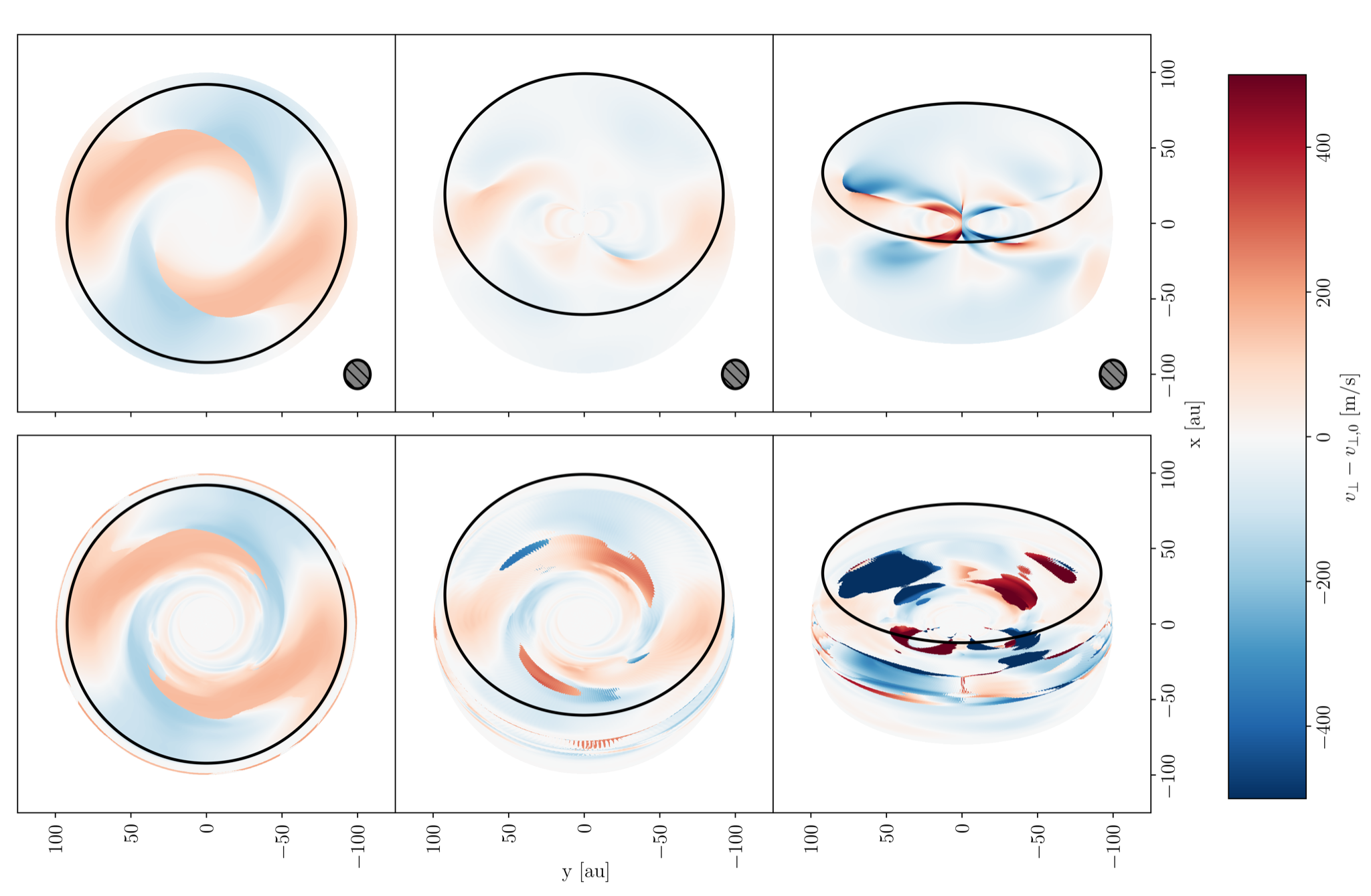}
    \put (3, 59) {\Large \molthirteenco \ $J = 3-2$, shadow}
    \end{overpic}
    \caption{Kinematic moment-1 (line-of-sight velocity) maps of shadow-driven spirals in \molthirteenco. Upper two rows are the $J = 2-1$ transition, while the lower two are $J = 3-2$. From left to right, we plot disk inclinations of 0$^\circ$, 30$^\circ$, and 60$^\circ$. }
    \label{fig:_13co_.shadow}
\end{figure*}

\subsection{\molceighteeno}
\begin{figure*}[h]
    \centering
    \begin{overpic}[width=0.9174364896\textwidth]{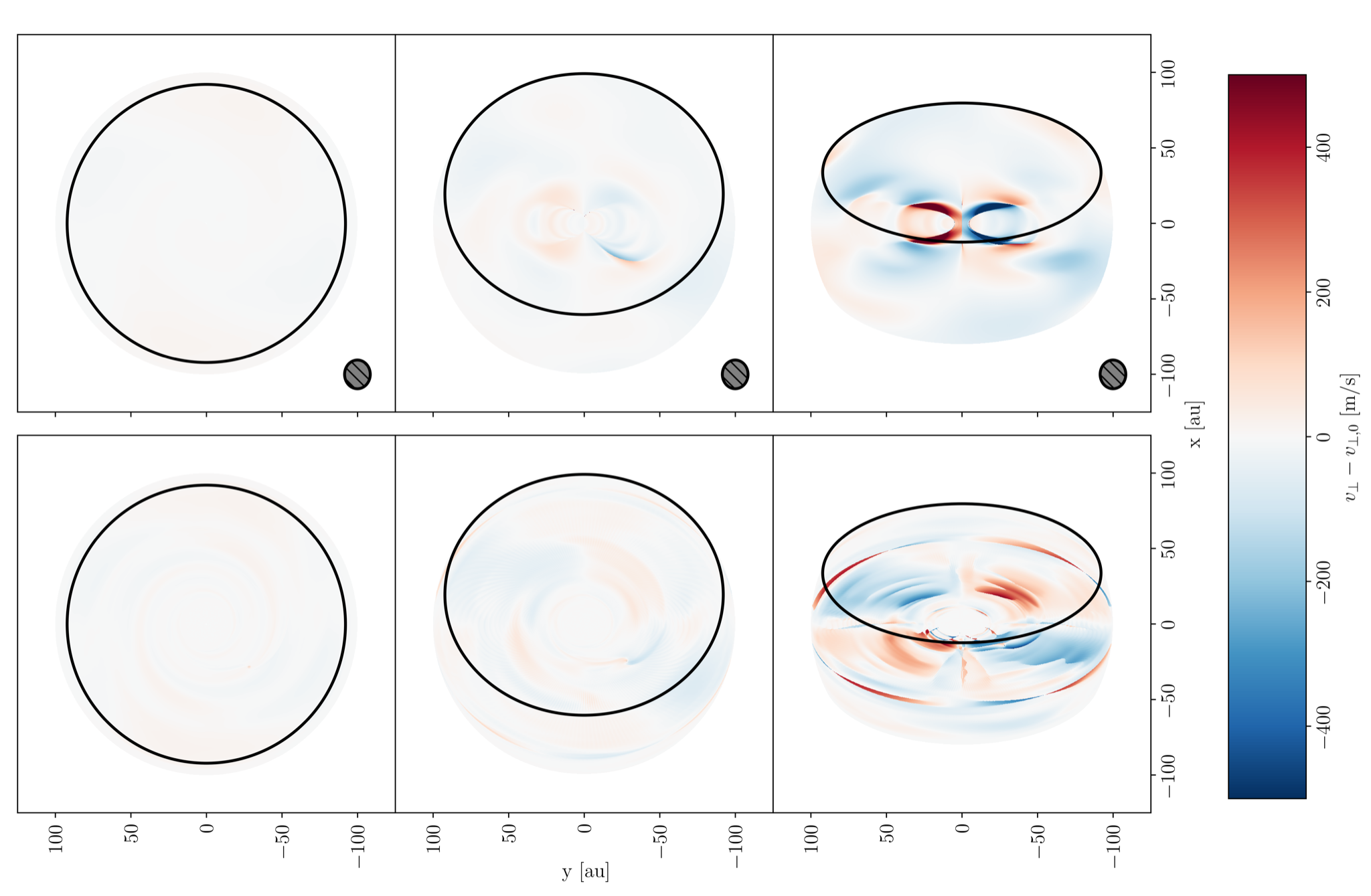}
    \put (3, 59) {\Large \molceighteeno \ $J = 2-1$, shadow}
    \end{overpic}
    \begin{overpic}[width=0.9174364896\textwidth]{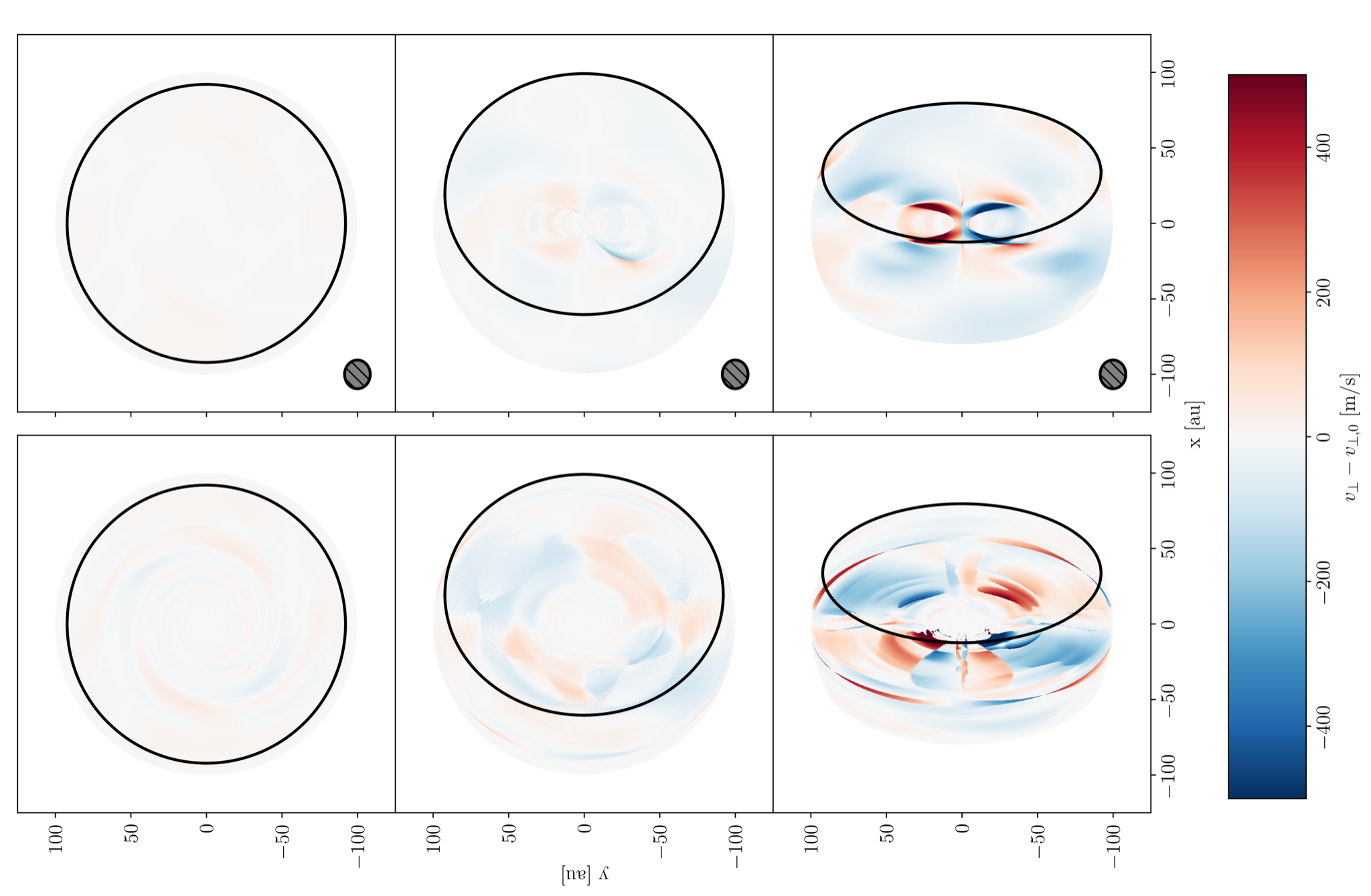}
    \put (3, 59) {\Large \molceighteeno \ $J = 3-2$, shadow}
    \end{overpic}
    \caption{Kinematic moment-1 (line-of-sight velocity) maps of shadow-driven spirals in \molceighteeno. Upper two rows are the $J = 2-1$ transition, while the lower two are $J = 3-2$. From left to right, we plot disk inclinations of 0$^\circ$, 30$^\circ$, and 60$^\circ$. }
    \label{fig:_c18o_.shadow}
\end{figure*}

\section{Planet-driven spirals}
\label{sec:app_c}
\subsection{\moltwelveco}
\begin{figure*}[h]
    \centering
    \begin{overpic}[width=0.9174364896\textwidth]{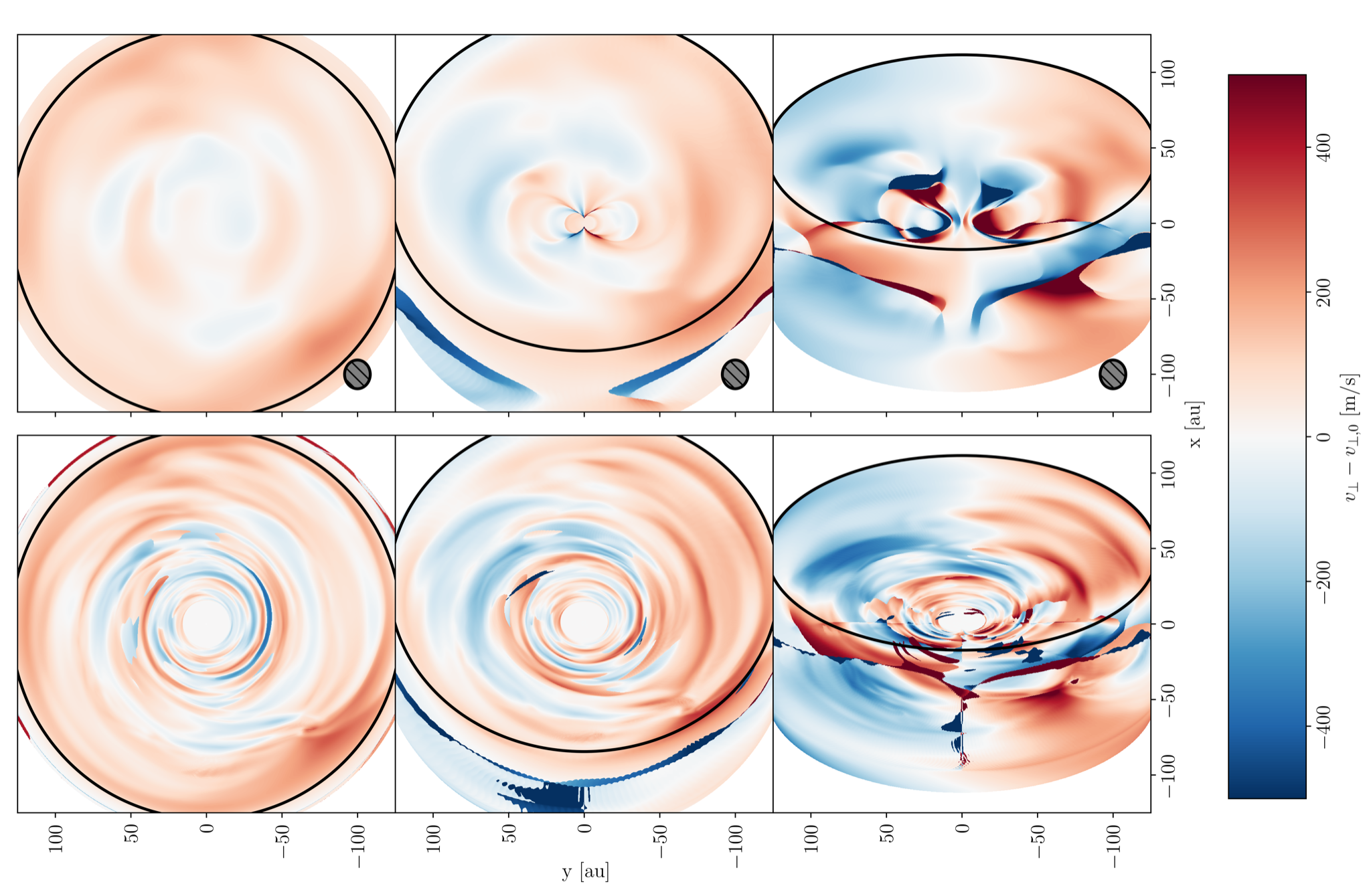}
    \put (3, 59) {\Large \moltwelveco \ $J = 2-1$, planet}
    \end{overpic}
    \begin{overpic}[width=0.9174364896\textwidth]{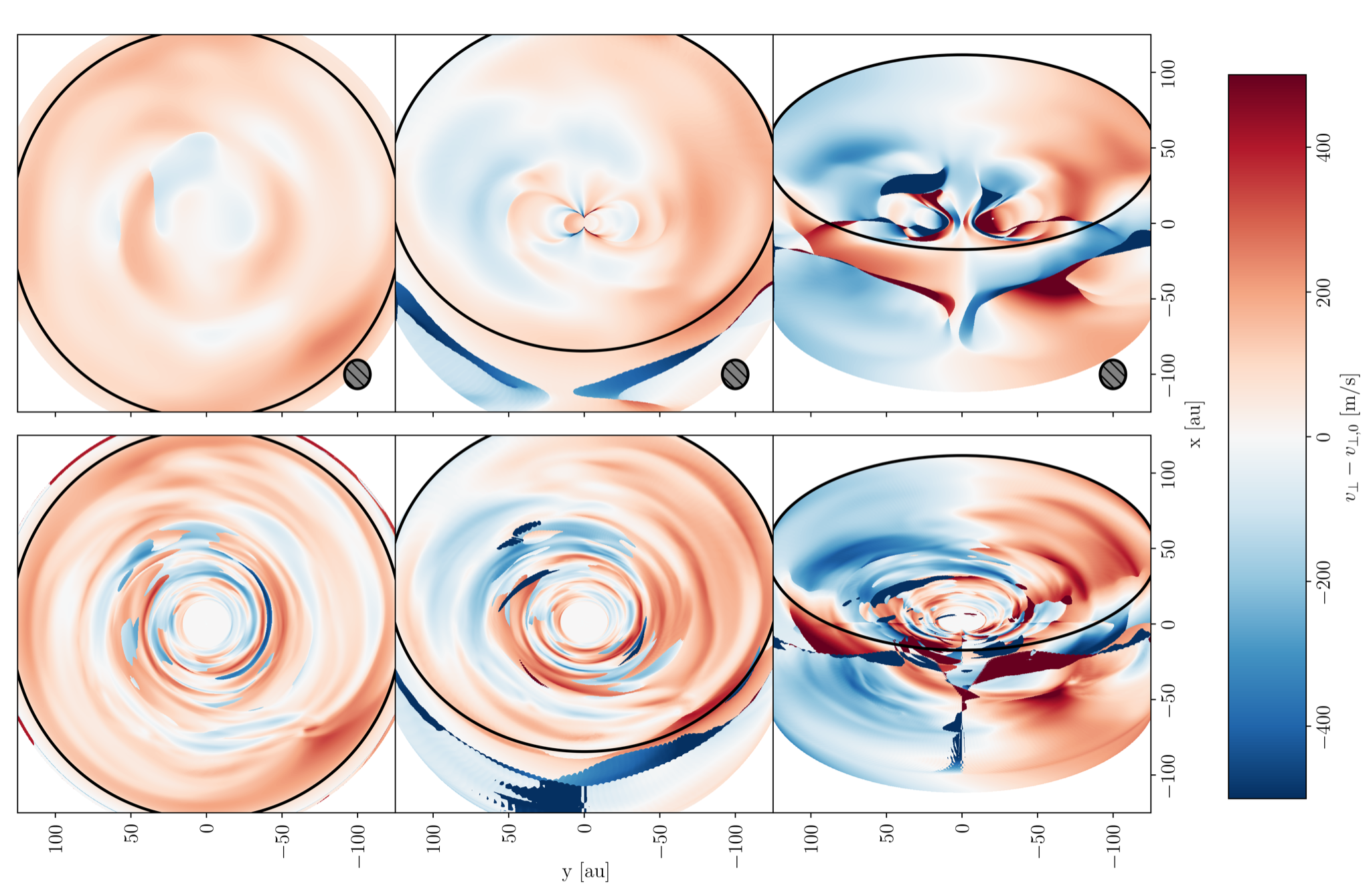}
    \put (3, 59) {\Large \moltwelveco \ $J = 3-2$, planet}
    \end{overpic}
    \caption{Kinematic moment-1 (line-of-sight velocity) maps of planet-driven spirals in \moltwelveco. Upper two rows are the $J = 2-1$ transition, while the lower two are $J = 3-2$. From left to right, we plot disk inclinations of 0$^\circ$, 30$^\circ$, and 60$^\circ$. }
    \label{fig:_12co_.planet}
\end{figure*}
\subsection{\molthirteenco}
\subsection{\molceighteeno}
\end{document}